\documentclass[a4paper,11pt]{article}
\usepackage{jcappub}
\usepackage{lineno}
\usepackage{subcaption}
\usepackage{array}      
\usepackage{multirow}   
\usepackage{booktabs}  

\newcommand{\tightcell}[2][1.0]{
    \renewcommand{\arraystretch}{#1}
    \begin{tabular}[t]{@{}r@{}} #2 \end{tabular}
}
\graphicspath{ {./figures/} }

\newcommand{\ctightcell}[2][1.0]{
    \renewcommand{\arraystretch}{#1}
    \begin{tabular}[t]{@{}c@{}} #2 \end{tabular}
}

\newcommand{\beq}{\begin{equation}}
\newcommand{\eeq}{\end{equation}}
\newcommand{\bea}{\begin{eqnarray}}
\newcommand{\eea}{\end{eqnarray}}

\title{\boldmath Neutrino non-radiative decay in matter: constraints and prospects}

\author[a]{Pilar Iváñez-Ballesteros}
\author[b,1]{and Maria Cristina Volpe\note{Corresponding author.}}
\affiliation[a]{Université Paris Cité, Astroparticule et Cosmologie,\\ F-75013 Paris, France}
\affiliation[b]{CNRS, Université Paris Cité, Astroparticule et Cosmologie,\\ F-75013 Paris, France}
\emailAdd{ivanez@apc.in2p3.fr,volpe@apc.in2p3.fr}

\abstract{Neutrinos, being massive, can decay. A heavier neutrino could decay into a lighter one and a massless scalar or pseudoscalar boson, such as the Majoron. Two-body non-radiative decay could occur in dense matter, such as in the inner dense regions of a core-collapse supernova. We first derive novel bounds on neutrino-Majoron couplings using the spectral distortions induced by neutrino non-radiative two-body decay in matter,
and two-dimensional likelihood analyses of the 24 $\bar{\nu}_e$ events from SN1987A. We then explore the prospects of neutrino-Majoron couplings from a future galactic core-collapse supernova, leaving either a neutron star or a black-hole. To this aim, we use information from detailed one-dimensional supernova simulations. We consider the supernova neutrino signal associated with inverse-beta decay in the upcoming JUNO and Hyper-Kamiokande detectors, with neutrino-argon scattering in DUNE, or with coherent neutrino-nucleus scattering in the DARWIN experiment. In a full 3$\nu$ framework, based on the spectral distortions induced by neutrino decay in matter, we perform two-dimensional likelihood analyses and provide prospects for the limits on neutrino-Majoron couplings.  Our results show that the observation of a future supernova will significantly improve on the current bounds, in particular from SN1987A and neutrinoless double-beta decay. Finally, we explore the impact of neutrino decay in matter
on the diffuse supernova neutrino background formed by past supernova explosions. We show for the first time that the effects on black-hole contributions are important and modify the DSNB number of events by several tens of percent in Hyper-Kamiokande.}

\begin{document}
\maketitle
\flushbottom

\section{Introduction} \label{sec:intro}
Neutrinos are unique elementary particles with mixing. Weakly interacting, they tell us about the interior of the Sun and of massive
stars that undergo gravitational core-collapse. They are tightly linked
to stellar nucleosynthesis processes and left an imprint on primordial nucleosynthesis, one second after the Big Bang. 
Milestone observations in astrophysics were the pioneering experiment of R. Davis on solar neutrinos \cite{Davis:1968cp} and the detection of SN1987A neutrino events, i.e., of the first neutrinos from the core-collapse of a massive star \cite{Kamiokande-II:1987idp,Bionta:1987qt,Alekseev:1988gp}. GW170817 was also unique, bringing the first measurement of gravitational waves concomitantly with a short gamma-ray burst and a kilonova \cite{LIGOScientific:2017ync}.
A new window in neutrino astrophysics was opened with the detection of the first PeV neutrinos in the IceCube experiment \cite{Aartsen:2014gkd} 
and, very recently, with the measurement of a cosmic neutrino event, with the highest energy ever observed, by the KM3NeT Collaboration \cite{KM3NeT:2025npi}.

After the breakthrough discovery of neutrino oscillations \cite{Fukuda:1998mi}, solar, reactor, and accelerators
have paved our knowledge of the neutrino oscillation parameters \cite{ParticleDataGroup:2024cfk}. Still, many important open questions remain.
Among the open questions is how neutrinos change flavor in dense environments. Indeed, neutral-current neutrino-neutrino interactions become sizable
in such sites, and make neutrino propagation a non-linear many-body problem \cite{Pantaleone:1992eq}. Moreover, collisions, shock wave, 
and turbulence effects also add to the complexity. In this context, important progress is being made to assess how neutrinos change flavor in dense environments (see \cite{Volpe:2023met,Volpe:2015rla,Tamborra:2020cul,Duan:2010bg} for reviews).
Astrophysical and cosmological neutrinos provide complementary avenues to put tight constraints
on unknown neutrino properties and new physics.  Neutrinos are also tightly connected to dark
matter searches in several ways and constitute a background for dark matter experiments, often referred to as the ``neutrino floor".
Interestingly, the PandaX-4T \cite{PandaX:2024muv} and XENONnT \cite{XENON:2024ijk} dark matter experiments found the first identification of $^8$B solar neutrinos.

Core-collapse supernovae are among the most powerful neutrino sources. During their explosions, about 3$\times 10^{53}$ erg of gravitational
binding energy is taken away by neutrinos of all flavors in a ten-second burst. Core-collapse supernovae are rare events in our Galaxy. A combined
analysis of different estimates of the core-collapse supernova rate gives a mean time of occurrence of $61 ^{+ 24}_{- 14}$ years \cite{Rozwadowska:2020nab},
Past supernovae in our Universe constitute the diffuse supernova neutrino background (DSNB) of all species, which is actively searched. A combined analysis of the ensemble of Super-Kamiokande data (SK) shows a hint at $1.8 \sigma$ from the model-dependent analysis \cite{Super-Kamiokande:2021jaq,Beauchene}. Significance is now at the level of  2.3$\sigma$ with the first results with the inclusion of Gadolinium, an idea first suggested in ref.~\cite{Beacom:2003nk} with the goal of suppressing backgrounds through better neutron tagging. 

\begin{figure*}[t]    \centering
    \includegraphics[scale=0.5]{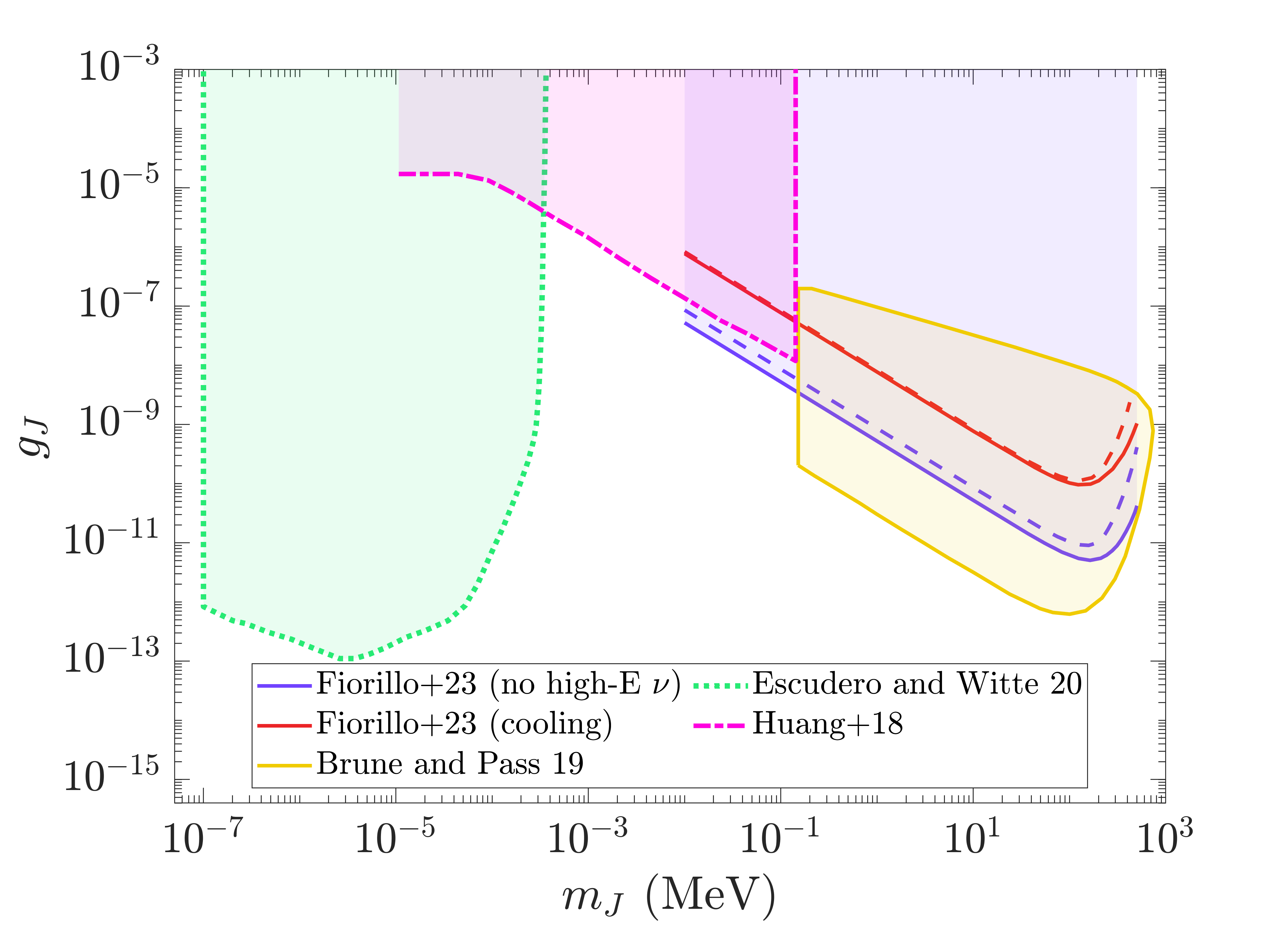}
     \caption{Current limits on the neutrino-Majoron coupling $g_J$ as a function of the Majoron mass $m_J$. The bounds shown are obtained from: i) the absence of 100 MeV neutrinos from the decay of Majoron-like bosons (more stringent), or from the luminosity argument (less stringent) using a ``hot" (solid line) and a ``cold" (dashed line) supernova model \cite{Fiorillo:2022cdq}; ii) the luminosity argument applied to SN1987A neutrino events (closed contour) \cite{Brune:2018sab}; iii) CMB observations (dotted lines) \cite{Escudero:2019gvw}; iv) Big Bang nucleosynthesis  (dot-dashed contour) \cite{Huang:2017egl}.}
  \label{fig:limits}
\end{figure*}

Unknown neutrino properties include neutrino non-radiative two-body decay. Core-collapse supernovae and cosmology have a unique sensitivity to long lifetime-to-mass ratios for this process (see figure 1 of ref.~\cite{Ivanez-Ballesteros:2023lqa}). 
In particular, neutrino decay in vacuum or in matter into a massive Majoron, or a Majoron-like particle, was investigated based on CMB observations 
\cite{Escudero:2019gvw}, using core-collapse supernovae as neutrino sources \cite{Brune:2018sab}, considering the DSNB produced by past supernovae \cite{Fogli:2004gy,DeGouvea:2020ang,Ivanez-Ballesteros:2022szu,Roux:2024zsv,Martinez-Mirave:2024hfd}, or the 24 $\bar{\nu}_e$ events from SN1987A \cite{Kolb:1987qy,Fiorillo:2022cdq,Ivanez-Ballesteros:2023lqa}. Current limits on neutrino decay into a massive Majoron are summarized in figure \ref{fig:limits}. 
Note that bounds for massless Majorons (not shown) are provided by analyses based on the neutrino events from SN1987A \cite{Kachelriess:2000qc,Tomas:2001dh,Farzan:2002wx,Ivanez-Ballesteros:2024nws}. 

Concerning the DSNB, ref.~\cite{Ivanez-Ballesteros:2023lqa} provided the first investigation in a full 3$\nu$ framework including astrophysical uncertainties from the evolving core-collapse supernova rate. DSNB predictions in the absence and in the presence of decay are found to be degenerate for normal neutrino mass ordering, while for inverted mass ordering, the DSNB rates are strongly suppressed by neutrino decay. Ref.~\cite{Roux:2024zsv} provided the first Bayesian analysis combining different
DSNB detection channels in SK, JUNO, DUNE, and Hyper-Kamiokande (HK) in a $3 \nu$ framework, 
with the goal of exploring the capability to discriminate DSNB predictions without and with
neutrino non-radiative two-body decay in vacuum. Such an analysis included the uncertainty in the evolving core-collapse
supernova rate, and used neutrino fluxes from the Garching and Nakazato one-dimensional core-collapse supernova simulations. The results
showed that, for the case of normal mass ordering, discriminating between the case with decay and without decay requires new avenues. 

Numerous constraints on neutrino decay were also derived using the 24 $\bar{\nu}_e$ events from SN1987A, recorded in the 
Kamiokande \cite{Kamiokande-II:1987idp}, Irvine-Michigan-Brookhaven (IMB) \cite{Bionta:1987qt} detectors and Baksan Scintillator Telescope (BST) \cite{Alekseev:1988gp}. These detectors were operating\footnote{Note that the Mont Blanc Liquid Scintillator Detector (LSD) recorded 5 events several hours before others \cite{Aglietta:1987it}.} when the supernova exploded, in the Large Magellanic Cloud, on the 27th of February 1987. Tight bounds were obtained concerning neutrino radiative decay \cite{ParticleDataGroup:2024cfk}. 
Moreover, ref.~\cite{Ivanez-Ballesteros:2023lqa} performed the first likelihood analysis of the SN1987A events, including the spectral distortion from non-radiative decay in vacuum.  For inverted neutrino mass ordering, the analysis brought the lower bound of $2.4~(1.2) \times 10^5$ eV/s at 68$\%$ ($90\%$) CL  for the $\nu_2$ and $\nu_1$ neutrino mass eigenstates. This limit is tighter than previous ones, and competitive with cosmological ones (see e.g. \cite{Chen:2022idm}).
 
Neutrinos could also decay in matter. This possibility was studied in, e.g., refs.~\cite{Kachelriess:2000qc,Tomas:2001dh,Fuller:1988ega,Suliga:2024nng}. Neutrino two-body decay can produce a pseudoscalar Goldstone boson, the Majoron, associated with total lepton number violation.
Majorons were first introduced in a singlet model \cite{Chikashige:1980qk} and then in a triplet model \cite{Gelmini:1980re}, now excluded. 
In relation to SN1987A, limits on neutrino-Majoron couplings were discussed either using the cooling argument, or based on a deleptonization argument that exploits the ruled-out supernova prompt explosion mechanism \cite{Baron:1987zz}, as well as using neutrino spectral distortions from neutrino decay \cite{Kachelriess:2000qc,Tomas:2001dh}. Ref. \cite{Farzan:2002wx} refined the limits based on the cooling argument, whereas ref.~\cite{Giunti:1992sy} discussed decay rates of neutrino-Majoron decay in matter. 
Moreover, in our previous work ref.~\cite{Ivanez-Ballesteros:2024nws}, we performed the first likelihood analysis of the spectral distortion of 
the SN1987A neutrino events due to neutrino-Majoron decay in matter. The analyses provided novel bounds on neutrino-Majoron couplings
that go beyond the existing limits from SN1987A, and are competitive with those obtained from neutrinoless double-beta decay
or several orders of magnitude better than the ones from lepton and meson decays.

In this manuscript, we present new bounds on neutrino-Majoron couplings, considering neutrino non-radiative two-body decay in matter, first considering
SN1987A neutrino events, and then focusing on future observations of neutrinos from the next galactic core-collapse supernova and the DSNB. 
Following the theoretical framework developed in our previous work \cite{Ivanez-Ballesteros:2024nws}, we consider the impact on the supernova neutrino signal of neutrino two-body non-radiative decay in matter, into a massless (or almost massless) Majoron or Majoron-like particle. To this aim, we consider that the matter and decaying eigenstates coincide, and use detailed information from one-dimensional supernova simulations to calculate the neutrinosphere, transportsphere, and energysphere radii. These quantities are necessary to determine the impact of neutrino decay and the consequent neutrino spectral distortion. 
We first perform a likelihood analysis of the neutrino spectral distortions induced by neutrino decay, considering the 24 $\bar{\nu}_e$ events coming from SN1987A detected in the Kamiokande, IMB, and Baksan experiments, thus extending our previous work. Such analyses include the detailed response functions for each of the detectors. 
We then focus on a future galactic supernova, for which we consider the following two cases. i) A core-collapse supernova giving a neutron star (NS) for which we use the simulations from the Garching group \cite{Garching,Fiorillo:2023frv} that leave 1.36 $M_{\rm \odot}$, 1.44 ${\rm  M}_{\rm \odot}$, 1.62 $M_{\rm \odot}$ with the DD2 or the SFHx neutron star equations of state (EOS). ii) A core-collapse supernova leaving a black-hole (BH) for which we employ Nakazato simulations for a 30 $M_{\odot}$ progenitor with the Lattimer-Swesty (L), the Shen (S), or the Togashi (T) EOS \cite{Nakazato:2012qf,Nakazato:2021gfi,data}. 
We present the impact of neutrino decay in matter on the core-collapse supernova neutrino fluxes, including also the Mikheev-Smirnov-Wolfenstein effect, and on the expected neutrino number of events in the upcoming JUNO, HK, and the more distant future DUNE detector, as well as the DARWIN experiment. We then perform likelihood analyses of the spectral distortions of such events due to decay, considering both the signal in detectors individually and then combining them. We present novel bounds on neutrino-Majoron couplings and discuss how the limits improve on current ones and vary with the core-collapse supernova distance, in the Milky Way.  
Finally, we investigate the role of neutrino decay in matter on the DSNB, and explore its impact in the JUNO, HK, DUNE, and DARWIN detectors.

The manuscript is structured as follows. Section~\ref{sec:theory} introduces the $3 \nu$ formalism employed to study neutrino non-radiative two-body decay in matter. 
Section~\ref{sec:models_fluxes} presents the core-collapse supernova neutrino fluxes used from the
Garching and Nakazato's one-dimensional simulations. Details are also given of the calculations of the neutrinospheres, the transportspheres, and
the energyspheres necessary to compute the fluxes when neutrinos decay in matter. Section~\ref{sec:limits_sn1987a} discusses the likelihood
analysis for the spectral distortion of SN1987A data in Kamiokande, IMB, and Baksan, and presents new results on the neutrino-Majoron
bounds. Section~\ref{sec:prospects} gives the expected number of events as well as the prospects on the neutrino-Majoron couplings 
using the neutrino signal from a future supernova in JUNO, HK, DUNE, and DARWIN.
Section~\ref{sec:dsnb} presents the theoretical framework used for the DSNB, and our predictions for the expected DSNB number of events in the presence and in the absence of decay. We conclude in section~\ref{sec:conclusions}. Finally, appendix~\ref{appendix:statistics} presents the statistical analysis we employ, and appendix~\ref{appendix:results} gives supplementary results.

\section{Theoretical framework} \label{sec:theory}
\noindent
Since neutrinos are massive particles, they could decay into a massless (or massive) scalar or pseudoscalar
boson such as a Majoron, or a Majoron-like particle. Thus, while we will refer to Majorons, our analysis also holds for any massless (or almost
massless) pseudoscalar boson. 

Neutrino-Majoron interactions can be described generally by the following Lagrangian

\begin{equation}\label{eq:lagrangian}
    \mathcal{L}_{\rm int} \propto \sum_{i,j} g_{ij}\bar\nu_i\gamma_5 \nu_j J  ~ ,
\end{equation}
in which $\nu_j$ (with $i, j = 1, 2, 3$) is the neutrino field and $J$ the Majoron field. In the simplest Majoron models \cite{Schechter:1981cv}, the neutrino-Majoron coupling matrix $g_{ij}$ is proportional to the neutrino mass matrix. Thus, the coupling matrix in the mass basis is diagonal and can be expressed as $g_{ij} \propto m_i  \delta_{ij}$ with $m_i$ the mass of the mass eigenstate $i$. Consequently, only one coupling is independent, and the remaining two are related to it by the values of the squared-mass differences measured in oscillation experiments. In our study, we will consider the theoretical framework of three active neutrino flavors. 

It is to be noted that the proportionality of the coupling matrix to the mass matrix implies a suppression of the decay $\nu_h \to \nu_l + J$ in vacuum, from a heavier $\nu_h$ to a lighter $\nu_l$ neutrino. Interestingly, in matter, the interactions of neutrinos with the particles in the medium induce effective neutrino masses breaking this proportionality~\cite{Berezhiani:1987gf}. Thus, off-diagonal terms $g_{ij}$ ($i \neq j$) appear in the coupling matrix, allowing new interaction channels that can enhance neutrino decay in matter to Majorons. 

Thus, in our analysis, there is only one independent coupling. For the case of normal neutrino mass ordering, $g_{11}$ and $m_1$ will be our fit parameters, with the other two couplings $g_{22}$ and $g_{33}$ determined following the relations

\begin{equation}\label{eq:g_NO}
    g_{22} = g_{11} \sqrt{1+ \frac{\Delta m^2_{21}}{m^2_1}},~~~
    g_{33} = g_{11} \sqrt{1+ \frac{\Delta m^2_{31}}{m^2_1}} \ .
\end{equation}

Similarly, in the case of inverted neutrino mass ordering, $g_{33}$ and $m_3$ will be our fit parameters, with $g_{11}$ and $g_{22}$ derived through analogous expressions

\begin{equation} \label{eq:g_IO}
    g_{11} = g_{33} \sqrt{1+ \frac{|\Delta m^2_{31}|}{m^2_3}},~~~
    g_{22} = g_{33} \sqrt{1+ \frac{|\Delta m^2_{32}|}{m^2_3}} \ .
\end{equation}
\noindent

\subsection{The neutrino Hamiltonian} \label{sec:interactions}
\noindent
We introduce here the 3$\nu$ theoretical framework describing neutrino evolution in matter.
The Hamiltonian describing neutrino propagation in a core-collapse supernova, including neutrino decay, can be expressed as

\begin{equation}\label{eq:H}
    H = H_0 + H_{\rm matter} + H_{\rm int} \ , 
\end{equation}
the first is the vacuum term, the second accounts for the interactions with the particles in the medium, and the last implements neutrino-Majoron interactions eq.~\eqref{eq:lagrangian}. In dense environments, such as a core-collapse supernova, the neutrino-neutrino interactions, shock wave effects, and turbulence might also influence the neutrino spectra (see, for example, \cite{Volpe:2023met,Volpe:2015rla,Tamborra:2020cul,Duan:2010bg} for reviews). In particular, the neutral-current neutrino-neutrino interactions make neutrino evolution a complex weakly interacting many-body problem, which is still under active investigation. Therefore, in the present analysis, we only consider neutrino-matter contributions responsible for the established 
Mikheev-Smirnov-Wolfenstein effect \cite{Wolfenstein:1977ue,Mikheev:1986wj}. 

The first contribution to the neutrino Hamiltonian is responsible for the well-established phenomenon of neutrino vacuum oscillations.
In the flavor basis, it reads  
\beq \label{eq:Hvac}
H_0 = U H_{\rm vac} U^{\dagger} \ , 
\eeq
with $H_{\rm vac} = diag(E_i)$ ($i=1,3$) being the propagation Hamiltonian of the mass eigenstates, and $E_i$ the corresponding neutrino energies.
Since neutrinos are ultrarelativistic (in the equal momentum approximation), the elements of the vacuum Hamiltonian read
\begin{equation}\label{eq:rel}
H_{vac, ij} = \left[E + \frac{m_i^2}{2E} \right] \delta_{ij} \ , 
\end{equation}
where $E = \vert \vec{p} \vert$, with $\vec{p} $ the neutrino momentum and $ \delta_{ij}$ the Kronecker delta.

The unitary matrix $U$ ($U^{\dagger} = U^{-1}$) is the Pontecorvo-Maki-Nakagawa-Sakata (PMNS) matrix relating the mass and flavor bases 
\begin{equation}\label{eq:flms}
|\nu_\alpha \rangle = U^*_{\alpha i} |\nu_i \rangle \ ,
\end{equation} 
(a summation on the i index is subtended).
For 3$\nu$ flavors, the PMNS matrix is parametrized as\footnote{Note that we do not consider here the two Majorana phases that are not relevant in our analyses.} \cite{ParticleDataGroup:2024cfk}
\begin{equation} \label{eq:PP}
U =  \left(
\begin{tabular}{ccc}
$c_{12} c_{13}$  & $s_{12} c_{13} $   &$ s_{13} e^{-i \delta}$ \\
$ - s_{12} c_{23} - c_{12} s_{13} s_{23}  e^{i \delta} $ & $ c_{12} c_{23} - s_{12}  s_{13} s_{23} e^{i \delta} $  &$ c_{13}  s_{23} $ \\
$s_{12} s_{23} - c_{12}  s_{13} c_{23} e^{i \delta}$  &$  - c_{12} s_{23} - s_{12} s_{13} c_{23} e^{i \delta}  $ &  $ c_{13}  c_{23}$ \\
\end{tabular}
\right) \, ,
\end{equation}
with $c_{ij} = \cos \theta_{ij}$, $s_{ij} = \sin \theta_{ij}$ ($i, j = 1, 3$) and $\delta$ the Dirac CP violating phase that we set to zero in our work.

 As neutrinos propagate in the supernova layers, they interact with the electrons, protons, and neutrons composing the 
 medium. Electron neutrinos $\nu_e$ and antineutrinos $\bar\nu_e$ interact through both charged-current (CC) and neutral-current (NC) interactions, whereas
 non-electron neutrinos $\nu_x$ and antineutrinos $\bar\nu_x$ ($x = \mu,~\tau$) mostly experience NC interactions\footnote{We neglect here the presence of a finite density of muons in the core-collapse supernova core, that breaks the degeneracy between muon and tau neutrinos (and antineutrinos) \cite{Bollig:2017lki}.}. The corresponding potentials, in the mean-field approximation, 
are given by
\begin{equation} \label{eq:potential}
\begin{split}
    V_{\rm CC} &= \sqrt{2} G_F n_B (Y_e + Y_{\nu_e}), \\
    V_{\rm NC} &= \sqrt{2} G_F n_B (-\frac{1}{2}Y_n + Y_{\nu_e}),
\end{split}
\end{equation}
\noindent
with $G_F$ the Fermi constant, $n_B$ the baryon density, and $Y_i = (n_i - \bar n_i)/n_B$ the particle number fraction\footnote{Note that in our computations we set $Y_{\nu}$ =0.} with $i = e$ (electrons), $\nu_e$, and $n$ (neutrons). For antineutrinos, these potentials have opposite signs. Thus, in the flavor basis, the mean-field neutrino-matter Hamiltonian is
\begin{equation}\label{eq:Vab}
    V_{\alpha\beta} = 
    \begin{pmatrix}
        V_{\rm CC} + V_{\rm NC} & 0 & 0 \\
        0 & V_{\rm NC} & 0 \\
        0 & 0 & V_{\rm NC}
    \end{pmatrix} \ .
\end{equation}
\noindent
Putting the different contributions together, the neutrino Hamiltonian eq.~\eqref{eq:H} becomes
\begin{equation}\label{eq:Ham}
    H = E \mathbb{I} + \frac{1}{2E} U \mathrm{M}^2 U^\dagger + V_{\alpha\beta} \ ,
\end{equation}
where $ \mathbb{I} $ is the identity matrix.

\subsection{Neutrino non-radiative two-body decay in matter}
\noindent
In our treatment of neutrino two-body non-radiative decay in matter, we extend the framework of Refs.\cite{Kachelriess:2000qc,Tomas:2001dh}.
To discuss the effect of neutrino decay in matter, we use the matter basis that instantaneously diagonalizes the neutrino Hamiltonian \eqref{eq:H}, including the vacuum and the matter contributions (only), through the following transformation
\begin{equation}\label{eq:Htilde}
    \Tilde{H} = \Tilde{U} H \Tilde{U}^{\dagger}  \ , 
\end{equation}
where $\Tilde{U}$ is the mixing matrix in matter, relating the flavor and matter bases 
\begin{equation}\label{eq:matterbas}
| \nu_{\alpha} \rangle = {\Tilde U}_{\alpha i}  | \Tilde{\nu}_i \rangle \ , 
\end{equation}
Note that $\Tilde{U}$ depends on the helicity of neutrinos, i.e. negative (positive) for neutrinos (antineutrinos) and on the neutrino mass ordering.

For illustration purposes, let us discuss the case of $2 \nu$ flavors. In this case, the matrices $U = R(\theta)$ and $\Tilde{U} = R(\tilde{\theta})$ are rotation matrices parametrized by the  $\theta$ and $\Tilde{\theta}$ angles. The diagonalization of $H$ eq.~\eqref{eq:Htilde} requires
\begin{equation} \label{eq:diagonal}
    \tan 2\Tilde{\theta} =  \frac{ \sin 2\theta}{\cos 2\theta - 2E V_{\rm CC} / \Delta m^2} \ , 
\end{equation}
with $\Delta m^2 = m^2_2 - m^2_1$ the squared-mass difference. At very high densities, $|V_{\rm CC}| \to \infty$, the matter mixing angle is "suppressed", i.e. $\Tilde{\theta} \to \pi/2~(0)$ for neutrinos (antineutrinos), so that the matter and the flavor eigenstates essentially coincide. 

For $3\nu$ flavors, the mixing matrix in matter can be parametrized as 
\begin{equation}
    \Tilde{U} = \Tilde{U}_{12}\Tilde{U}_{13}\Tilde{U}_{23},
\end{equation}
where $\Tilde{U}_{ij} = \Tilde{U}_{ij}(\Tilde{\theta}_{ij})$ are rotation matrices in the $i$-$j$ plane by an angle $\Tilde{\theta}_{ij}$ (we fixed $\delta = 0^{\circ}$). Since the neutrino mass ordering can be normal or inverted, because the sign of $\Delta m^2_{31}$ is unknown, the sign of the second term in the denominator of eq.~(\ref{eq:diagonal}) can be positive or negative, thus affecting the value of $\Tilde{\theta}_{13}$. 
Table~\ref{tab:eigenstates} shows the correspondence, at high matter densities, between the matter and flavor eigenstates, considering both normal and inverted mass ordering. 

In general, the coupling matrix in the matter basis is related to the one in the flavor basis through

\begin{equation}\label{eq:gij}
    \Tilde{g}_{ij}^{h_i \to h_j} =
    \Tilde{U}_{i \alpha}^{(h_i) \dagger} g_{\alpha\beta} \Tilde{U}_{\beta j}^{(h_j)},
\end{equation}
where we added the notation $h_{i,j}$ here to indicate the neutrino helicity. At high densities, the coupling matrix in the matter basis can be approximated by the one in the rotated flavor basis \cite{Kachelriess:2000qc, Tomas:2001dh}:

\begin{equation}\label{eq:gijrel}
    \Tilde{g}_{ij} \to U_{23}(\theta_{23}) g_{\alpha\beta} U_{23}(-\theta_{23}) = g_{\alpha^\prime\beta^\prime} \ .
\end{equation}

\begin{table}[t]
    \setlength{\tabcolsep}{8pt} 
    \renewcommand{\arraystretch}{1.5} 
    \centering
\begin{tabular}{ccccccc}
\toprule
                         & $\Tilde{\nu}_1^{(-)}$ & $\Tilde{\nu}_2^{(-)}$ & $\Tilde{\nu}_3^{(-)}$ & $\Tilde{\nu}_1^{(+)}$   & $\Tilde{\nu}_2^{(+)}$  & $\Tilde{\nu}_3^{(+)}$   \\ \cmidrule(r){2-4}\cmidrule(l){5-7}
NO & $\nu_{\mu^\prime}$    & $\nu_{\tau^\prime}$   & $\nu_e$               & $\bar\nu_e$             & $\bar\nu_{\mu^\prime}$ & $\bar\nu_{\tau^\prime}$ \\ 
IO & $\nu_{\mu^\prime}$    & $\nu_e$               & $\nu_{\tau^\prime}$   & $\bar\nu_{\tau^\prime}$ & $\bar\nu_{\mu^\prime}$ & $\bar\nu_e$             \\ \bottomrule
\end{tabular}
\caption{Correspondence between the matter and the flavor eigenstates in a $3 \nu$ flavor framework, for both normal (NO) and inverted (IO) neutrino mass orderings. The superscript in the matter eigenstates indicates the helicity, namely $(-)$ for neutrinos and $(+)$ for antineutrinos. The prime indicates a rotated basis in the $\nu_\mu$--$ \nu_\tau$ subspace.}
\label{tab:eigenstates}
\end{table}

From Eqs.\eqref{eq:gij}-\eqref{eq:gijrel}, we observe that the coupling matrix in matter is no longer diagonal and that, deep in the stellar interior, a Majoron-like particle couples to flavor neutrino eigenstates. 
In the relativistic approximation, only helicity-flipping neutrino decays are allowed with the total decay rate of the process $\nu_\alpha^{(\pm)} \to \nu_\beta^{(\mp)} + J$ given by \cite{Kachelriess:2000qc,Berezhiani:1987gf}

\begin{equation}\label{eq:decayrate}
    \Gamma_{\alpha\beta} = \frac{g_{\alpha\beta}^2}{16\pi}\Delta V_{\alpha\beta} \ ,    
\end{equation}
with $\alpha, \beta = e, \mu^\prime, \tau^\prime$ (the prime indicates the rotated basis). This process is only allowed if 
\begin{equation} \label{eq:decaycond}
\Delta V_{\alpha\beta} = V_\alpha - V_\beta > 0 \ .
\end{equation}
The energy distribution of the daughter neutrinos is given by the following expression \cite{Berezhiani:1989za}

\begin{equation} \label{eq:energy}
    \psi(E_\alpha, E_\beta) = \frac{2}{E_\alpha} \left( 1 - \frac{E_\beta}{E_\alpha} \right) \ ,
\end{equation}
where $E_\alpha$ and $E_\beta$ correspond to the energies of the initial and final neutrino, respectively.

\subsection{Neutrino fluxes in presence of neutrino-Majoron interactions}
\noindent
After being produced in the supernova core, neutrinos can undergo decay in matter when the condition \eqref{eq:decaycond} is fulfilled. 
To treat the influence of decay,
in contrast with ref.~\cite{Kachelriess:2000qc} we employ a theoretical framework that implements the time dependence of the neutrino emission in the core-collapse supernova. Interestingly, the computation of the impact of neutrino decay includes a term in the survival probability, first pointed out in ref.~\cite{Kachelriess:2000qc} 
(see eq.~\eqref{eq:survival_nux}) that involves an effective neutrino velocity and dominates over the constant term previously considered. 
The early work \cite{Kachelriess:2000qc} evaluated its contribution by assuming time-independent density profiles using inputs from ref.~\cite{Nunokawa:1997ct} (at time $t=6$ s).
In our framework, we perform a detailed calculation of this term, of the associated mean-free paths for the different microscopic processes that influence the neutrino species.

In the models we employ and the relevant radii, the condition  \eqref{eq:decaycond} is fulfilled only for antineutrinos, and not for neutrinos.
Therefore, neutrinos will be treated as stable. 
The survival probability of $\bar\nu_e$ with energy $E_\nu$ emitted at the neutrinosphere is given by \cite{Kachelriess:2000qc}
\begin{equation} \label{eq:survival_nue}
    N_{\bar\nu_e} (E_\nu) = \exp\Biggl\{ \int_{R_{\rm NS}}^{\infty} 
    dr^\prime~\Gamma_{\bar\nu_e}(r^\prime)
    \Biggr\} \ .
\end{equation}
For $\bar\nu_x$ emitted at the energysphere, the survival probability reads
\begin{equation} \label{eq:survival_nux}
    N_{\bar\nu_x} (E_\nu) = \exp\Biggl\{ 
    - \int_{R_{\rm ES}}^{R_{\rm TS}} 
    \frac{dr^\prime}{v(E_\nu,r^\prime)} \Gamma_{\bar\nu_x}(r^\prime)
    - \int_{R_{\rm TS}}^{\infty} 
    dr^\prime~\Gamma_{\bar\nu_x}(r^\prime)
    \Biggr\} \ .
\end{equation}
In these expressions, $r^\prime$ is the radial distance to the core-collapse supernova core. The term $\Gamma_{\bar\nu_\alpha} = \sum_{\beta = e, \mu, \tau} \Gamma_{\alpha\beta}$ represents the total decay rate.
In eq.~(\ref{eq:survival_nux}), the first integral inside the exponential accounts for the decay occurring between the energysphere $R_{\rm ES}$ and the transportsphere $R_{\rm TS}$. In this region, $\bar\nu_x$ are no longer in thermal equilibrium, but they remain trapped and diffuse outwards with an effective velocity given by $v(E_\nu,r) = \lambda(E_\nu,r) / (R_{\rm TS} - R_{\rm ES})$. The second term in the exponential represents the decay once $\bar\nu_x$ start free-streaming. 
In contrast, in eq.~(\ref{eq:survival_nue}), there is only one term inside the exponential. In this case, $\bar\nu_e$ fall out of equilibrium and start free-streaming at the same point. Thus, only decay occurring beyond the neutrinosphere $R_{\rm NS}$ affects the spectrum.

\subsubsection*{Energysphere, transportsphere, and neutrinosphere}
The neutrino flux suppression due to neutrino decay in matter requires the computation of the neutrinosphere for $\bar{\nu}_e$ \eqref{eq:survival_nue}, the energysphere and the transportsphere for $\bar{\nu}_x$ and $\nu_x$ \eqref{eq:survival_nux}. The main processes that keep $\nu_e$ and $\bar\nu_e$ in thermal equilibrium are beta processes \cite{Raffelt:2001kv}: 
\begin{equation}\label{eq:beta}
    \nu_e + n \leftrightarrow p + e^-  \ , ~~
    \bar\nu_e + p \leftrightarrow n + e^+ \ .
\end{equation}
These processes freeze out at a region called the neutrinosphere, where $\nu_e$ and $\bar\nu_e$ fall out of thermal equilibrium and start free streaming. The radius $r_{\rm NS}$ of the neutrinosphere is defined as the energy-dependent location where the optical depth is equal to $2/3$, i.e.,
\begin{equation} \label{eq:rNS}
    \tau(r_{\rm NS}) = \int^\infty_{r_{\rm NS}} dr \lambda_{\beta}^{-1}(r) = \frac{2}{3} \ , 
\end{equation}
where $\lambda_{\beta}^{-1}(r)$ is the inverse mean free-path\footnote{Not to overburden the text, the explicit dependence on the neutrino energy in eq.~(\ref{eq:rNS}) (coming from the energy-dependence of the mean free-path) was dropped.} of neutrinos associated to the beta processes\footnote{Generally, the mean free path of a particle that interacts through several processes is calculated as

\begin{equation}
    \lambda^{-1} (r, E) = \sum_i \lambda_i^{-1}(r, E) = \sum_i n_i(r) \sigma_i(E) \ ,
\end{equation}
where the index $i$ runs over all the processes, the cross section of these processes is denoted by $\sigma_i$, and $n_i$ indicates the target number.} eq.~\eqref{eq:beta}.

Regarding the non-electronic flavors, neutral-current collisions on nucleons $N$
\begin{equation}
    \nu+N \leftrightarrow \nu + N \ , 
\end{equation}
are the main processes through which $\nu_x$ and $\bar\nu_x$ interact with the medium \cite{Raffelt:2001kv}. The energy exchange in this process is inefficient due to the large mass of the nucleons ($m_N \sim 940$~MeV) compared to the neutrino energies ($E_\nu\sim 10$~MeV). Consequently, other processes are instead responsible for keeping $\nu_x$ and $\bar\nu_x$ in thermal equilibrium: nucleon-nucleon bremsstrahlung and leptonic processes, respectively
\begin{equation}
    N + N \leftrightarrow N + N + \nu +\bar\nu \ , ~~~~
    e^+ + e^- \leftrightarrow \nu + \bar\nu \ , ~~~~
    \nu + e \leftrightarrow \nu + e \ .
\end{equation}
Non-electronic neutrinos and antineutrinos are kept in thermal equilibrium until these interactions freeze out at the energysphere. The radius of the energysphere $r_{\rm ES}$ is found using the thermal optical depth \cite{Raffelt:2001kv,Shapiro:1983du}:
\begin{equation}
    \tau(r_{\rm ES}) = \int^\infty_{r_{\rm ES}} dr \sqrt{\lambda_E^{-1}(r)\left[ \lambda_T^{-1}(r) + \lambda_E^{-1}(r)\right]} = \frac{2}{3} \ .
\end{equation}
In this expression, $\lambda_E^{-1}$ and $\lambda_T^{-1}$ denote the ``energy" and ``transport" inverse mean free paths. For the calculation of the ``energy" mean free path, we consider only nucleon bremsstrahlung, $\lambda_E = \lambda_{\rm brem}$, which dominates over the leptonic processes \cite{Raffelt:2001kv}. For the ``transport" mean free path, we consider NC collisions on nucleons, $\lambda_T^{-1} = \lambda_{\nu p}^{-1} + \lambda_{\nu n}^{-1}$.

Finally, the NC interactions of neutrinos with nucleons keep $\nu_x$ and $\bar\nu_x$ trapped up to the transportsphere. The radius of the transportsphere $r_{\rm TS}$ can be obtained using the following expression
\begin{equation}
    \tau(r_{\rm TS}) = \int^\infty_{r_{\rm TS}} dr \lambda_T^{-1}(r) = \frac{2}{3} \ .
\end{equation}

\subsection{Neutrino fluxes with decay: from the supernova core to the Earth}
\label{sec:flux}
\noindent
In order to calculate the core-collapse supernova neutrino fluxes\footnote{Note that we omit the dependence of the core-collapse supernova neutrino fluxes on the progenitor mass, not to overburden the text.} including neutrino decay, one needs $\phi^0_{\bar\nu_\alpha}(E_\nu,t)$, namely
the flux of neutrinos $\nu_\alpha$ with energy $E_\nu$ at a post-bounce time $t$ produced in the deep regions of the supernova core, before decoupling.  

\subsubsection*{Supernova neutrino fluxes before decay}
Since we consider the possibility that the next galactic core-collapse supernova leaves either a NS or a BH, we use two sets of simulations from the literature, one from the Garching group for the NS case, and one from Nakazato's work for the BH case.
The inputs for the neutrino fluxes for the two sets of core-collapse supernova simulations differ.

\vspace{.3cm}

\noindent
{\it NS case} \\
For the Garching simulations \cite{Fiorillo:2023frv,Garching}, the neutrino fluxes are well accounted for by the usual power-law distribution \cite{Keil:2002in}, namely

\begin{equation}
    \phi^0_{\nu_\alpha}(E_\nu, t) = 
        \frac{L_\nu (t)}{\langle E_{\nu}(t) \rangle} \varphi_{\nu_\alpha}(E_\nu, t)\ ,
\end{equation}
where $L_\nu (t)$ and $\langle E_{\nu}(t) \rangle$ indicate the neutrino luminosity and average energy, respectively. The neutrino energy distribution is given by

\begin{equation} \label{eq:PL}
     \varphi_{\nu_\alpha}(E_\nu, t) =  \frac{1}{\langle E_{\nu}(t) \rangle}
     \frac{\left[\alpha(t) + 1\right]^{\alpha(t) + 1}}{ \Gamma\left(\alpha(t) + 1\right)}  
     \left(
     \frac{E_\nu}{{\langle E_{\nu}(t) \rangle}} 
     \right)^{\alpha(t)} \exp\left\{-\frac{\left[1+ \alpha(t)\right] E_\nu}{\langle E_{\nu}(t) \rangle}\right\} \ .
\end{equation}
In this expression, $\alpha(t)$ represents the pinching parameter related to the first and second moments of the energy through the relation

\begin{equation}
    \alpha(t) = \frac{\langle E_{\nu}(t)^2 \rangle - 2 \langle E_{\nu}(t) \rangle ^2}{\langle E_{\nu}(t) \rangle ^2 - \langle E_{\nu}(t)^2 \rangle} \ .
\end{equation}

\noindent
{\it BH case} \\
For Nakazato's simulations \cite{Nakazato:2012qf,Nakazato:2021gfi,Nakazato}, we directly use the numerical output and perform a log-linear interpolation of the data.
An example of the numerical fit is shown in figure~17 of ref.~\cite{Roux:2024zsv}. 

\subsubsection*{Supernova neutrino fluxes with decay}
In order to implement the time dependence, our theoretical framework employs time-dependent information from detailed supernova simulations. To this aim, the core-collapse supernova evolution was split into six intervals to optimize the computation. We chose a representative time point for each interval $\Tilde{t}_i \in [t_{i-1}, t_i]$ and calculated the survival probabilities Eqs.\eqref{eq:survival_nue}-\eqref{eq:survival_nux} at that time $N_{\bar\nu_\alpha}^i(E_\nu) = N_{\bar\nu_\alpha}(E_\nu, \Tilde{t}_i)$. We then applied this survival probability to the fluxes in the corresponding time ranges. Namely, the antineutrino fluxes after decay at time $t \in [t_{i-1}, t_i]$ ($i = 1, ..., 6$) become
\begin{equation}
   \phi^d_{\bar\nu_\alpha}(E_\nu,t) = N_{\bar\nu_\alpha}^i(E_\nu) \phi^0_{\bar\nu_\alpha}(E_\nu,t) \ .
\end{equation}
where the first factor is given by \eqref{eq:survival_nue}-\eqref{eq:survival_nux}.

If neutrinos can decay, the neutrino fluxes as a function of time, when neutrinos start free-streaming, can be expressed as follows
\begin{equation}
    \phi^d_{\nu_\alpha}(E_\nu,t) = \phi^0_{\nu_\alpha}(E_\nu,t) + \phi^\Delta_{\nu_\alpha}(E_\nu,t) \ .
\end{equation}
\noindent
Here $\phi^\Delta_{\nu_\alpha}(E_\nu,t)$ is the contribution to the neutrino flux from the antineutrinos that underwent decay. The latter can be determined using the expression

\begin{equation}
    \phi^\Delta_{\nu_\alpha}(E_\nu,t) = \int_{E_\nu}^{\infty}
    dE'_\nu \psi(E'_\nu, E_\nu)
    \sum_{\beta = e, \mu, \tau} \left[1 - N^i_{\bar\nu_\beta \to \nu_\alpha} (E'_\nu) \right] \phi^0_{\bar\nu_\beta}(E_\nu,t) \ .
\end{equation}
The factor $\psi(E'_\nu, E_\nu)$ represents the energy distribution of the daughter neutrino eq.~(\ref{eq:energy}), and $N^i_{\bar\nu_\beta \to \nu_\alpha}$ are the partial survival probabilities of $\bar\nu_\beta$ to $\nu_\alpha$ in the $i$-th time interval. We remind that only antineutrinos satisfy the condition for decay in matter given by the
relation \eqref{eq:decaycond}.

In order to verify that our numerical results do not depend on the procedure we use, we checked that they do not depend on the representative points $\Tilde{t}_i $ in each interval, and we increased the number of intervals from six to ten. While two intervals were sufficient for the accretion phase, the cooling phase required at least three intervals to achieve an accurate representation. It is to be noted that we also checked the consistency of our results with those of ref.~\cite{Kachelriess:2000qc} by considering a single time interval and using the core-collapse supernova profiles from the Wilson model \cite{Nunokawa:1997ct}, and obtained good agreement.

Once the neutrinos reach the MSW region, their fluxes are modified by neutrino-matter interaction.
Thus, depending on the sign of the squared mass differences $\Delta m^2_{31} = m^2_{3} - m^2_{1} $ (that still remains unknown), the neutrino yields at the supernova surface become\footnote{Note that here we neglect the explicit dependence of the neutrino fluxes on energy and time, not to overburden the text.} 
\begin{align}\label{eq:MSWNO}
  \phi_{\nu_1} & =   \phi^d_{\nu_x} ~~ \phi_{\nu_2} =   \phi^d_{\nu_x} ~~ \phi_{\nu_3} =   \phi^0_{\nu_e} ~~~{\rm (NO)}  \ ,  \\
 \phi_{\nu_1} & =   \phi^d_{\nu_x}~~  \phi_{\nu_2} =   \phi^0_{\nu_e} ~~ \phi_{\nu_3} =   \phi^d_{\nu_x} ~~~{\rm (IO)}  \nonumber \ , 
\end{align}
and the antineutrino yields are
\begin{align}\label{eq:MSWIO}
 \phi_{\bar{\nu}_1} & =  \phi^d_{\bar{\nu}_e} ~~ \phi_{\bar{\nu}_2} =  \phi^d_{\nu_x} ~~ \phi_{\bar{\nu}_3} = \phi^d_{\nu_x} ~~~{\rm (NO)} \ , \\
 \phi_{\bar{\nu}_1} & =  \phi^d_{{\nu}_x~~}  \phi_{\bar{\nu}_2} =  \phi^d_{\nu_x} ~~ \phi_{\bar{\nu}_3} =  \phi^d_{\bar{\nu}_e} ~~~{\rm (IO)}  \nonumber \ , 
\end{align}
with NO standing for normal (i.e. $\Delta m^2_{31} > 0$) and IO for inverted (i.e. $\Delta m^2_{31} < 0$) $\nu$ mass ordering.

Finally, neutrinos propagate in vacuum from the core-collapse supernova surface to the Earth. In a detector on Earth, the fluxes in the flavor basis are related to the ones in the mass basis through the relation

\begin{equation}\label{eq:nuflux}
    \phi_{\nu_\alpha} (E_\nu,t) = \frac{1}{4\pi L^2}
    \sum_{i = 1,3} |U_{\alpha i}|^2 \phi_{\nu_i} (E_\nu,t) \ , 
\end{equation}
where $L$ is the core-collapse supernova distance. Here, $U_{\alpha i}$ are the elements of the PMNS matrix. We use the following values of the neutrino mixing angles, $\sin^2\theta_{12} = 0.307$, $\sin^2\theta_{23} = 0.547~(0.534)$, and $\sin^2\theta_{13} = 0.02$ for normal (inverted) neutrino mass ordering \cite{ParticleDataGroup:2022pth}.

\section{Supernova models and neutrino fluxes} \label{sec:models_fluxes}
\noindent
The evolution of a core-collapse supernova during the explosion and the corresponding neutrino emission can be divided into three main phases. 
At early times, as the shock breaks out from the neutrinosphere, large amounts of $\nu_e$ are produced through the neutronization of the core, leading to the so-called neutronization burst (figures~\ref{fig:sn-param-NS} and \ref{fig:sn-param-BH}, left panels). During this phase, the $\nu_e$ luminosity is very high, around $L_{\nu_e} \sim 10^{53}$~erg/s, though it lasts only about 50~ms. Following this, matter keeps falling and accreting the core in what is called the accretion phase (figures~\ref{fig:sn-param-NS} and \ref{fig:sn-param-BH}, middle panels), which lasts up to several hundred ms. In this phase, neutrinos and antineutrinos of all flavors are emitted with luminosities of $\sim 10^{52}$~erg/s (or higher if a BH is formed). The shock wave stalls at this stage, and a core-collapse supernova explosion is only possible if some mechanism revives the shock. 

Currently, the shock revival is thought to happen via neutrino energy deposition, combined with neutrino-induced convection and turbulence, in the so-called {\it delayed neutrino heating mechanism} first introduced by Bethe and Wilson \cite{Bethe:1984ux}. Finally, the last stage of the neutrino emission during the death of a massive star is the cooling of the newly formed proto-neutron star (if the core-collapse supernova does not leave a black-hole). So, the luminosity significantly decreases, and neutrinos are emitted with a quasi-thermal spectrum and almost luminosity equipartition across flavors (figure~\ref{fig:sn-param-NS}, right panels). This last phase lasts for several seconds up to $\sim 10$~s if a NS is formed.
Interestingly, if the core-collapse supernova is close enough, one could also detect the pre-supernova neutrinos from the last few days of the Si-burning phase preceding the core-collapse \cite{Kato:2020hlc}, and also measure the late-time cooling of the proto-neutron star that would be informative on the neutron star EOS, the fate of the core-collapse supernova, and possible non-standard cooling mechanisms \cite{Li:2023ulf}.

\subsection{Supernova models}
\noindent
Our investigation considers that the core-collapse supernova leaves either a NS or a BH. To this aim, we exploit the outcome of detailed one-dimensional core-collapse supernova simulations either from the Garching group for the NS case \cite{Garching, Fiorillo:2023frv}, or from Nakazato for the BH case \cite{Nakazato:2012qf,Nakazato:2021gfi,Nakazato}.
Note that Garching simulations account well for SN1987A observations, except for the late neutrino events. 

In contrast with ref.~\cite{Kachelriess:2000qc}, in our framework, the neutrino emission is time-dependent and the transportspheres, the energyspheres, and the neutrinospheres 
are determined from microscopic processes, namely nucleon-nucleon bremsstrahlung, neutrino-nucleon NC interaction, and inverse-beta decay.  
Thus, to compute the neutrino decay rates in matter eq.~\eqref{eq:decayrate} and the survival probabilities
for $\bar{\nu}_e$ eq.~\eqref{eq:survival_nue} and $\nu_x$ eq.~\eqref{eq:survival_nux} we use the neutrino fluxes, the matter densities $\rho(t, r)$ as a function of position and time, and the electron fraction $Y_e(t,r)$ from detailed core-collapse supernova simulations. 
 
\begin{figure*}[t]
    \centering
     \includegraphics[scale=0.33]{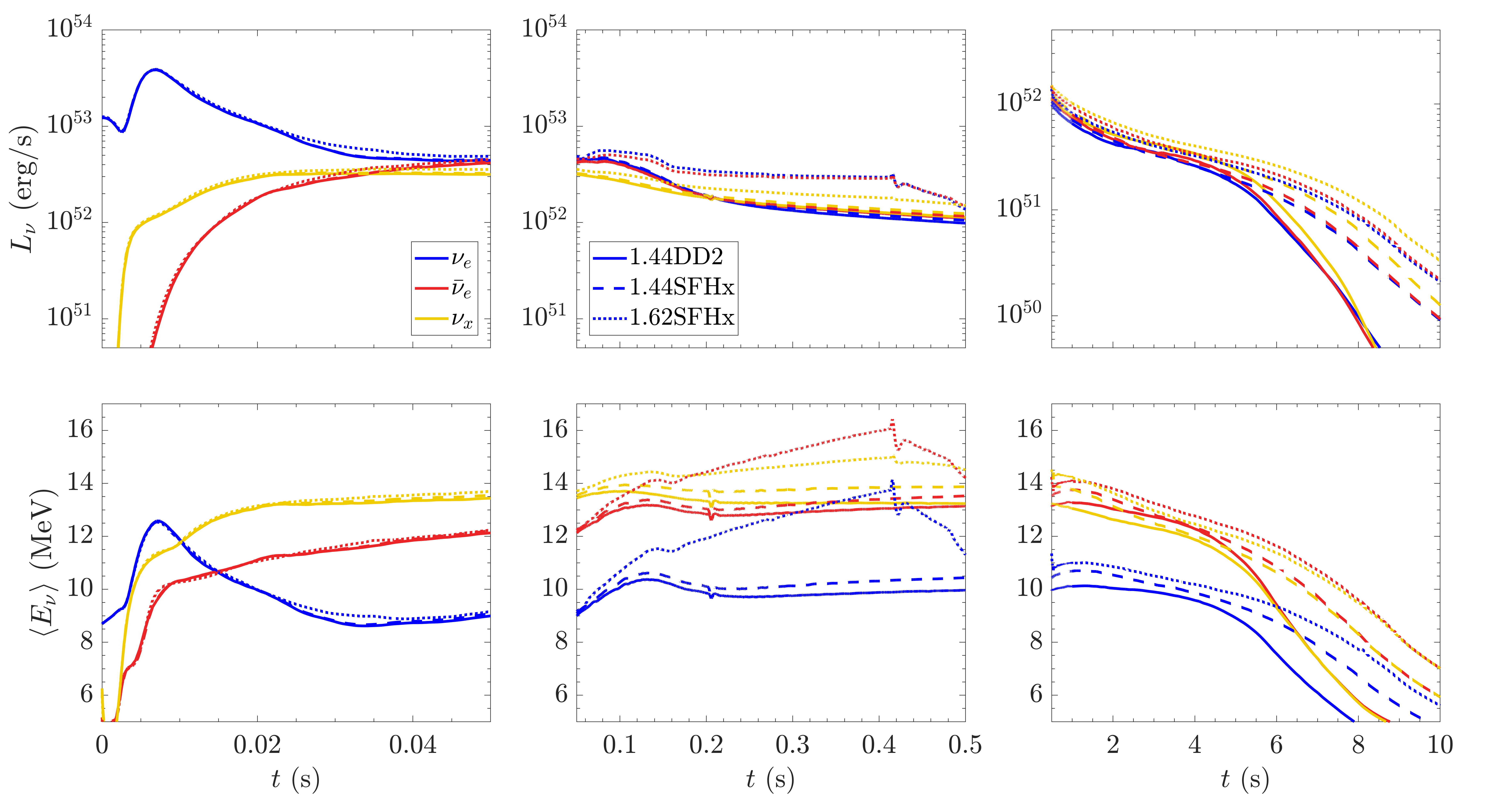}
         \caption{{\bf NS case:} Parameters defining the neutrino fluxes, as a function of time, for a core-collapse supernova at 10 kpc leaving a NS. The three phases of the core-collapse supernova explosion are shown, namely the neutronization burst (left), the accretion (middle), and the cooling phase of the newly born neutron star (right figures). As a function of time, the figures present the neutrino luminosity or the neutrino average energies. The curves correspond to the models 1.44DD2 (solid lines), 1.44SFHx (dashed lines), and 1.62SFHx (dotted lines) from one-dimensional simulations of the Garching group. For each model the results are for  $\nu_e$ (blue), $\bar\nu_e$ (red), and $\nu_x$ (yellow lines). (With courtesy from \cite{Garching, Fiorillo:2023frv}.)}
    \label{fig:sn-param-NS}
\end{figure*}

\begin{figure*}[t]
    \centering
   \includegraphics[scale=0.4]{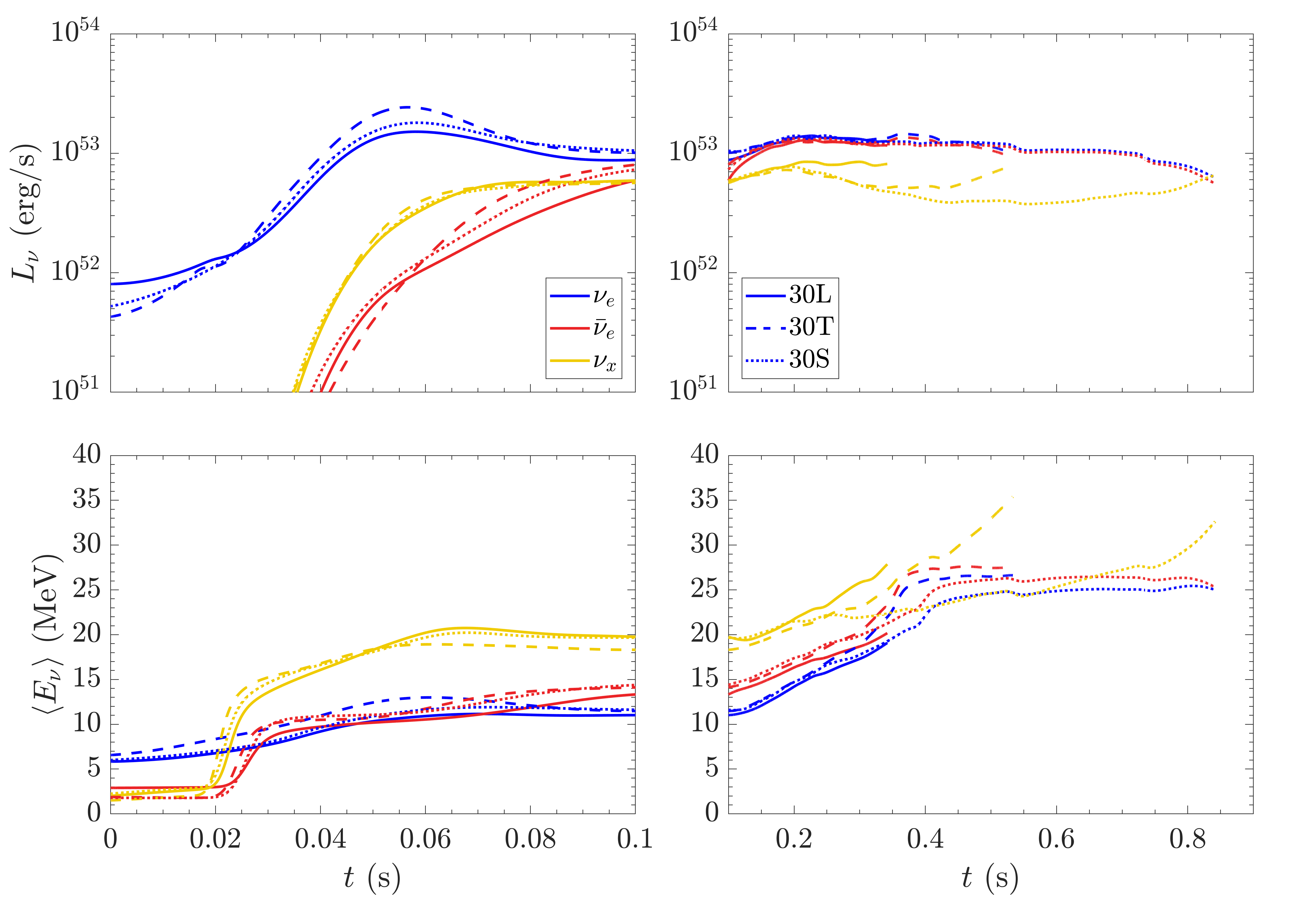}
    \caption{{\bf BH case:} Parameters defining the neutrino fluxes, as a function of time, for a core-collapse supernova at 10 kpc leaving a BH. The figures show the neutrino luminosity and the neutrino average energies, as functions of time, for the 30 $M_{\odot}$, with metallicity $Z=0.004$ and the LS220, the Shen or the Togashi EOS, from Nakazato's simulations: 30L (solid lines), 30S (dotted lines), and 30T (dashed lines). (With courtesy from \cite{Nakazato:2012qf,Nakazato:2021gfi,Nakazato}.)}
    \label{fig:sn-param-BH}
\end{figure*}

Figure \ref{fig:sn-param-NS} illustrates the evolution of two of the parameters that characterize the core-collapse supernova neutrino fluxes, i.e., the luminosity, the average energy for the different neutrino flavors. The results correspond to the two sets of supernova simulation results used in our study. For the NS case we considered the supernova simulation inputs corresponding to the two neutron star masses 1.44 $M_{\rm \odot}$ and 1.62 $M_{\rm \odot}$ with either the SFHx \cite{Steiner:2012rk} or the DD2 \cite{Hempel:2009mc} NS EOS. 
On the other hand, figure \ref{fig:sn-param-BH} shows the evolution of the neutrino luminosities and average energies for the three neutrino species for the BH
that corresponds to a 30 $M_{\odot}$ progenitor, with metallicity $Z=0.004$ and three different EOS, namely the LS220, the Shen, or the Togashi one \cite{Nakazato:2012qf,Nakazato:2021gfi,Nakazato}. 

\subsection{Results on the neutrino fluxes with decay}
\begin{figure*}[t]
\begin{center}
\includegraphics[scale=0.33]{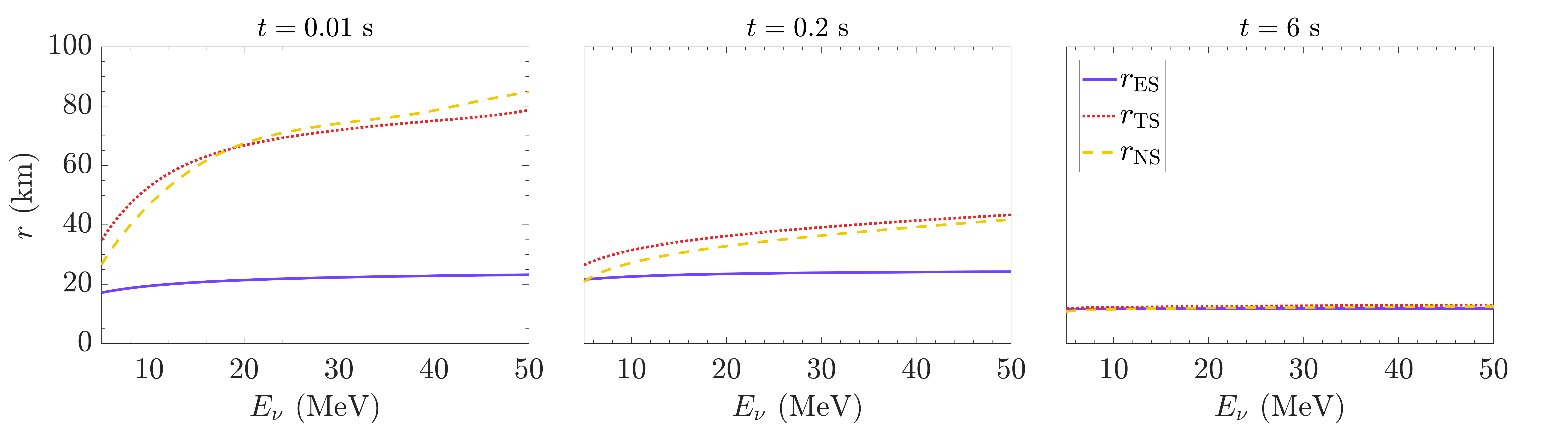}
\caption{The figure shows examples of the results for the neutrinosphere, the transportsphere, and the energysphere at three different time snapshots, during the neutronization burst, the accretion phase, and the cooling of the newly formed proto-neutron star. The results correspond to the 1.44SFHx model for the NS case.}
\label{fig:spheres}
\end{center}
\end{figure*}

\begin{figure*}[t]
    \centering
    \includegraphics[width=0.45\textwidth]{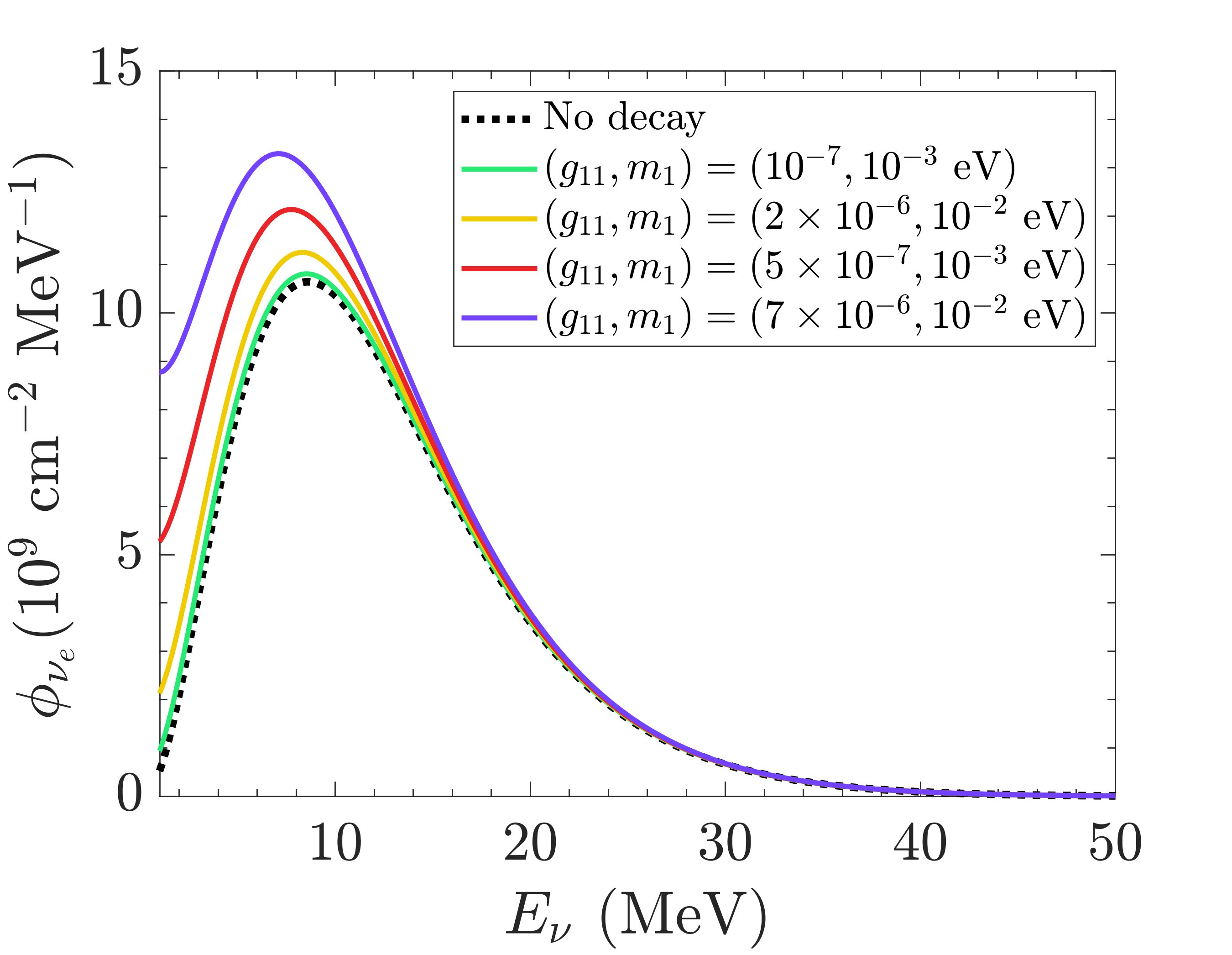}
    \includegraphics[width=0.45\textwidth]{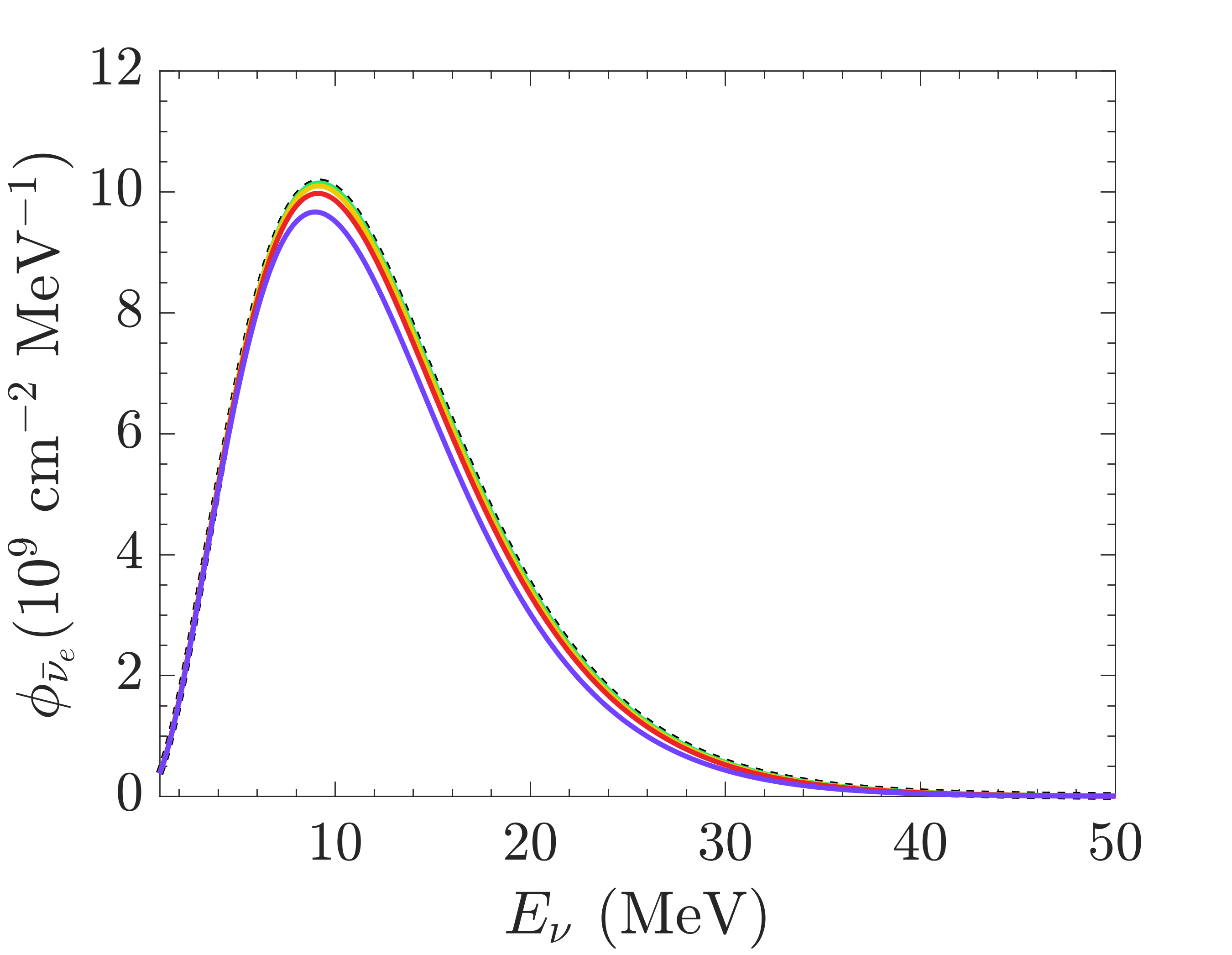}
    \includegraphics[width=0.45\textwidth]{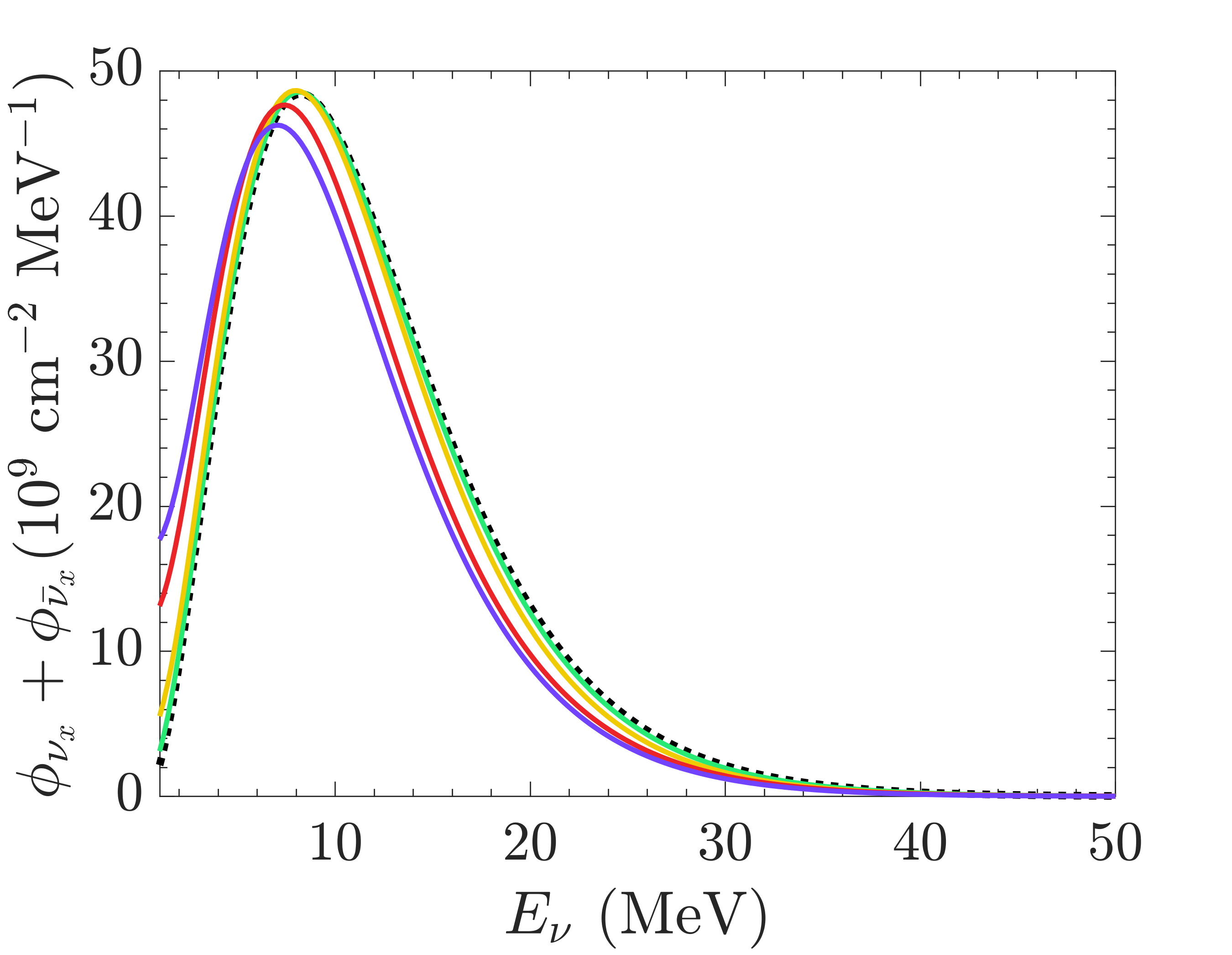}
    \caption{{\bf NS case:} Neutrino fluxes from a future core-collapse supernova, located at 10 kpc, as a function of neutrino energy, and impact of neutrino non-radiative decay for the case where a NS is formed.  The one-dimensional core-collapse supernova model used here is for a NS with 1.44 $M_{\rm \odot}$ mass and the SFHx EOS (named 1.44SFHx) from the Garching simulations \cite{Garching, Fiorillo:2023frv}. The results shown are valid for normal neutrino mass ordering and include neutrino-Majoron interactions with different values of the lightest neutrino mass $m_{1}$ and of the $g_{11}$ neutrino-Majoron coupling. The flux predictions in the absence of neutrino-Majoron interactions are also shown for comparison (black dotted line).}
    \label{fig:fluxesNS}
\end{figure*}

\begin{figure*}[t]
    \centering
    \includegraphics[width=0.45\textwidth]{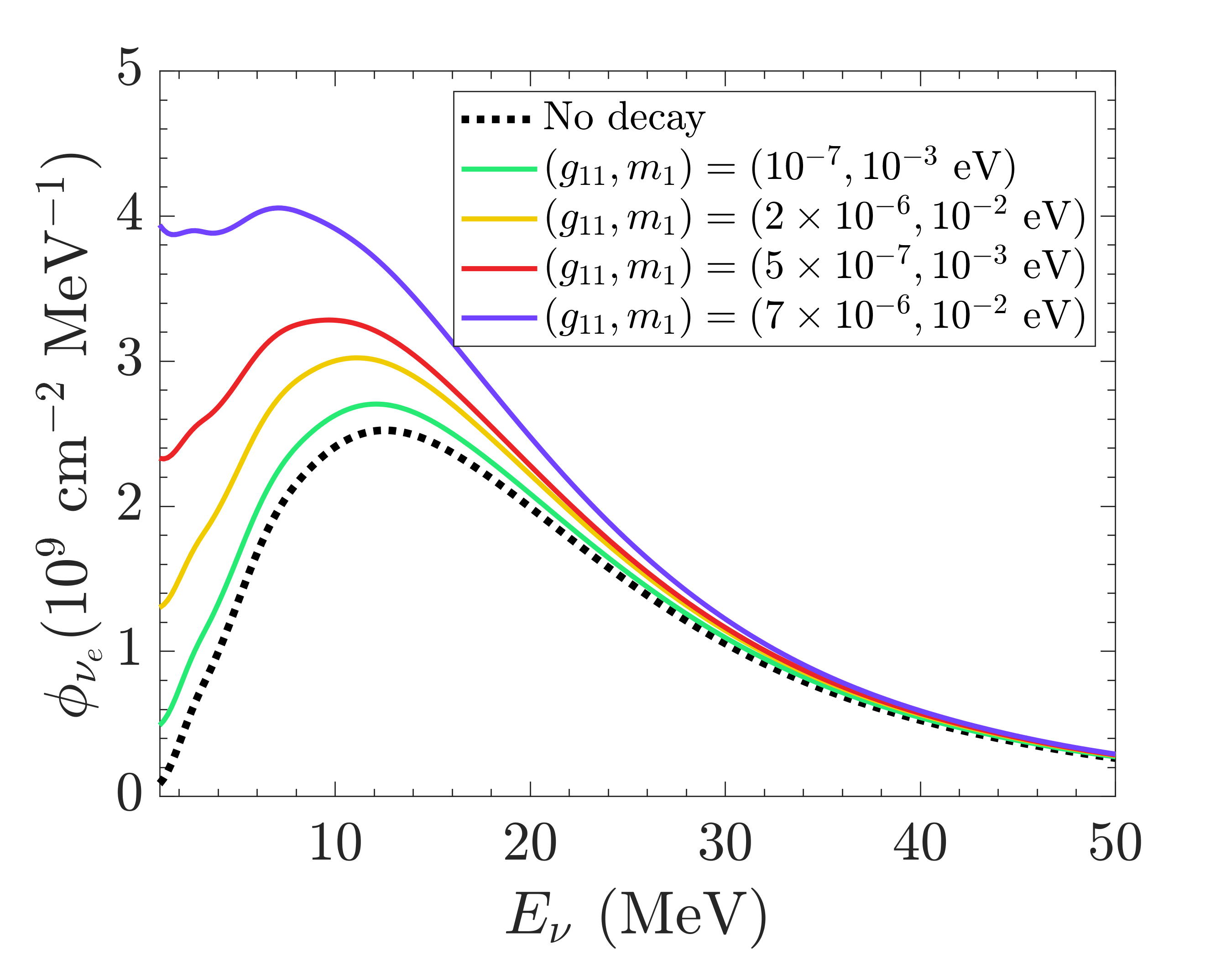}
    \includegraphics[width=0.45\textwidth]{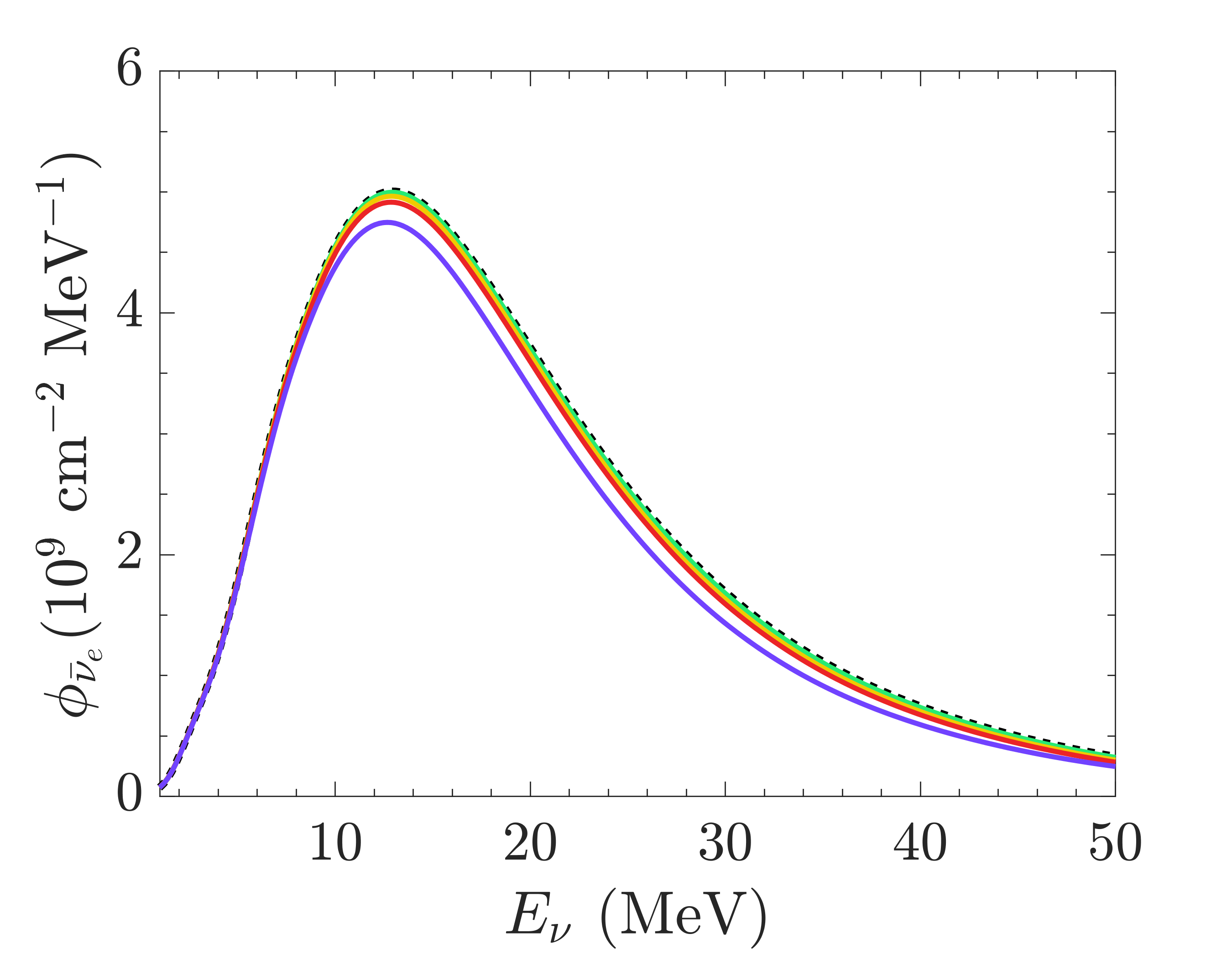}
    \includegraphics[width=0.45\textwidth]{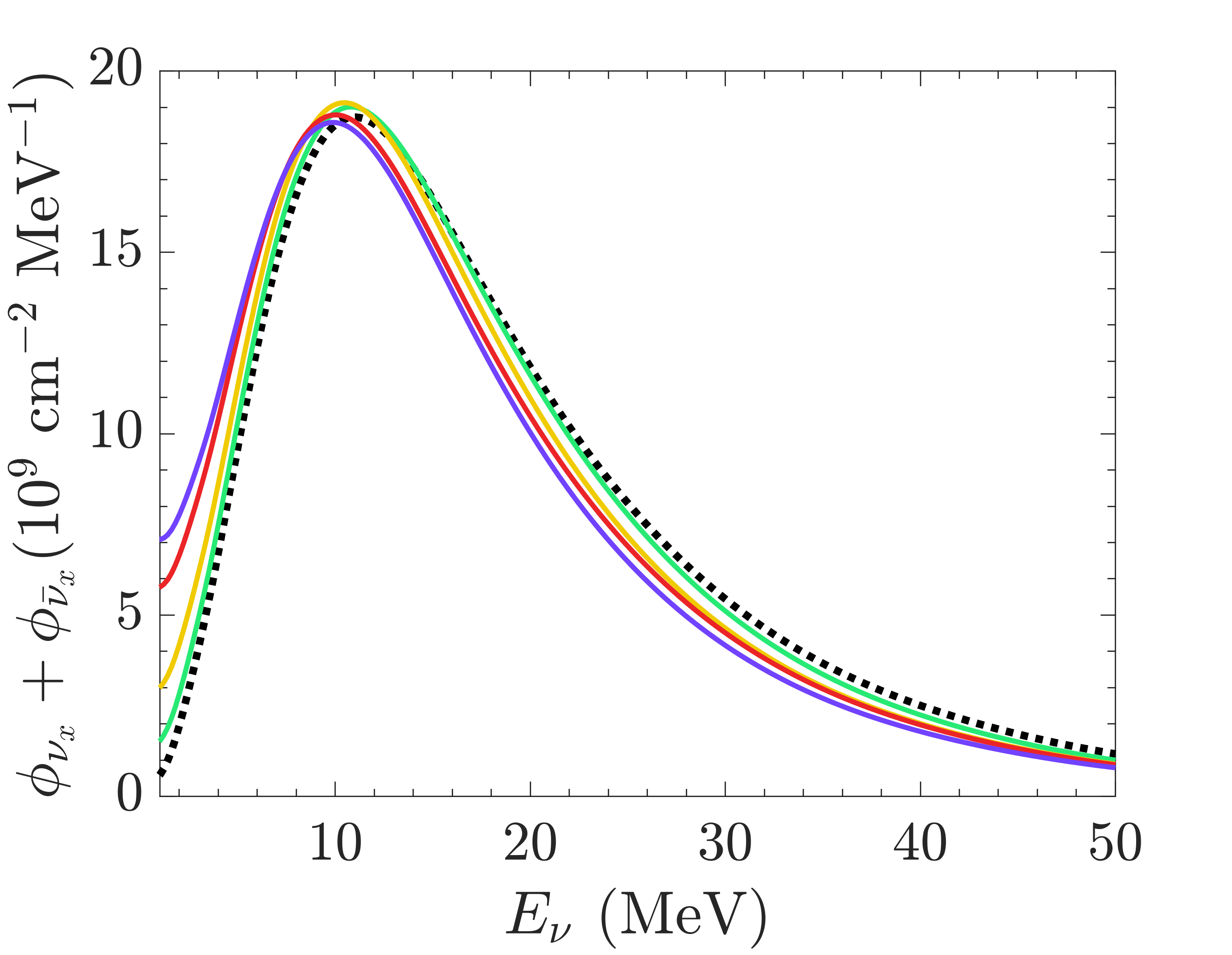}
    \caption{{\bf BH case:} Neutrino fluxes from a future core-collapse supernova, located at 10 kpc, as a function of neutrino energy, and impact of neutrino non-radiative decay for the case where a BH is formed. The one-dimensional core-collapse supernova model used here is the 30 $M_{\odot}$ for the BH case from Nakazato simulations, with metallicity $Z=0.004$ and the Togashi EOS \cite{Nakazato:2012qf,Nakazato:2021gfi,Nakazato}. The results are valid for normal neutrino mass ordering and include neutrino-Majoron interactions with different values of the lightest neutrino mass $m_{1}$ and of the $g_{11}$ neutrino-Majoron couplings. The flux predictions in the absence of neutrino-Majoron interactions are also shown for comparison (black dotted line).}
    \label{fig:fluxesBH}
\end{figure*}

\noindent 
It is interesting to look at the impact of neutrino decay in matter on the neutrino fluxes for different values of the neutrino-Majoron couplings. In figure \ref{fig:spheres}, we give three examples of the average energyspheres and transportspheres for the three phases of neutronization burst, accretion phase, and cooling of the newly born proton-neutron star for information.
Figures~\ref{fig:fluxesNS} and \ref{fig:fluxesBH} present the neutrino fluxes as a function of neutrino energy for the NS and the BH cases, respectively. The results correspond to $\nu_e$, $\bar{\nu}_e$ and to the sum of non-electron flavors ($\nu_{\mu}, \nu_{\tau}, \bar{\nu}_{\mu}, \bar{\nu}_{\tau}$) denoted as $\nu_x$.  As one can see from the figure, while the decay of $\bar{\nu}_e$ produces a depletion of the corresponding flux, the $\nu_e$ fluxes increase due to the $\bar{\nu}_e$ and $\bar{\nu}_x$ decay. Similarly, the sum of the $\nu_x$ fluxes shows both an increase and a decrease due to neutrino decay. Indeed, it presents an increase at low energies, and a decrease at high energies, due to the decay of the $\bar{\nu}_x$ into $\nu_e$, $\bar{\nu}_e$, and $\nu_x$. 

In the following, we shall use the simulations from the Garching group first to extract from SN1987A neutrino events new bounds on neutrino-Majoron bounds from neutrino decay in matter. Then, we shall use Nakazato simulations and the same simulations from the Garching group to make prospects on the limits for neutrino-Majoron couplings from neutrino decay in matter if a future galactic core-collapse supernova leaves a newly born NS as remnant. The supernova location will be either 10 kpc or 0.2~kpc, having in mind Betelgeuse as an example. Finally, we shall use the same models as templates to perform DSNB predictions in the presence and absence of neutrino nonradiative two-body decay in matter.

\section{New bounds on neutrino-Majoron couplings from SN1987A} \label{sec:limits_sn1987a}
\noindent
The observation of SN1987A, located 50~kpc away, marked the first and, so far, only detection of neutrinos emitted from a core-collapse supernova. Three detectors, Kamiokande-II (2.14~kton), IMB (6.8~kton), and Baksan (0.28~kton), recorded 11, 8, and 5 $\bar\nu_e$ events, respectively \cite{Kamiokande-II:1987idp, Bionta:1987qt, Alekseev:1988gp}. The dominant process in these detectors was inverse beta decay (IBD). More information about these events can be found in ref.~\cite{Ivanez-Ballesteros:2023lqa}.
For the likelihood analysis, we follow \cite{Ivanez-Ballesteros:2023lqa,Vissani:2014doa}, and implement detailed information on the response of each of the detectors included in the analysis. 
A complete description of the statistical analysis and precise inputs can be found in appendix~\ref{appendix:statistics}. 

Before presenting the results, we would like to emphasize that
in the present work, we determine limits on neutrino-Majoron couplings considering that Majorons are free-streaming. Thus, our bounds are valid only in the free-streaming regime.
It is to be noted, however, that in the very dense regions of the inner core, Majorons and neutrinos can be trapped so that neutrino-Majoron and Majoron-Majoron
scattering can be important, as well as Majoron decay to neutrinos. Moreover, neutrino scattering mediated by Majorons is possible.   

A complete treatment would require self-consistent simulations solving neutrino transport in the stellar core, including Majorons 
which is beyond the scope of the present work.
It is to be noted that bounds on neutrino-Majoron couplings, available in the literature, do not use such a theoretical framework,
and are always given in the free-streaming regime. Refs.~\cite{Fuller:1988ega} and \cite{Suliga:2024nng} are an exception, since the authors performed 
core-collapse supernova simulations with Majorons, but with the goal of assessing their impact on the stellar collapse, for the latter in the case of a massive Majoron. 
In this respect, it is to be noted that our bounds lie in a region where weak processes dominate over neutrino scattering mediated by Majorons
(see eq.(6b) of ref.~\cite{Fuller:1988ega}). So neutrino scattering mediated by Majorons should not impact the core-collapse supernova dynamics 
in the range of couplings of our bounds.

\subsection{Limits for $(g, m)$}

\subsubsection*{Neutrino-Majoron bounds in the mass basis}
We will now present results using the simulation inputs from the 1.44SFHx supernova model, as a reference model for SN1987A. As discussed in ref.~\cite{Fiorillo:2023frv}, this model is the one that better accounts for SN1987A observations. For the likelihood analysis, the numerical results are obtained using the simplex method. We employ the mass-squared differences  $\Delta m^2_{21} = 7.53\times10^{-5}$~eV$^2$ and $\Delta m^2_{32} = 2.437~(-2.519) \times10^{-3}$~eV$^2$ for normal (inverted) neutrino mass orderings~\cite{ParticleDataGroup:2022pth}.

\noindent
\begin{figure}[t]
\begin{center}
\includegraphics[width=0.48\textwidth]{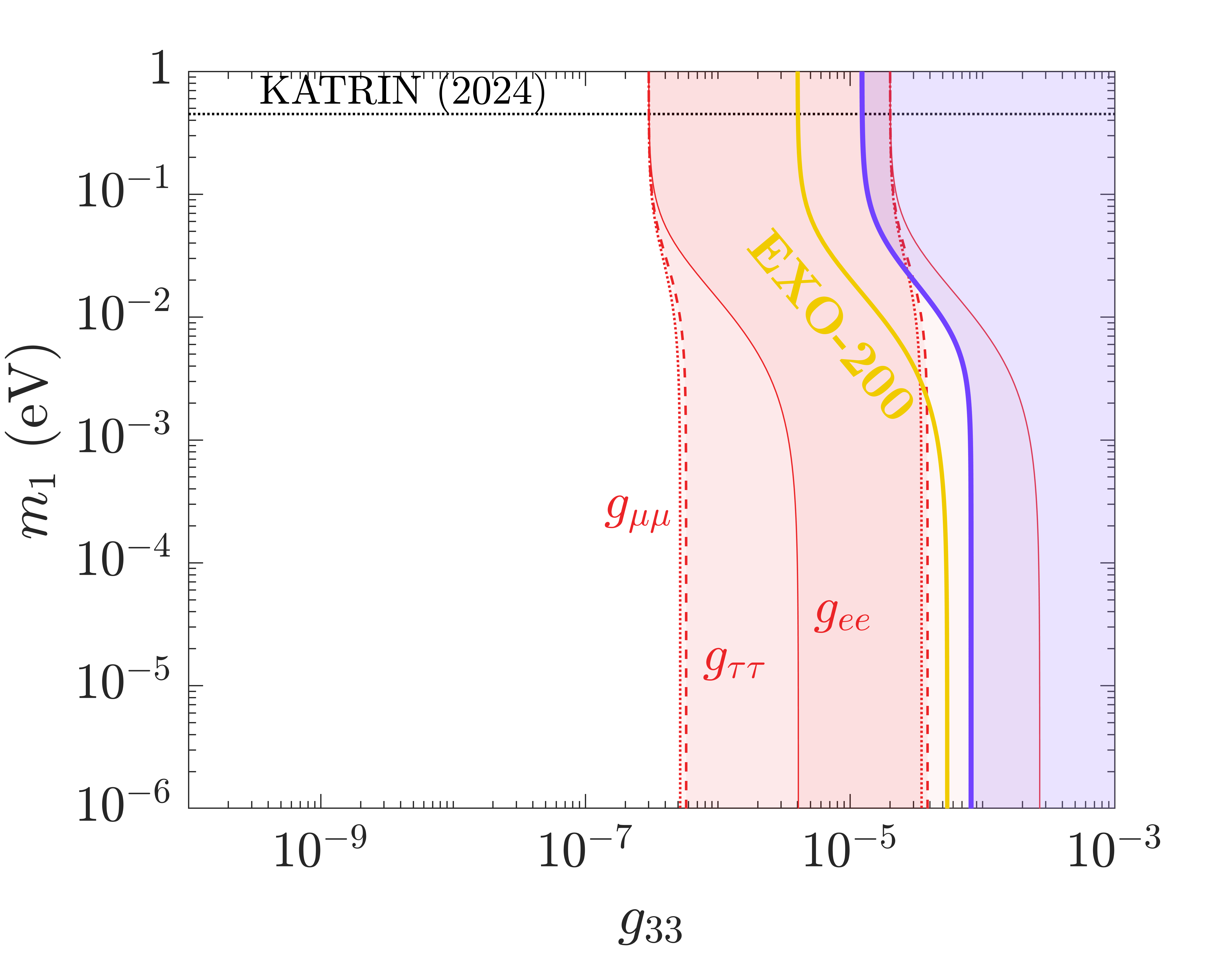}
\includegraphics[width=0.48\textwidth]{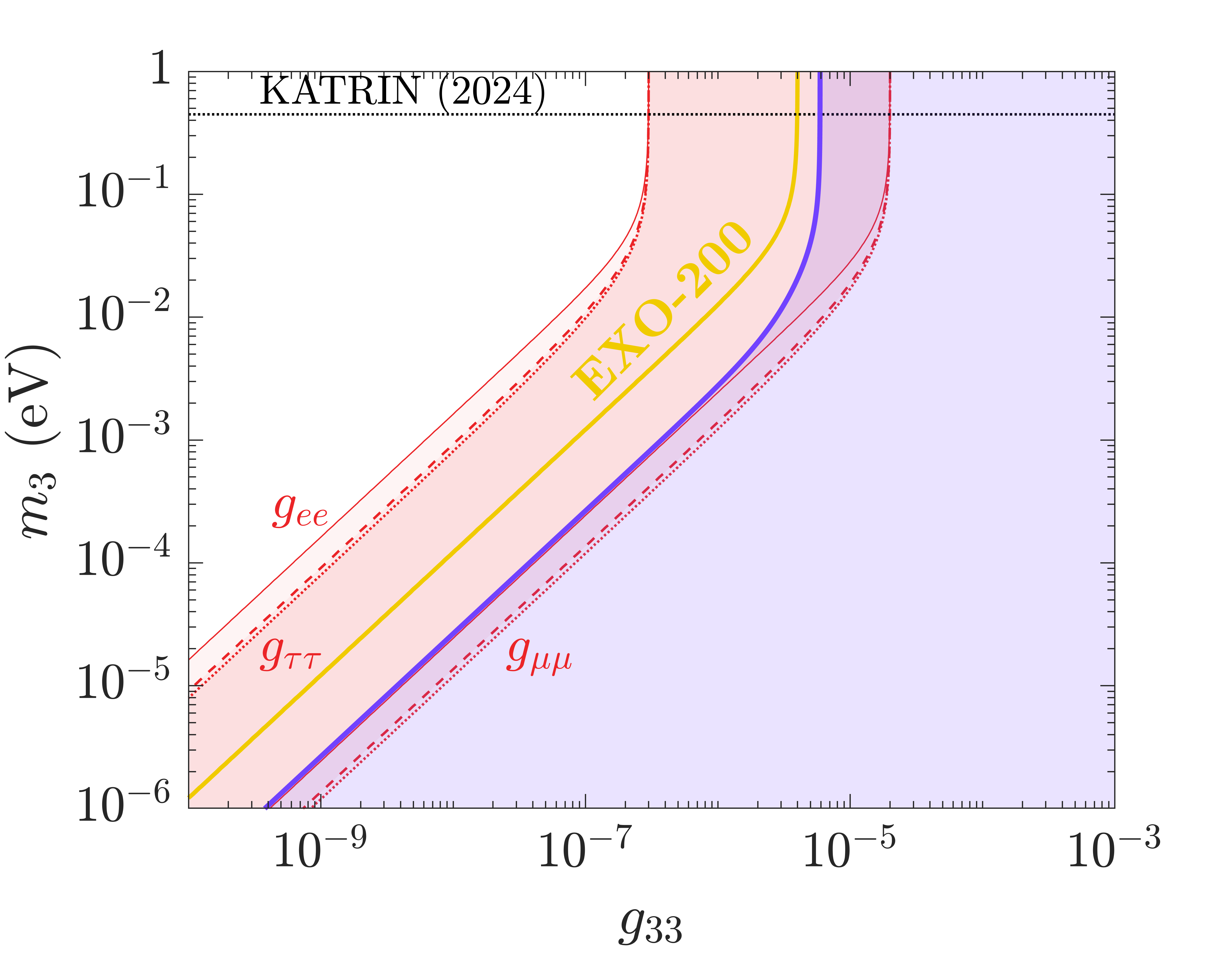}
\caption{Constraints on the neutrino-Majoron coupling in the $g_{33}$--$m_{1, 3}$ plane from our SN1987A analysis (blue-shaded area) for normal (left) and inverted ordering (right panel) at 90\% CL. The limits are obtained using the 1.44SFHx model. For comparison, we show the constraints obtained by the Majoron luminosity argument (red), which usually considers only one neutrino flavor at a time from ref.~\cite{Kachelriess:2000qc}, along with the tightest limits from $0\nu\beta\beta$ decay (yellow solid line) \cite{Kharusi:2021jez}. The red-shaded areas are obtained by requiring the limit from ref.~\cite{Kachelriess:2000qc} on $g_{ee}$ (red solid line), $g_{\mu\mu}$ (red dotted line), and $g_{\tau\tau}$ (red dashed line). Additionally, the current limit on the absolute neutrino mass obtained by the KATRIN experiment is also shown \cite{Katrin:2024tvg}.}
\label{fig:g33-m}
\end{center}
\end{figure}

Figure~\ref{fig:g33-m} shows our constraints in the neutrino mass basis, and in particular the ones on $g_{33}$, which were never shown before, as a function of the lightest neutrino mass, for normal and inverted ordering. The blue area represents the region excluded by our SN1987A likelihood analysis, considering the neutrino spectral distortions due to neutrino decay. The areas excluded by the Majoron luminosity argument ($3\times10^{-7} < |g_{\alpha\beta}| < 2\times10^{-5}$)  \cite{Kachelriess:2000qc} are shown in red for comparison. It is to be noted that the luminosity constraints are commonly obtained by assuming only one coupling at a time. For that reason, we show the bounds requiring the limit only on $g_{ee}$ (red solid line), $g_{\mu\mu}$ (red dotted line), and $g_{\tau\tau}$ (red dashed line). 
Searches for Majoron-emitting neutrinoless double beta decay ($0\nu\beta\beta J$) have been performed for many years. Here, we show the tightest limit obtained from $0\nu\beta\beta$ decay translated into the $g_{33}$--$m_{1}$, or $g_{33}$--$m_3$ plane. This limit, currently coming from the EXO-200 \cite{Kharusi:2021jez}, is represented by the yellow solid line.

Our results can be directly compared to those from \cite{Kachelriess:2000qc, Tomas:2001dh}. Our bounds for $g_{11}$ improve on the ones from the spectrum analysis in \cite{Kachelriess:2000qc} by more than an order of magnitude in normal ordering and by almost 2 orders of magnitude for $m_3 \gtrsim 10^{-2}$~eV in inverted ordering. Additionally, the limits we obtain for $g_{22}$ are over an order of magnitude stronger than those reported in ref.~\cite{Tomas:2001dh}. Note that a comparison for $g_{33}$ is not possible since this coupling was set to zero in ref.~\cite{Tomas:2001dh}.

\subsubsection*{Neutrino-Majoron bounds in the flavor basis}
\begin{figure}[t]
    \centering
    \includegraphics[width=0.48\textwidth]{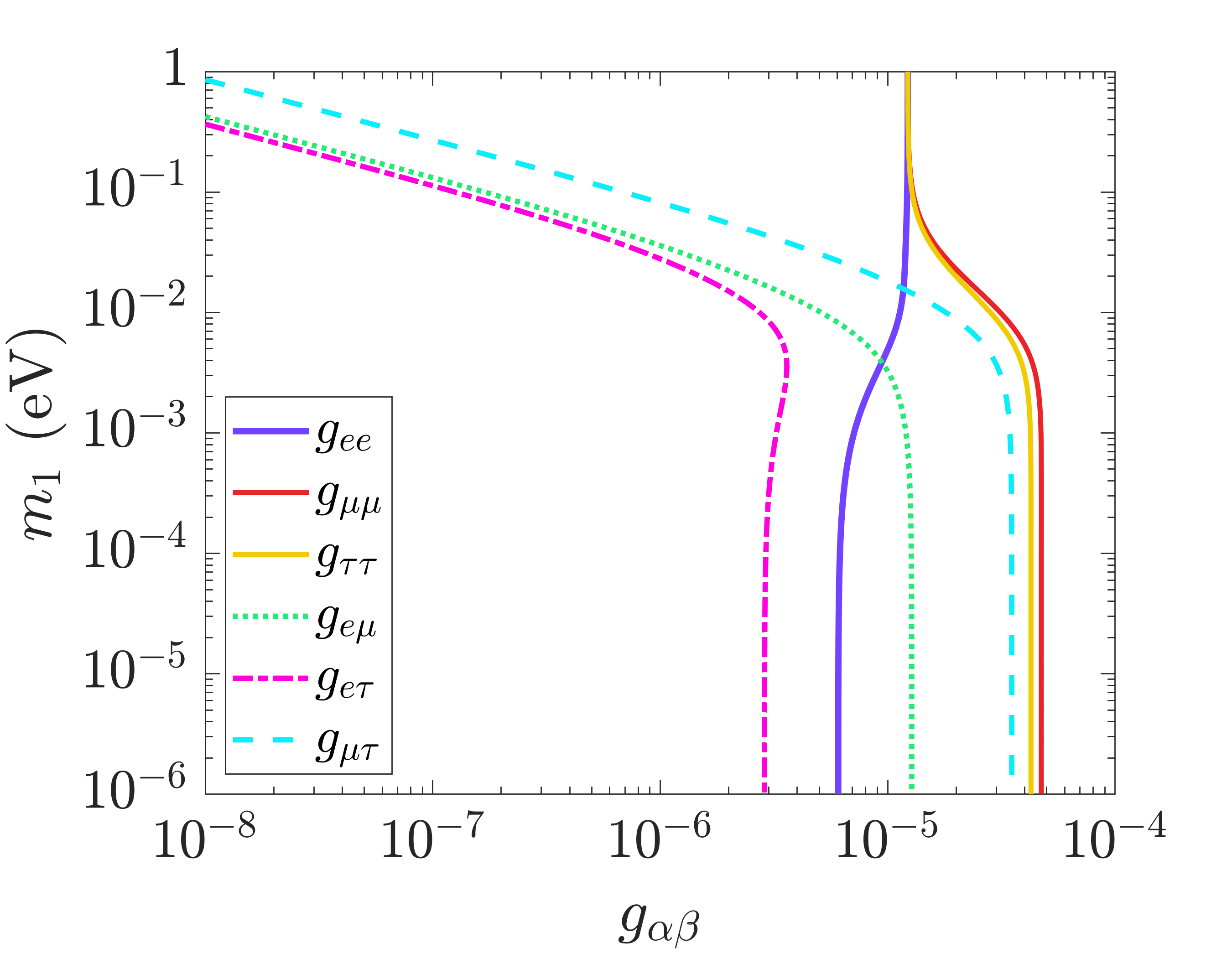}
    \includegraphics[width=0.48\textwidth]{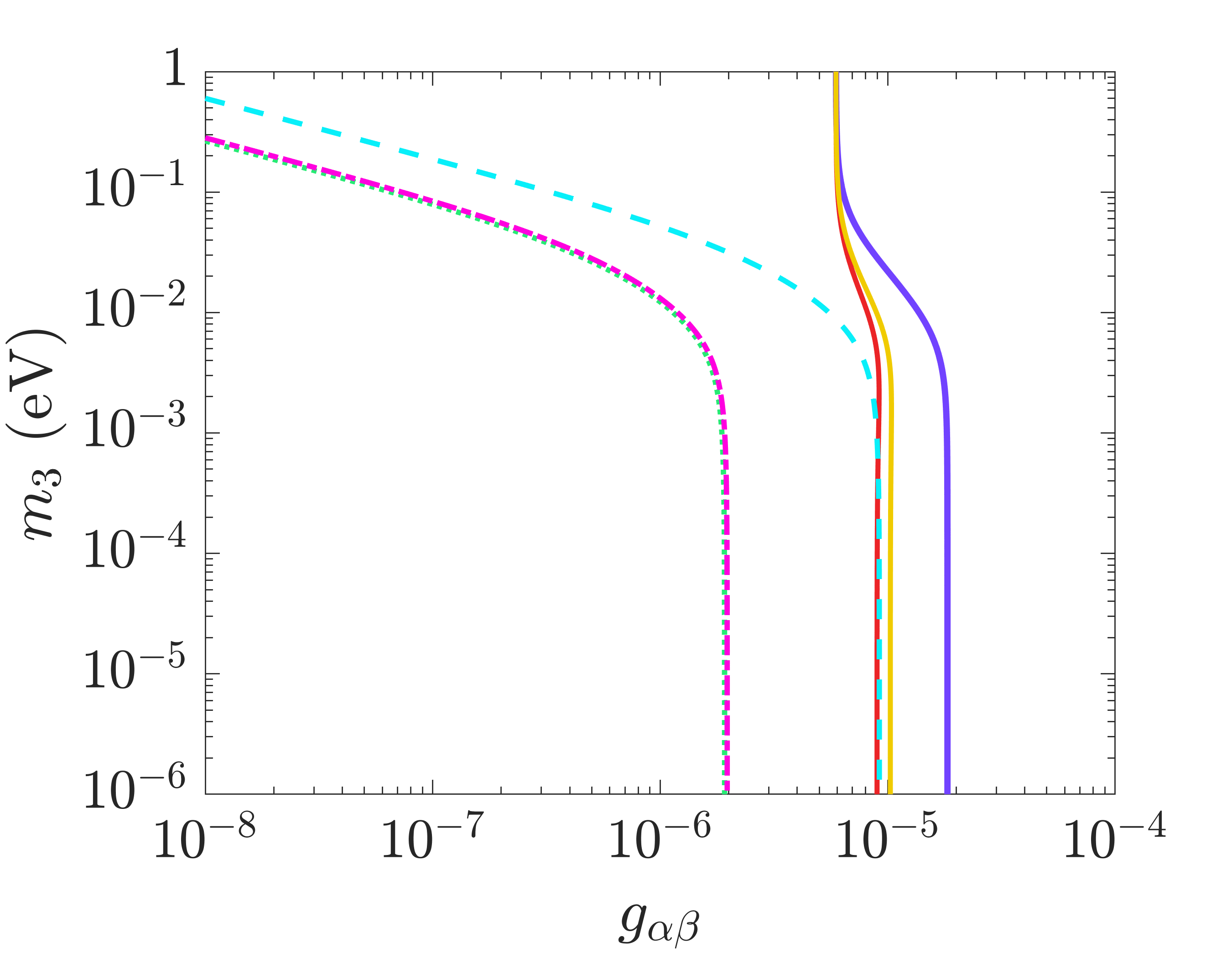}
    \caption{Constraints on the neutrino-Majoron couplings in the flavor basis at 90\% CL, as a function of the lightest neutrino mass, namely $m_1$ for normal (left panel) and $m_3$ for inverted (right panel) neutrino mass ordering. The results shown are obtained from our likelihood analysis of the spectral distortion of SN1987A neutrino events using the 1.44SFHx model. The excluded areas are to the right of the lines.}
    \label{fig:gab-m}
\end{figure}

The neutrino-Majoron coupling matrix in the flavor basis is related to the one in the mass basis through the relation
\begin{equation}
    g_{\alpha\beta} = U_{\alpha i} g_{ij}  U^\dagger_{j \beta} \ .
\end{equation}
Our results on the neutrino-Majoron bounds, presented in the previous section, translated to the flavor basis, are shown in figure~\ref{fig:gab-m}, both in normal and inverted neutrino mass ordering. 

For the $g_{ee}$ coupling, our limits can be compared to the ones established by $0\nu\beta\beta$ decay experiments. Table~\ref{tab:gee_limits} summarizes the upper bounds on $|g_{ee}|$ obtained from our SN1987A analysis and those from such experiments. The ranges shown for our results reflect the dependence on $m_1$ for normal ordering and $m_3$ for inverted ordering. For the $0\nu\beta\beta$ results, the indicated ranges account for the uncertainties in the nuclear matrix elements. From Table~\ref{tab:gee_limits}, one can see that our limits are competitive with those from $0\nu\beta\beta$ decay experiments, and are surpassed by EXO-200 results only~\cite{Kharusi:2021jez}. It is important to note that the results presented in Table~\ref{tab:gee_limits} are the tightest limits obtained at each experiment, corresponding to $0\nu\beta\beta J$ models with one Majoron emission. Other models considered in their analyses yield significantly looser bounds (see ref.~\cite{Kharusi:2021jez}).

\begin{table}
    \setlength{\tabcolsep}{10pt} 
    \renewcommand{\arraystretch}{1.5} 
    \centering
\begin{tabular}{cccc}
\toprule
Analysis                                    & Reference  &                                    & $|g_{ee}|$ bounds     \\ \midrule
\textbf{SN1987A (NO)} & \textbf{This work}      &     & $\mathbf{(0.6-1.2)\times 10^{-5}}$      \\
\textbf{SN1987A (IO)} & \textbf{This work}      &     & $\mathbf{(0.6-1.8)\times 10^{-5}}$      \\ 
$0\nu\beta\beta J$ ($^{136}$Xe) & EXO-200       & \cite{Kharusi:2021jez}     & $(0.4-0.9)\times 10^{-5}$             \\
$0\nu\beta\beta J$ ($^{136}$Xe) & KamLand-Zen   & \cite{KamLAND-Zen:2012uen} & $(0.8-1.6)\times 10^{-5}$             \\
$0\nu\beta\beta J$ ($^{100}$Mo) & NEMO-3        & \cite{NEMO-3:2015jgm}      & $(1.6-3.0)\times 10^{-5}$             \\
$0\nu\beta\beta J$ ($^{82}$Se)  & CUPID-0       & \cite{CUPID-0:2022yws}     & $(1.8-4.4)\times 10^{-5} $            \\
$0\nu\beta\beta J$ ($^{76}$Ge)  & GERDA         & \cite{GERDA:2022ffe}      & $(1.8-4.4)\times10^{-5}$              \\
$0\nu\beta\beta J$ ($^{100}$Mo) & CUPID-Mo      & \cite{CUPID:2024qnd}       & $(4.0-6.9)\times 10^{-5}$ \\
\bottomrule
\end{tabular}
\caption{Upper bounds on $|g_{ee}|$ from different analyses. The limits derived from SN1987A were obtained using the 1.44SFHx supernova model of ref.~\cite{Fiorillo:2023frv}. The corresponding ranges reflect the dependence on the lightest neutrino mass. In contrast, the ranges in the $0\nu\beta\beta J$ results account for the uncertainties in the nuclear matrix elements.}
\label{tab:gee_limits}
\end{table}

It is to be noted that $0\nu\beta\beta$ decay experiments can only probe the $g_{ee}$ element of the flavor coupling matrix. For the other couplings, we can only compare them to the limits from the luminosity argument or meson and lepton decays.  On the other hand, the limits obtained from meson and lepton decays constrain certain combinations of the coupling elements rather than individual ones. A detailed comparison to these limits was presented in our previous work \cite{Ivanez-Ballesteros:2024nws} where we showed that our bounds on the $\sum_\beta g^2_{\alpha\beta}$ ($\alpha, \beta = e, \mu, \tau$) improve by $4-7$ orders of magnitude the bounds derived from meson and lepton decay data \cite{PIENU:2021clt, Lessa:2007up}.

\subsection{Comparison of results with different models}
\begin{figure}[t]
    \centering
    \includegraphics[width=0.49\textwidth]{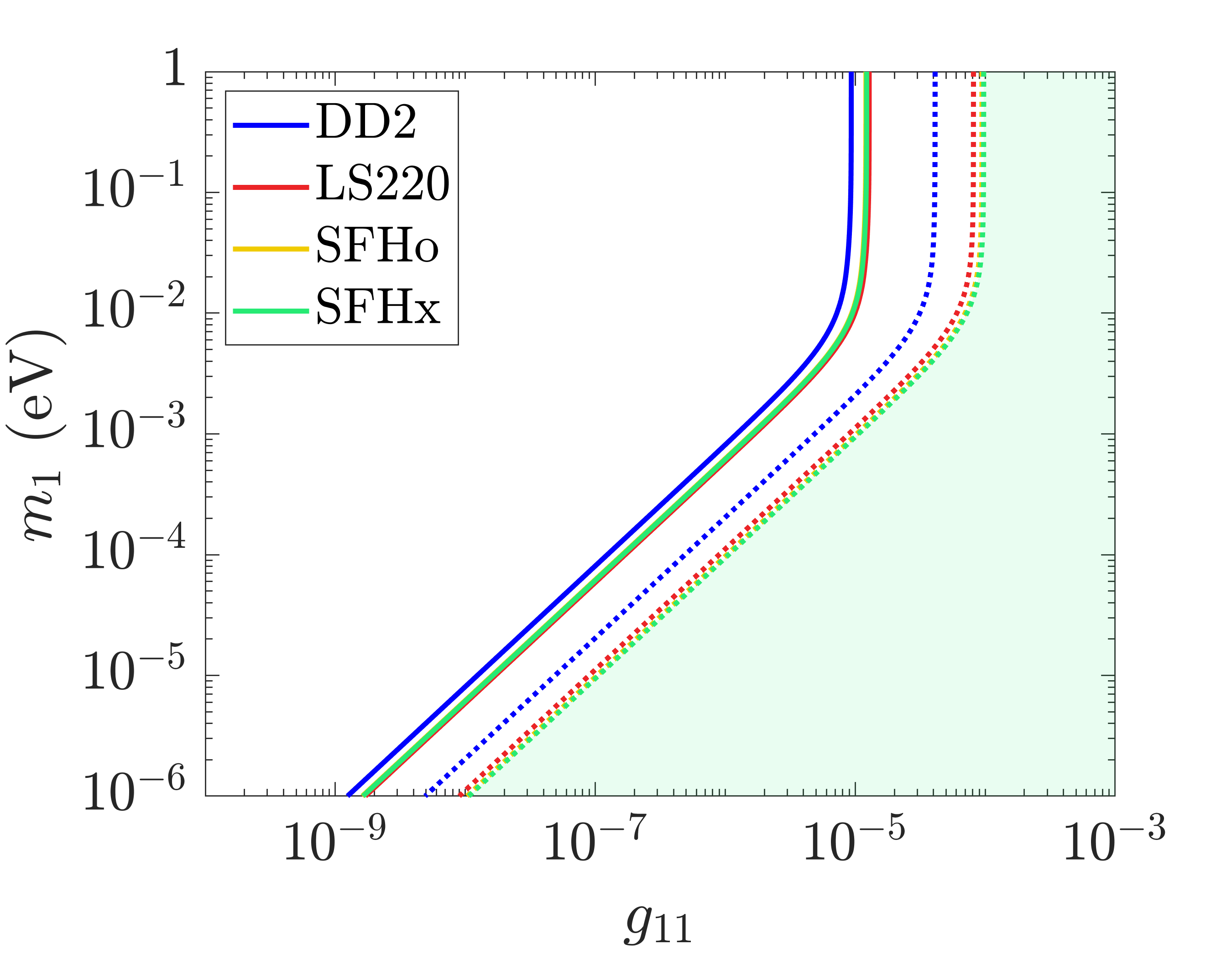}
    \includegraphics[width=0.49\textwidth]{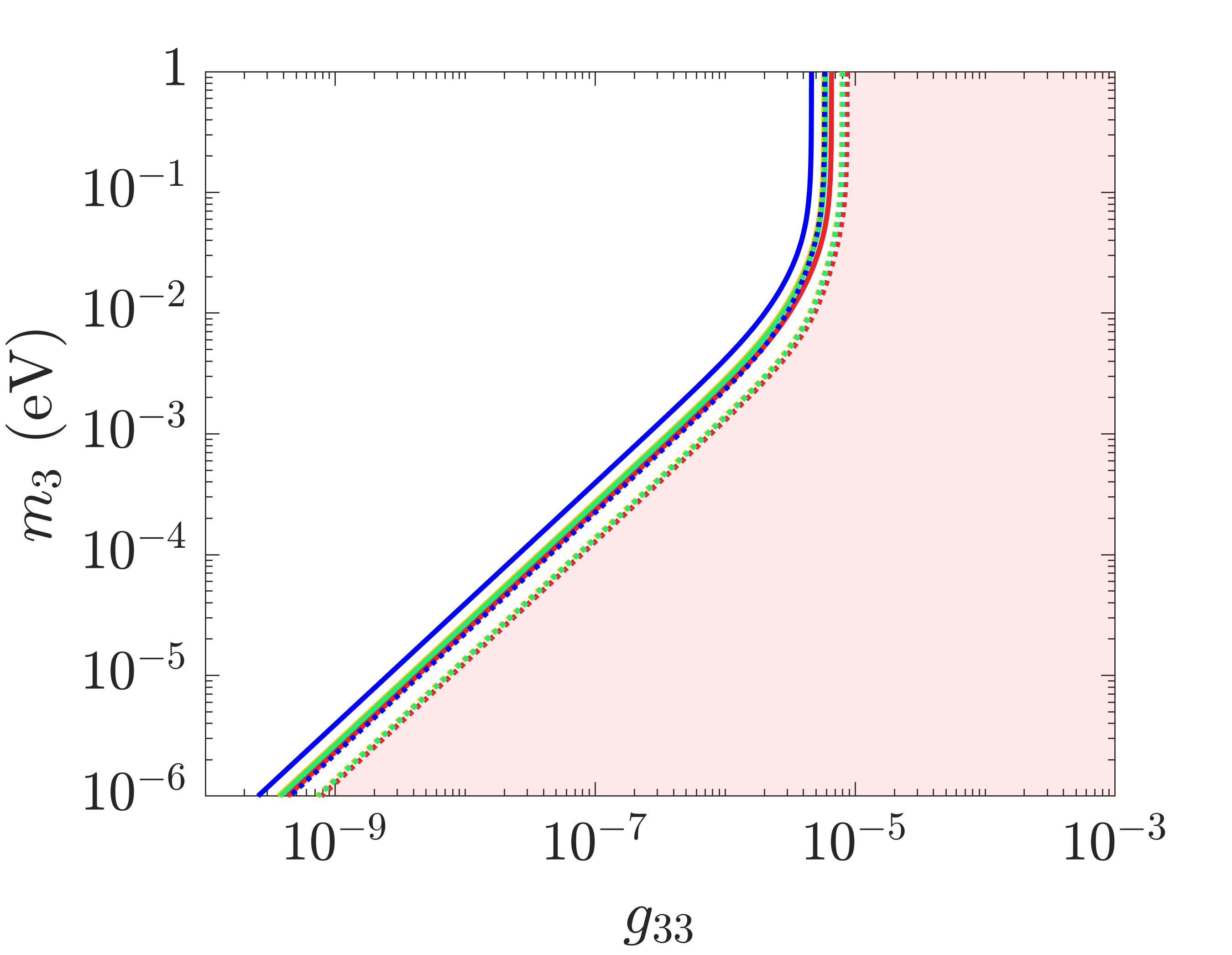}
    \caption{Comparison of our 90\% CL results for the eight models considered: 1.44$M_{\rm \odot}$ (solid lines) and 1.62$M_{\rm \odot}$ (dotted lines), each with four different equations of states (see legend). The left panel displays results in the $g_{11}$--$m_1$ plane for normal ordering, while the right panel shows results in the $g_{33}$--$m_3$ plane in the inverted ordering case. Shaded areas represent the most conservative excluded regions. Results from equations of state SFHo and SFHx are nearly identical, the corresponding lines being almost indistinguishable.}
    \label{fig:models}
\end{figure}

\noindent
We investigated the dependence of our bounds on the SN1987A model used. To this aim, figure~\ref{fig:models} compares the bounds derived using different core-collapse supernova models. The shaded areas indicate the most conservative excluded regions. The figure presents the results in the $g_{11}$--$m_1$ plane for normal mass ordering. In this case, the 90\% CL limits span approximately an order of magnitude. All the $1.44~M_{\rm \odot}$ models (solid lines) produce similar results, while the $1.62~M_\odot$ models (dotted lines) yield looser limits. Additionally, the EOS SFHo and SFHx lead to nearly indistinguishable predictions. We note, however, that the 1.44SFHx model, which we used in our reference analysis, provides the best agreement with SN1987A data according to ref.~\cite{Fiorillo:2023frv}.
Figure~\ref{fig:models} also shows the bounds in the $g_{33}$--$m_3$ plane obtained for the inverted mass ordering. Here, the bounds spread over approximately half an order of magnitude. This suggests that our limits in the inverted ordering case are practically model-independent (for the ensemble of models considered in the present work). We remind though that model 1.44SFHx was found to have the best agreement with SN1987A data.

\section{Prospects on the neutrino-Majoron couplings from a future supernova} \label{sec:prospects}
\noindent
Thanks to their larger fiducial volumes, the next generation of neutrino detectors will be able to observe the neutrino flux from a future galactic core-collapse supernova with significantly higher statistics compared to SN1987A. Additionally, as pointed out by ref.~\cite{Lang:2016zhv}, dark matter detectors will also be sensitive to the total supernova neutrino fluxes, offering complementary observations. We consider here some of the supernova neutrino observatories included in the upgraded version of the Supernova Neutrino Early Warning System (SNEWS 2.0). These are the HK experiment, expected to start running in 2028 \cite{Hyper-Kamiokande:2018ofw}, the JUNO experiment that will start in 2025-2026 \cite{JUNO:2015zny}, the DUNE experiment that will be operating in a staged approach between 2029 and 2032 \cite{DUNE:2020lwj}, and finally the DARWIN experiment \cite{DARWIN:2016hyl}. 

After presenting the details of the response functions and the inputs used for each of these supernova neutrino observatories, we will present here our predictions for the expected number of events associated with a galactic core-collapse supernova neutrino signal, both in the presence and in the absence of neutrino decay. 
Then we will show the outcome of our 2-dimensional likelihood analyses of the spectral distortions induced by neutrino decay in an exploding core-collapse supernova, and the bounds on the neutrino-Majoron couplings in case such a rare event takes place.

\subsection{Experiments} \label{sec:detectors}
\noindent
The expected signal rate at each of the future detectors considered, as a function of the reconstructed energy $E_r$, is calculated through the following expression
\begin{equation}\label{eq:true}
    \frac{dS}{dE_r} = \eta \int_0^\infty  G\big(E_t - E_r, \delta(E_r)\big) \frac{dS_t}{dE_t} dE_t \ ,
\end{equation}
where $dS_t/dE_t$ represents the true signal distribution as a function of the true energy $E_t$, and $\eta$ denotes the efficiency, assumed here to be energy-independent. For all experiments, we considered the idealized case of 100$\%$ efficiency, i.e., $\eta = 1$. The smearing function $G$ is considered Gaussian with uncertainty function $\delta(E_r)$, specific to each detector. 

We now present the experiments and the main detection channels of each detector under consideration. Table \ref{tab:detector_parameters} shows the corresponding number of targets, energy thresholds, and the references for the associated cross sections used in our computations. 
\begin{table}[htbp]
\centering
    \setlength{\tabcolsep}{8pt} 
    \renewcommand{\arraystretch}{1.5} 
    \begin{tabular}{ccccc}
        \toprule
        Experiment & Channel & $N_{\rm t}$ & $E_{th}$ (MeV) & ref. \\ \midrule
        HK & IBD & $1.25 \times 10^{34}$  & 5 & \cite{Strumia:2003zx, Ricciardi:2022pru} \\ 
          & ES & $6.25 \times 10^{34}$   & 5 &\cite{deGouvea:2019wav} \\
          JUNO & IBD & $1.21 \times 10^{33}$ & 5 & \cite{Strumia:2003zx, Ricciardi:2022pru} \\
          & pES & $1.21 \times 10^{33} $ & 0.3 &   \cite{Beacom:2002hs} \\
        DUNE & $\nu_e$-$^{40}$Ar & $6.02 \times 10^{32}$ &   5 &   \cite{snowglobes:webpage} \\
        DARWIN &  CE$\nu$NS & $1.83 \times 10^{29}$&  $10^{-3}$ &  \cite{Lang:2016zhv} \\  \bottomrule
    \end{tabular}
    \caption{Parameters characterizing the experiments and detection channels. The quantities $N_\text{t}$ and $E_{th}$ refer to the number of targets and the energy threshold, respectively, while the last column gives the reference for the cross section associated with the main detection channel of the corresponding experiment. For all experiments, an idealized efficiency of 100$\%$ was applied.}
    \label{tab:detector_parameters}
\end{table}

\subsubsection{Hyper-Kamiokande}
\noindent
HK \cite{Hyper-Kamiokande:2018ofw}, currently under construction in Japan, is set to be the successor of SK. This water Cherenkov detector will have a fiducial volume of 187~kton, more than eight times that of SK. HK is expected to start taking data in 2028.

The main detection channel at HK will be IBD, making this detector predominantly sensitive to the supernova $\bar\nu_e$ flux. The true signal for this detection channel was calculated with eq.~\eqref{eq:true}. 
For the reconstructed signal, we applied the uncertainty function given by \cite{Martinez-Mirave:2024hfd}
\begin{equation}
    \delta(E_r) = 0.1\sqrt{E_r {\rm[MeV]}} \ .
\end{equation}

Additionally, HK will be capable of detecting supernova neutrinos via elastic scattering on electrons (ES):
\begin{equation}
    \overset{(-)}{\nu_\alpha} + e^{-} \to \overset{(-)}{\nu_\alpha} + e^{-},~
    {\rm with}\ \alpha = e, \mu, \tau \ .
\end{equation}
Although its cross-section is several orders of magnitude smaller than that of IBD, this detection channel is still valuable as it probes all neutrino flavors. The true signal for ES can be expressed as
\begin{equation}
    \frac{dS_t}{dE_t} = N_t \int_{E_\nu^{\rm min}}^\infty \sum_{\alpha} \Big[ 
    \sigma_{{\rm ES}-\nu_\alpha}(E_\nu, E_t) \phi_{\nu_\alpha}(E_\nu) + \Big.    \Big. \sigma_{{\rm ES}-\bar\nu_\alpha}(E_\nu, E_t) \phi_{\bar\nu_\alpha}(E_\nu)
    \Big] dE_\nu \ ,
\end{equation}
where $E_t$ here represents the true kinetic recoil energy of the electron, and $E_\nu^{\rm min} \simeq E_t + m_e/2$ is the minimum energy necessary to induce a recoil $E_t$. The cross-section $\sigma_{{\rm ES}-\nu_\alpha}$ depends on the neutrino or antineutrino flavor. 
Finally, to compute the reconstructed signal for this channel, we used the following uncertainty function \cite{Super-Kamiokande:2010tar, Martinez-Mirave:2024hfd}:
\begin{equation}
    \frac{\delta(E_r)}{E_r} = 0.0349 + \frac{0.376}{\sqrt{E_r {\rm[MeV]} }} - \frac{0.123}{E_r {\rm[MeV]}} \ .
\end{equation}
Other detection channels at HK include charged-current interactions of $\nu_e$ and $\bar\nu_e$ on oxygen nuclei that we will neglect since these channels are subdominant \cite{Hyper-Kamiokande:2021frf}.

\subsubsection{JUNO}
\noindent
The Jiangmen Underground Neutrino Observatory (JUNO) \cite{JUNO:2015zny} is a liquid-scintillator neutrino experiment under construction in China. This detector will have a fiducial volume of 17~kton. 
As HK, the main detection channel of JUNO is IBD. We took the energy resolution
\begin{equation}
    \delta(E_r) = 0.03\sqrt{E_r {\rm[MeV]}} \ .
\end{equation}

In addition, JUNO is sensitive to all neutrino flavors through elastic scattering of neutrinos on protons (pES):
\begin{equation}
    \overset{(-)}{\nu_\alpha} + p \to \overset{(-)}{\nu_\alpha} + p,~
    {\rm with}\ \alpha = e, \mu, \tau \ .
\end{equation}
The cross-section of this interaction is around 3 times smaller than that of IBD. However, this lower cross-section is compensated for by the contribution of all flavors. The signal distribution as a function of the proton recoil energy $E'_t$ is
\begin{equation}
    \frac{dS_t}{dE_t'} = N_t \int_{E_\nu^{\rm min}}^\infty \Big[
    \sigma_{{\rm pES}\text{-}\nu}(E_\nu, E_t') \sum_{\alpha} \phi_{\nu_\alpha}(E_\nu) + \Big. \Big.
    \sigma_{{\rm pES}\text{-}\bar\nu}(E_\nu, E_t') \sum_{\alpha} \phi_{\bar\nu_\alpha}(E_\nu) \Big]
    dE_\nu \ ,
\end{equation}
where we distinguish the pES cross-section for neutrinos $\sigma_{{\rm pES}\text{-}\nu}$ and antineutrinos $\sigma_{{\rm pES}\text{-}\bar\nu}$. The minimum neutrino energy needed to produce a recoil energy $E_t'$ is 
\begin{equation}
    E_\nu^{\rm min} = \frac{E'_t + \sqrt{E'_t(E'_t + 2m_p)}}{2} \ .
\end{equation}

One should take into account that the visible energy $E_t$ in the detector is strongly quenched with respect to $E'_t$ \cite{Beacom:2002hs, Dasgupta:2011wg, vonKrosigk:2013sa}. To convert from the true proton energy to the visible one, we used the quenching factor from \cite{JUNO:2015zny}, obtained using the results of \cite{vonKrosigk:2013sa}. The observed signal was obtained using the same uncertainty function as for the IBD channel.

\subsubsection{DUNE}
\noindent
The upcoming Deep Underground Neutrino Experiment (DUNE) \cite{DUNE:2020lwj}, in the United States, will also be able to detect supernova neutrinos. DUNE will consist of four 10-kton liquid argon time projection chambers, resulting in a total fiducial volume of 40~kton. This detector will be built in a staged approach starting from early 2029 up to 2032.

The main detection channel at DUNE will be CC interactions on liquid argon, which will allow probing the $\nu_e$ component of the supernova flux 
\begin{equation}
    \nu_e + ^{40}{\rm Ar} \to e^- + ^{40}{\rm K}^* \ .
\end{equation}
The signal as a function of the true neutrino energy is
\begin{equation}
    \frac{dS_t}{dE_\nu} = N_t \sigma_{\nu\text{-}\rm Ar}(E_\nu) \phi_{\nu_e}(E_\nu) \ .
\end{equation}
The cross-section $\sigma_{\nu\text{-}\rm Ar}$ was taken from \texttt{SNOwGLoBES} \cite{snowglobes:webpage}.
To obtain the reconstructed signal, we used the following uncertainty function \cite{DUNE:2020zfm}
\begin{equation}
    \delta(E_r) = 0.2 E_r .
\end{equation}

\subsubsection{DARWIN}
\noindent
DARk matter WImp search with liquid xenoN (DARWIN) \cite{DARWIN:2016hyl} will be an experiment whose main purpose will be the direct detection of dark matter. In addition, it will also serve as a neutrino detector. DARWIN will consist of 40~tons of Xenon.
This detector will be sensitive to neutrinos and antineutrinos of all flavors through Coherent Elastic Neutrino-Nucleus Scattering (CE$\nu$NS) measured for the first time in 2017 by the COHERENT Collaboration \cite{COHERENT:2017ipa}, more than 40 years after its predictions by Freedman \cite{Freedman:1973yd}.

The true signal as a function of the true nuclear recoil energy $E_t$ is
\begin{equation}
    \frac{dS_t}{dE_t} = N_t \int_{E_\nu^{\rm min}}^\infty  \sigma_{\rm CE \nu NS} (E_\nu, E_t) \\
    \sum_\alpha \left[  \phi_{\nu_\alpha}(E_\nu) + \phi_{\bar\nu_\alpha}(E_\nu)   \right]
    dE_\nu \ .
\end{equation}
The minimum neutrino energy necessary to produce a recoil energy $E_t$ is $E_\nu^{\rm min} \simeq \sqrt{m_N E_t /2}$, where $m_N$ is the mass of the nucleus.
For the reconstructed signal, we assumed an uncertainty function given by \cite{Schumann:2015cpa, Martinez-Mirave:2024hfd}
\begin{equation}
    \frac{\delta(E_r)}{E_r} = 0.077 + \frac{0.232}{\sqrt{E_r {\rm[keV]} }} - \frac{0.069}{E_r {\rm[keV]}} \ .
\end{equation}

\begin{figure}[ht]
    \centering
    \begin{subfigure}{\textwidth}
        \includegraphics[width=0.49\textwidth]{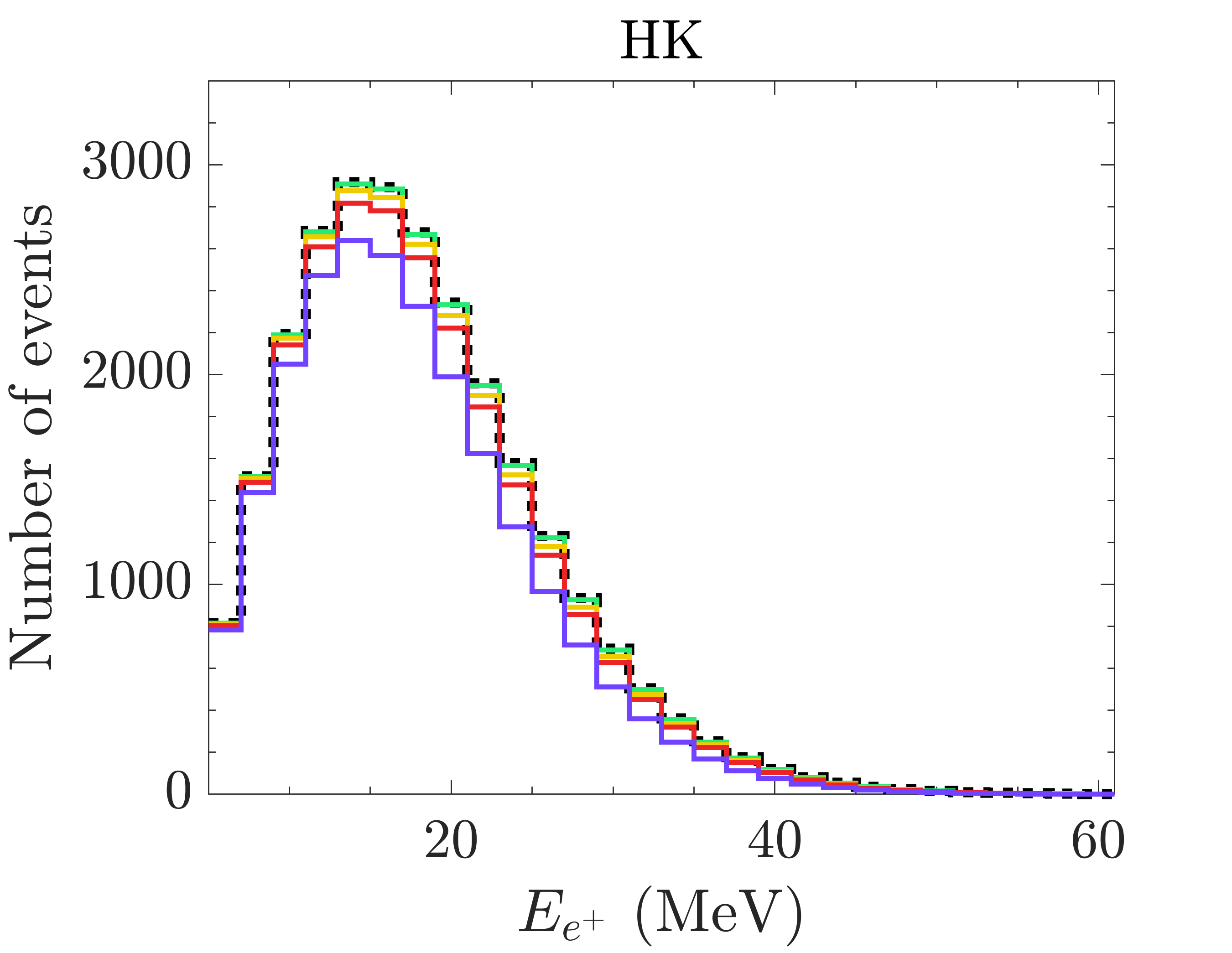}
        \includegraphics[width=0.49\textwidth]{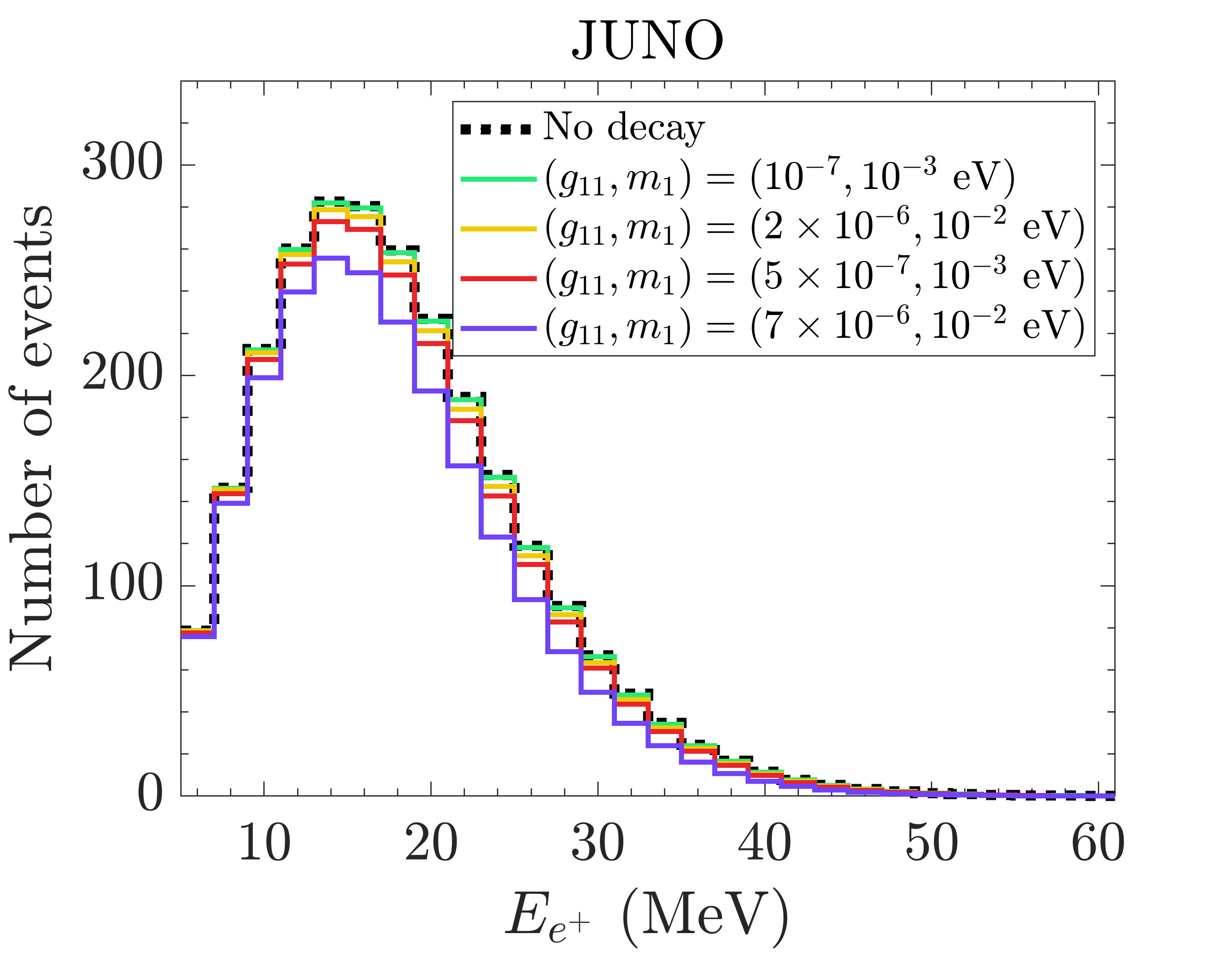}
        \caption{1.44SFHx (NS)}
    \end{subfigure}
    \begin{subfigure}{\textwidth}
        \includegraphics[width=0.49\textwidth]{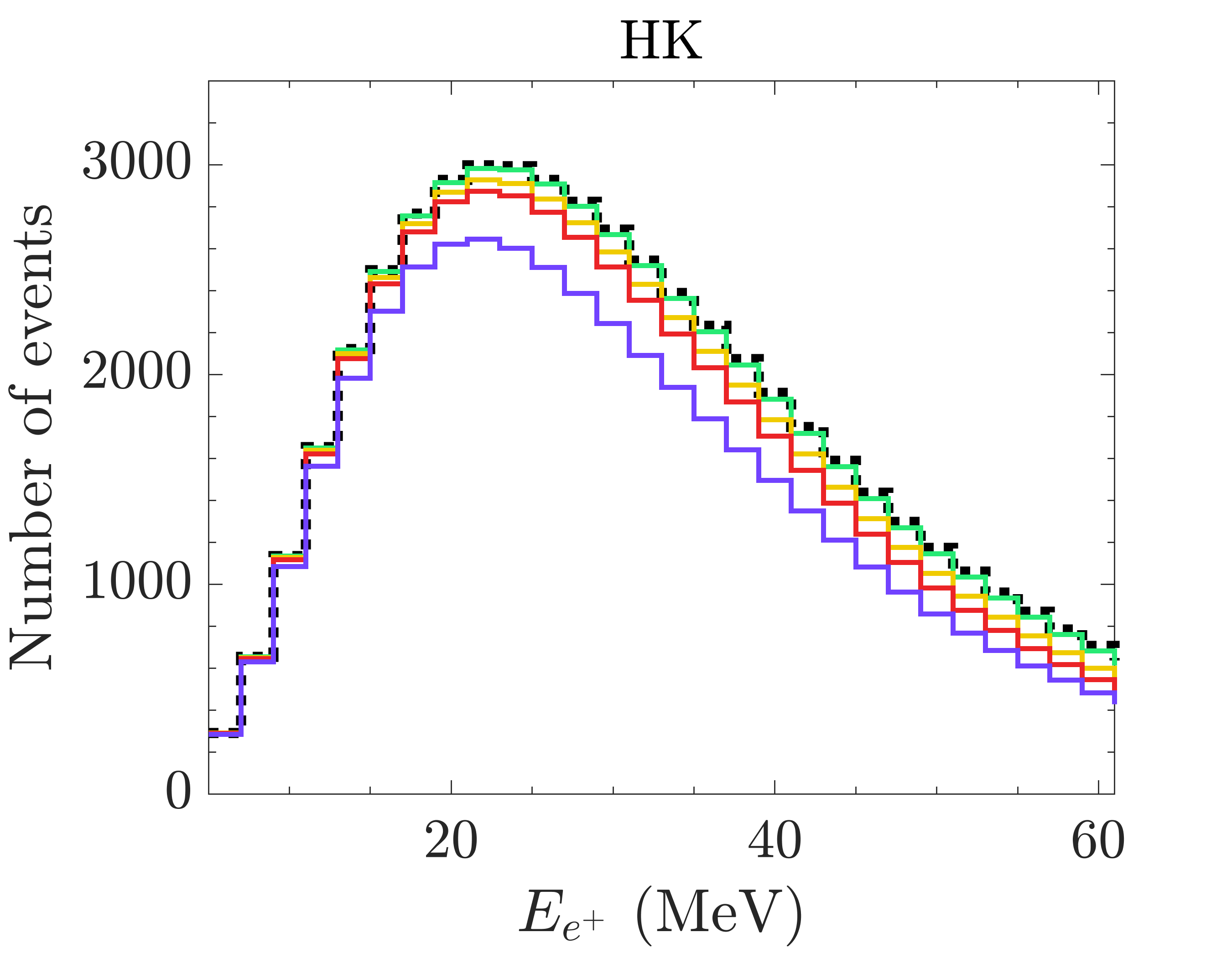}
        \includegraphics[width=0.49\textwidth]{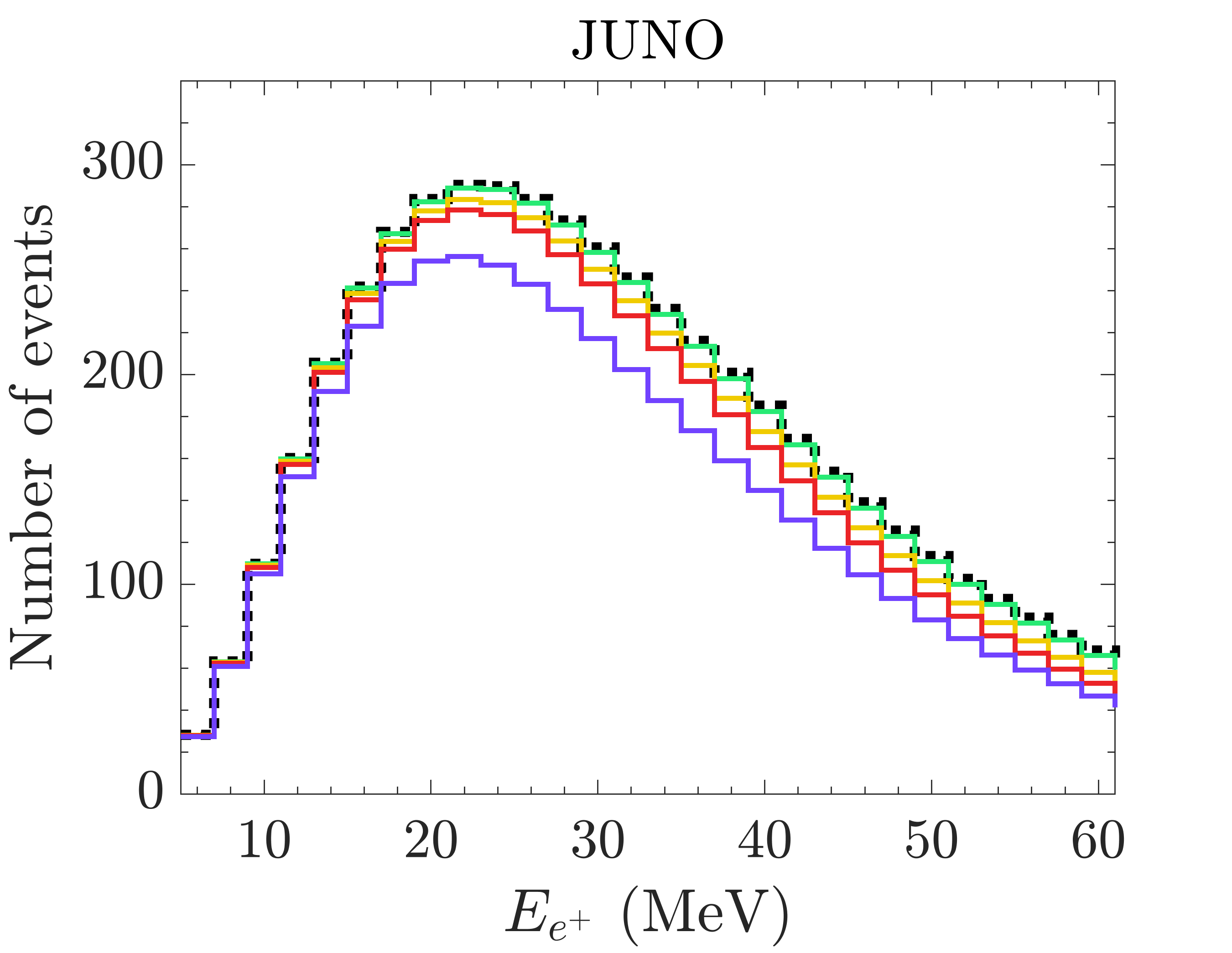}
        \caption{30T (BH)}
    \end{subfigure}
    \caption{Expected IBD events from a future core-collapse supernova at 10~kpc, assuming normal ordering and including neutrino-Majoron interactions for different values of $g_{11}$ and $m_{1}$ compatible with current constraints. Predictions are given for HK and JUNO for the NS case using the 1.44SFHx model (top panels) and for the BH case using the 30T model (bottom panels). For comparison, the black dotted line shows the prediction in the absence of neutrino-Majoron interactions.}
    \label{fig:eventsIBD}
\end{figure}

\subsection{Predictions on the neutrino number of events}
We present here our predictions for the number of events in the upcoming JUNO experiment, the near-future HK, as well as the more distant-future DUNE and DARWIN experiments. The computations were performed by considering a galactic core-collapse supernova at a nominal distance of 10~kpc, 8~kpc (corresponding to the galactic center), and 0.2~kpc (approximate distance to the supernova candidate Betelgeuse). While the distance simply acts as a scaling factor for the number of events, it is still interesting to examine its influence on the projected bounds on the neutrino-Majoron couplings.

\begin{figure}
    \centering
    \begin{subfigure}{\textwidth}
        \includegraphics[width=0.49\textwidth]{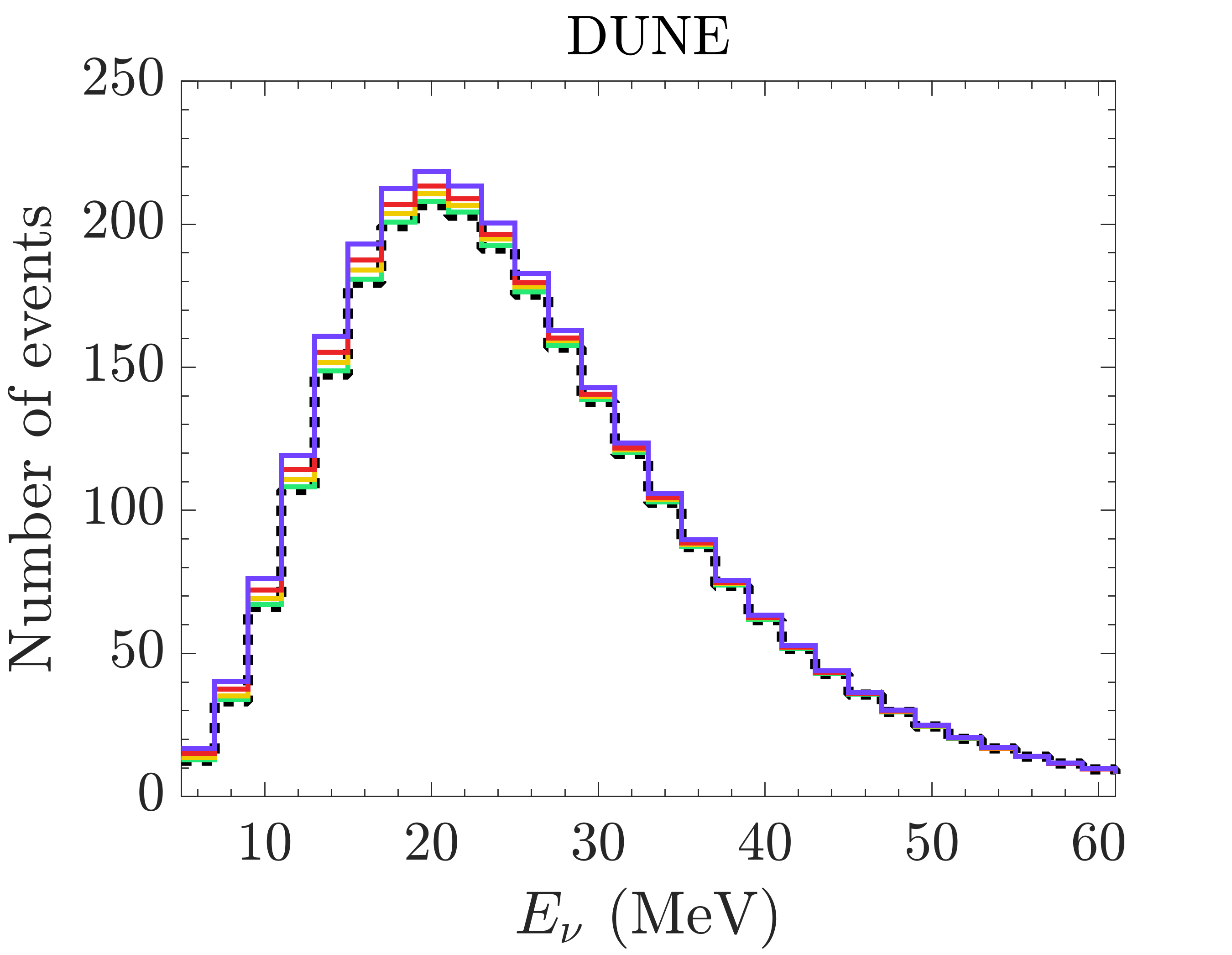}
        \includegraphics[width=0.49\textwidth]{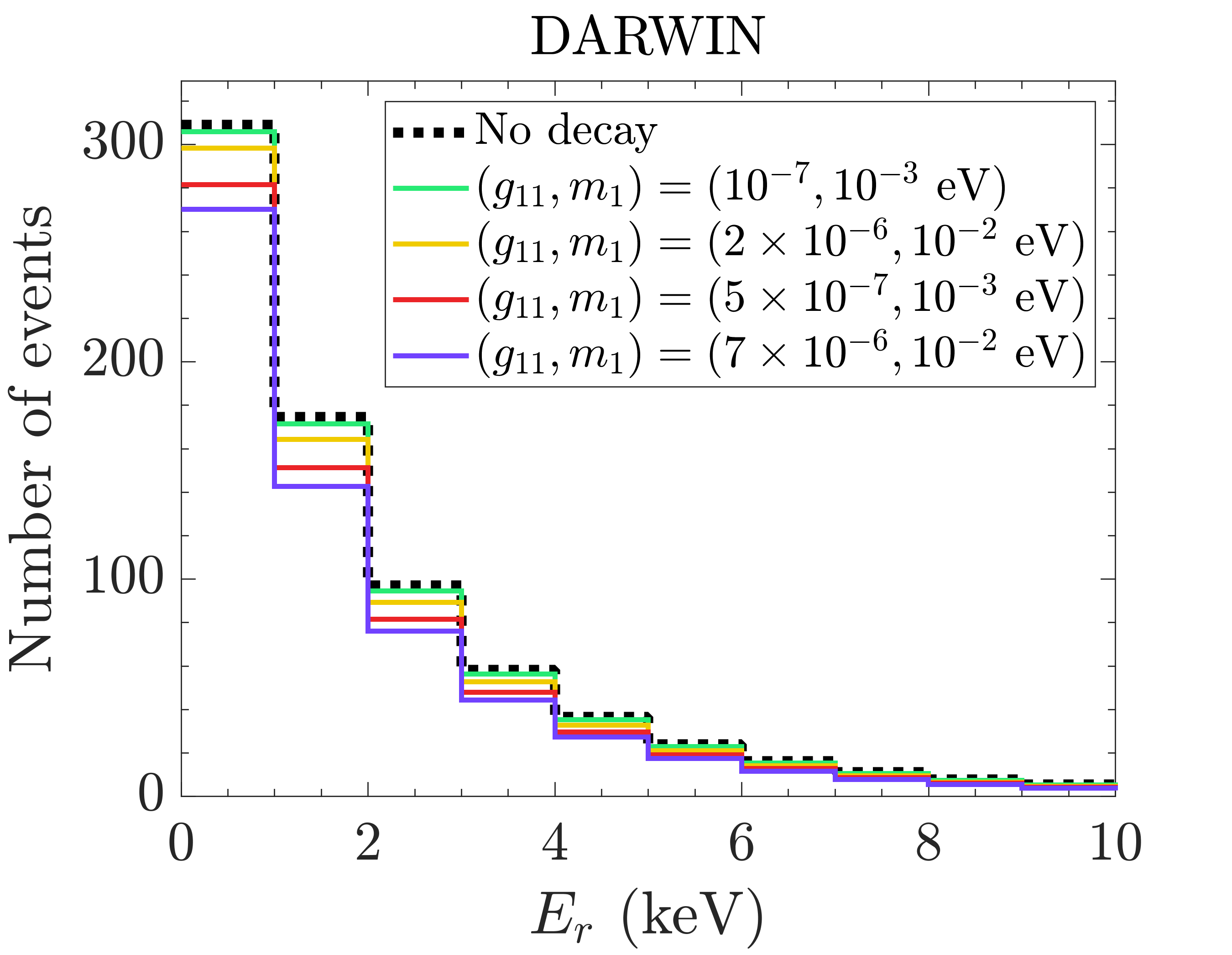}
        \caption{1.44SFHx (NS)}
    \end{subfigure}
    \begin{subfigure}{\textwidth}
        \includegraphics[width=0.49\textwidth]{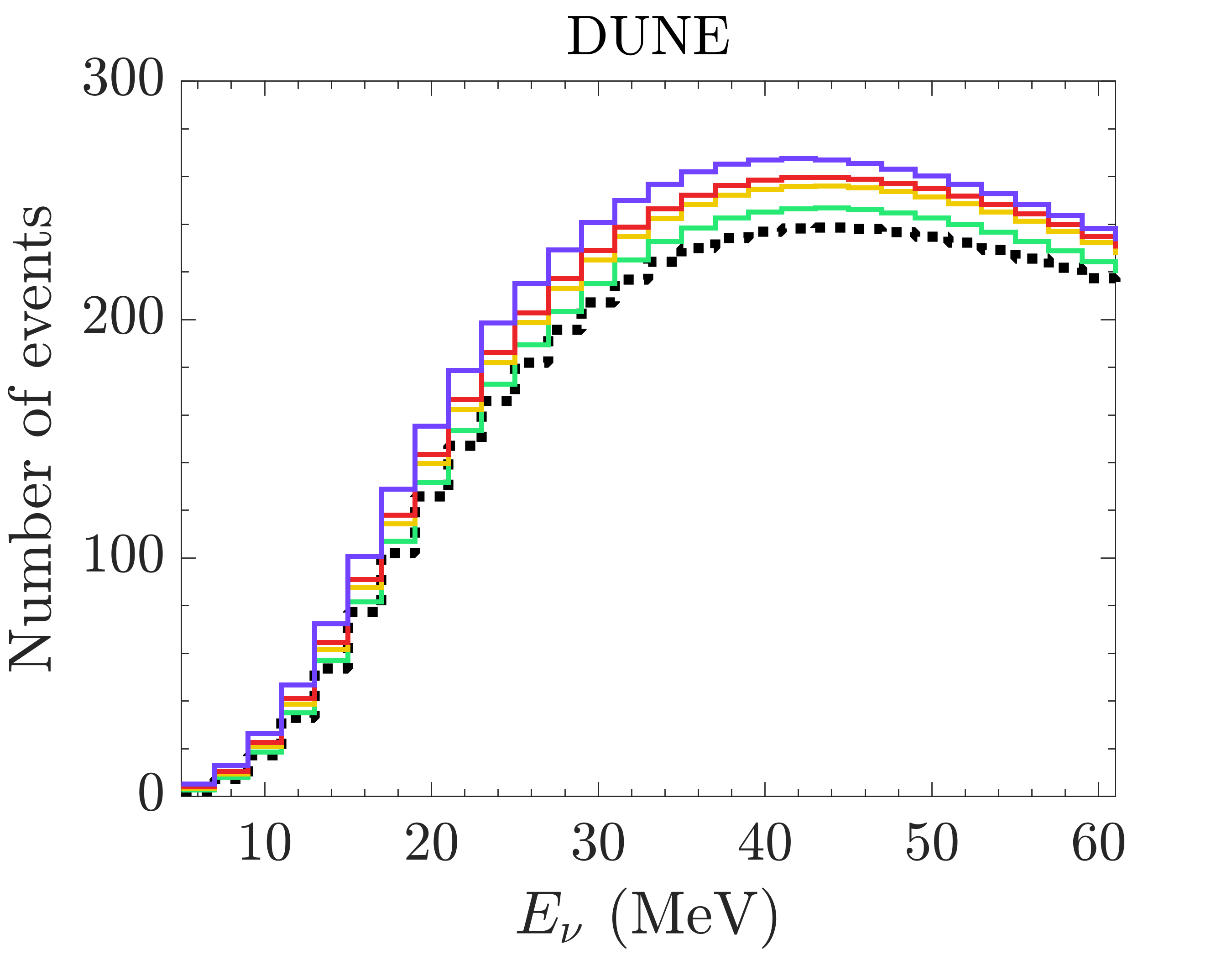}
        \includegraphics[width=0.49\textwidth]{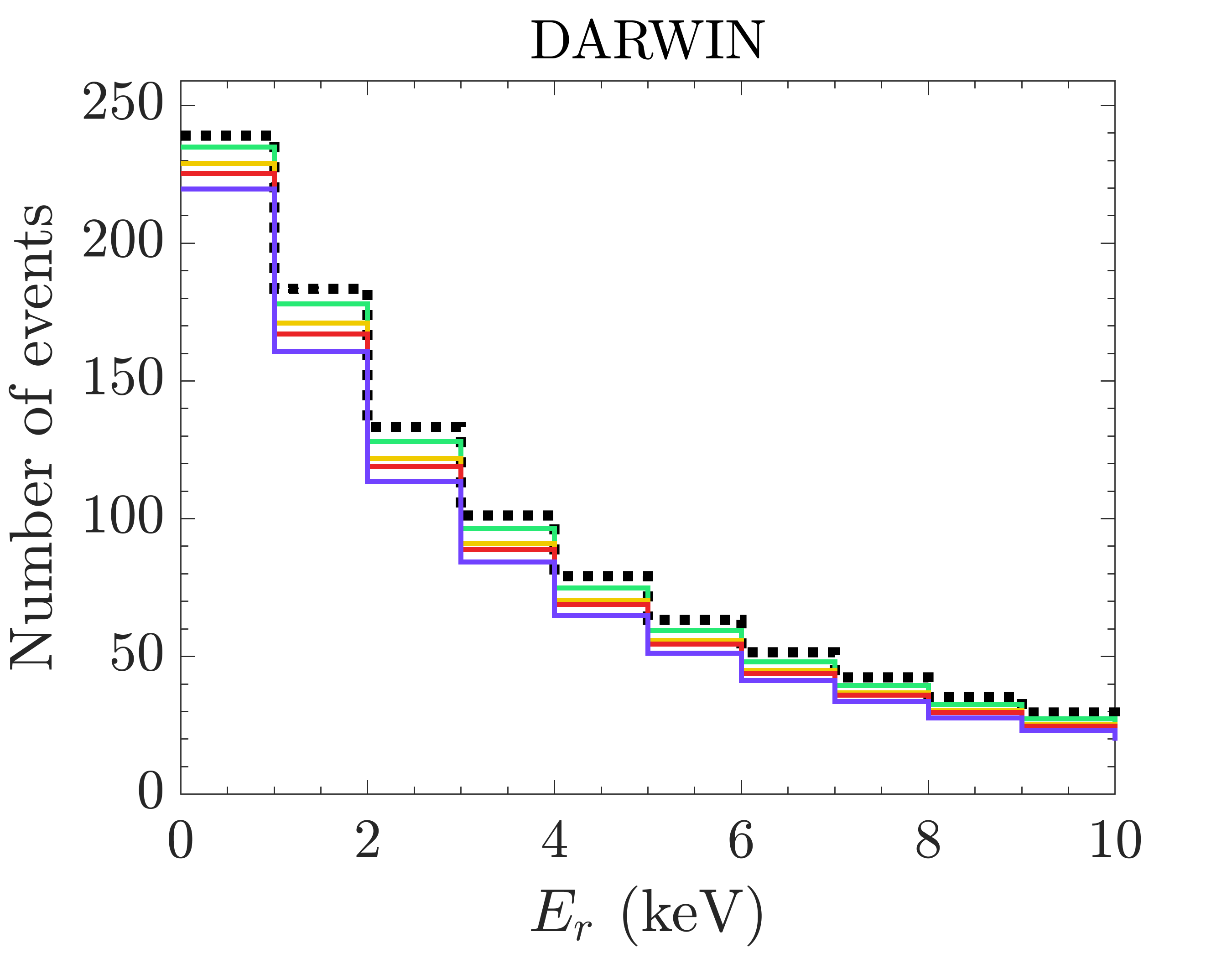}
        \caption{30T (BH)}
    \end{subfigure}
    \caption{Expected events from a future core-collapse supernova located at 10~kpc, assuming normal ordering and including neutrino-Majoron interactions for different values of $g_{11}$ and $m_{1}$ compatible with current constraints. Predictions correspond to $\nu$-Ar scattering in DUNE and to coherent neutrino-nucleus scattering in DARWIN for the NS case using the 1.44SFHx model (top panels) and for the BH case using the 30T model (bottom panels). For comparison, the black dotted line shows the prediction in the absence of neutrino-Majoron interactions.}
    \label{fig:eventsDUNE-DARWIN}
\end{figure}

Figure~\ref{fig:eventsIBD} presents the expected IBD events as a function of positron energy in HK and JUNO, assuming normal mass ordering. The results correspond to the 1.44SFHx model from the Garching simulations for the NS case and the 30T model from Nakazato’s simulations for the BH case. The predictions are shown for the no-decay scenario and for neutrino decay with several values of the $g_{11}$ coupling and the lightest neutrino mass $m_1$. Similarly, figure~\ref{fig:eventsDUNE-DARWIN} shows the expected $\nu$–$^{40}$Ar events at DUNE as a function of neutrino energy, along with the CE$\nu$NS events at DARWIN. These results are also shown for normal ordering and for the NS and BH cases using the 1.44SFHx and 30T models, respectively. Note that the effect of decay in matter is modest in the ES signal at HK and in the pES signal at JUNO (see appendix~\ref{appendix:results}). For the NS case, we find that decay effects are negligible in inverted ordering across all detectors and detection channels. In contrast, for the BH case, the dominant channels exhibit visible effects due to decay in inverted ordering (see appendix~\ref{appendix:results}). In this case, the IBD rates are overall smaller than in normal ordering, although the decay-induced suppression is more pronounced in inverted ordering for the BH case. Conversely, for DUNE and DARWIN, the impact of decay is reduced in inverted ordering compared to normal ordering.

Table~\ref{tab:events-NO} shows the total number of events in normal ordering, assuming fixed values of $g_{11}$ and $m_1$. (The results in inverted mass ordering can be found in appendix~\ref{appendix:results}.) The results are given for the NS case using the 1.44SFHx model, and for the BH case with the 30T model. To compare the impact of decay in different scenarios, table~\ref{tab:events-combined} presents the number of events for the no-decay case and for the case with the largest decay effects, along with the corresponding ratio. These values are provided for both normal and inverted ordering, and for the NS and BH cases using the 1.44SFHx and 30T models, respectively. Note that the largest decay effects are obtained for $g_{11} = 7 \times 10^{-6}$ and $m_1 = 10^{-2}$~eV for normal ordering and $g_{33} = 10^{-6}$ and $m_3 = 10^{-2}$~eV for inverted ordering. 

\begin{table}
\centering
\setlength{\tabcolsep}{5.5pt} 
\renewcommand{\arraystretch}{1.2} 
\begin{tabular}{lrrrrr}
\toprule
        & No decay & \begin{tabular}[c]{@{}c@{}}$g_{11} = 10^{-7}$ \\ $m_1 = 10^{-3}$~eV \end{tabular} & \begin{tabular}[c]{@{}c@{}}$g_{11} = 2 \times10^{-6}$ \\ $m_1 = 10^{-2}$~eV \end{tabular} & \begin{tabular}[c]{@{}c@{}}$g_{11} = 5 \times10^{-7}$ \\ $m_1 = 10^{-3}~$eV \end{tabular} & \begin{tabular}[c]{@{}c@{}}$g_{11} = 7 \times 10^{-6}$ \\ $m_1 = 10^{-2}~$eV \end{tabular} \\ \midrule
HK-IBD  & \tightcell{26085 \\ (58362)} & \tightcell{25957 \\ (57298)} & \tightcell{25444 \\ (53858)} & \tightcell{24794 \\ (51777)} & \tightcell{22441 \\ (47047)} \\
HK-ES    & \tightcell{1190 \\ (1516)}   & \tightcell{1185 \\ (1506)}   & \tightcell{1174 \\ (1498)}   & \tightcell{1154 \\ (1497)}   & \tightcell{1143 \\ (1494)}   \\
JUNO-IBD & \tightcell{2525 \\ (5650)}   & \tightcell{2513 \\ (5547)}   & \tightcell{2463 \\ (5214)}   & \tightcell{2400 \\ (5012)}   & \tightcell{2173 \\ (4554)}   \\
JUNO-pES & \tightcell{459 \\ (3997)}    & \tightcell{426 \\ (3637)}    & \tightcell{377 \\ (3363)}    & \tightcell{348 \\ (3291)}    & \tightcell{304 \\ (3060)}    \\
DUNE     & \tightcell{2452 \\ (10489)}  & \tightcell{2467 \\ (10821)}  & \tightcell{2498 \\ (11212)}  & \tightcell{2530 \\ (11351)}  & \tightcell{2588 \\ (11580)}  \\
DARWIN   & \tightcell{406 \\ (870)}     & \tightcell{394 \\ (816)}     & \tightcell{369 \\ (767)}     & \tightcell{337 \\ (749)}     & \tightcell{313 \\ (708)}     \\ 
\bottomrule
\end{tabular}
\caption{Expected total number of events from a future supernova located at 10~kpc, including neutrino-Majoron interactions producing neutrino nonradiative two-body decay in matter. The results are for the case of normal mass ordering. The first column shows the experiment and the detection channel considered. The second column shows the results in the absence of decay, while the other columns give the expected values for different values of the neutrino-Majoron couplings $g_{11}$ and the lightest neutrino mass $m_{1}$. The values correspond to the NS case with the 1.44SFHx model and to the BH case with the 30T model (in parentheses).}
\label{tab:events-NO}
\end{table}

\begin{table}
\centering
    \setlength{\tabcolsep}{6pt} 
    \renewcommand{\arraystretch}{1.5}
\begin{tabular}{lrrrrrr}
\toprule
& \multicolumn{3}{c}{\textbf{NS}} & \multicolumn{3}{c}{\textbf{BH}} \\
\cmidrule(r){2-4} \cmidrule(l){5-7}
& No decay & Largest effect & Ratio & No decay & Largest effect & Ratio \\ 
\cmidrule(r){2-4} \cmidrule(l){5-7} 
HK-IBD   & \tightcell{26085 \\ (27884)} & \tightcell{22441 \\ (26304)} & \tightcell{0.86 \\ (0.94)} &
           \tightcell{58362 \\ (42560)} & \tightcell{47047 \\ (31888)} & \tightcell{0.81 \\ (0.75)} \\
HK-ES    & \tightcell{1190 \\ (1169)}   & \tightcell{1143 \\ (1164)}   & \tightcell{0.96 \\ (1.00)} &
           \tightcell{1516 \\ (1631)}   & \tightcell{1494 \\ (1617)}   & \tightcell{0.99 \\ (0.99)} \\
JUNO-IBD & \tightcell{2558 \\ (2736)}   & \tightcell{2204 \\ (2583)}   & \tightcell{0.86 \\ (0.94)} &
           \tightcell{5659 \\ (4126)}   & \tightcell{4563 \\ (3093)}   & \tightcell{0.81 \\ (0.75)} \\
JUNO-pES & \tightcell{459 \\ (459)}     & \tightcell{304 \\ (423)}     & \tightcell{0.66 \\ (0.92)} &
           \tightcell{3997 \\ (3997)}     & \tightcell{3060 \\ (3597)}     & \tightcell{0.77 \\ (0.90)} \\
DUNE     & \tightcell{2452 \\ (2136)}   & \tightcell{2588 \\ (2170)}   & \tightcell{1.06 \\ (1.02)} &
           \tightcell{10489 \\ (10982)}   & \tightcell{11580 \\ (11631)}   & \tightcell{1.10 \\ (1.06)} \\
DARWIN   & \tightcell{406 \\ (406)}     & \tightcell{313 \\ (393)}     & \tightcell{0.77 \\ (0.97)} &
           \tightcell{870 \\ (870)}     & \tightcell{708 \\ (810)}     & \tightcell{0.81 \\ (0.93)} \\
\bottomrule
\end{tabular}
\caption{Expected total number of events for each experiment and detection channel (first column) from a future supernova located at 10 kpc, including neutrino–Majoron interactions. The second and third columns show the results for the no-decay case and for the scenario with the largest decay effect, respectively. The fourth column gives the ratio between the largest effect and the no-decay case. Results are given for normal ordering and inverted ordering (in parentheses). The largest decay effects are obtained for $g_{11} = 7 \times 10^{-6}$ and $m_1 = 10^{-2}~$eV for normal ordering and $g_{33} = 10^{-6}$ and $m_3 = 10^{-2}~$eV for inverted ordering. Results were obtained for the NS case using the 1.44SFHx model and for the BH case using the 30T model (shown in columns fifth to seven).}
\label{tab:events-combined}
\end{table}

\subsection{Prospects on the neutrino-Majoron couplings} It is interesting to determine the sensitivity to the neutrino-Majoron couplings that will be obtained from the observation of the next galactic core-collapse supernova. To that aim, we perform a likelihood analysis of simulated data (for details on the analysis, see appendix~\ref{appendix:statistics}). Considering the IBD events in HK, the $\nu$-$^{40}$Ar events in DUNE, and the CE$\nu$NS events in DARWIN, we simulate the events at these detectors assuming a \textit{true} model without neutrino-Majoron interactions, and extract constraints by comparing it to a set of \textit{test} models including such interactions and profiling over this set of test models. 

We assume that the large signal from the next galactic event will allow identification of the remnant (NS or BH), and accordingly, restrict the profiling to models with the same outcome as the true model. For the NS case, we consider two true models: 1.36DD2 and 1.62SFHx, chosen from the Garching group simulations for their distinct standard fluxes and decay effects. These “extreme” cases help assess the model dependence of the results. The profiling includes these two and an intermediate model, 1.44SFHx. For the BH case, the true models considered are 30T and 30S, selected among the three models from Nakazato's simulations. The profiling is performed over all three: 30L, 30T, 30S.

\begin{figure}[t]
    \centering
    \begin{subfigure}{\textwidth}
        \includegraphics[width=0.49\textwidth]{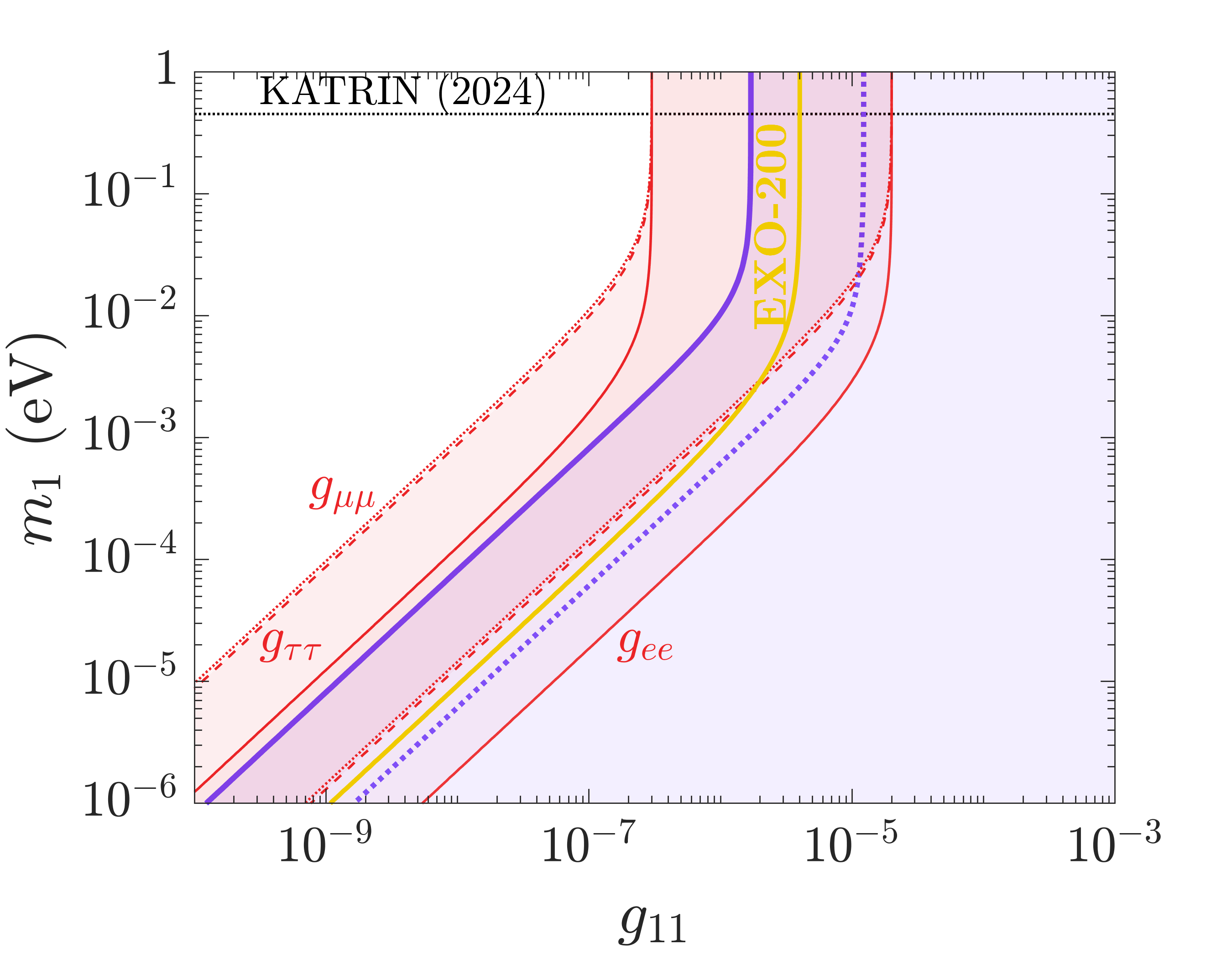}
        \includegraphics[width=0.49\textwidth]{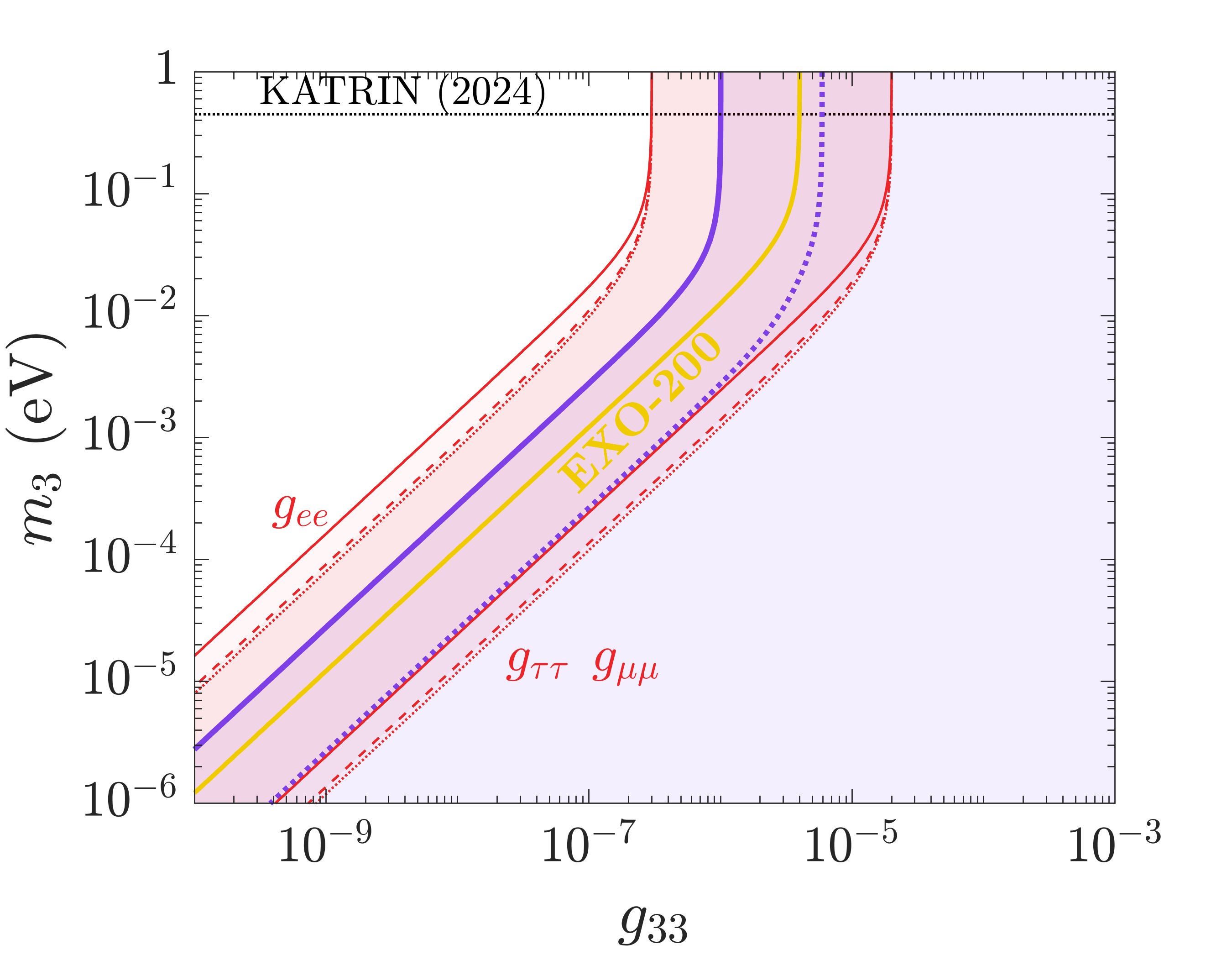}
        \caption{True model: 1.62SFHx (NS)}
        \vspace{1em}
    \end{subfigure}
    \begin{subfigure}{\textwidth}
        \includegraphics[width=0.49\textwidth]{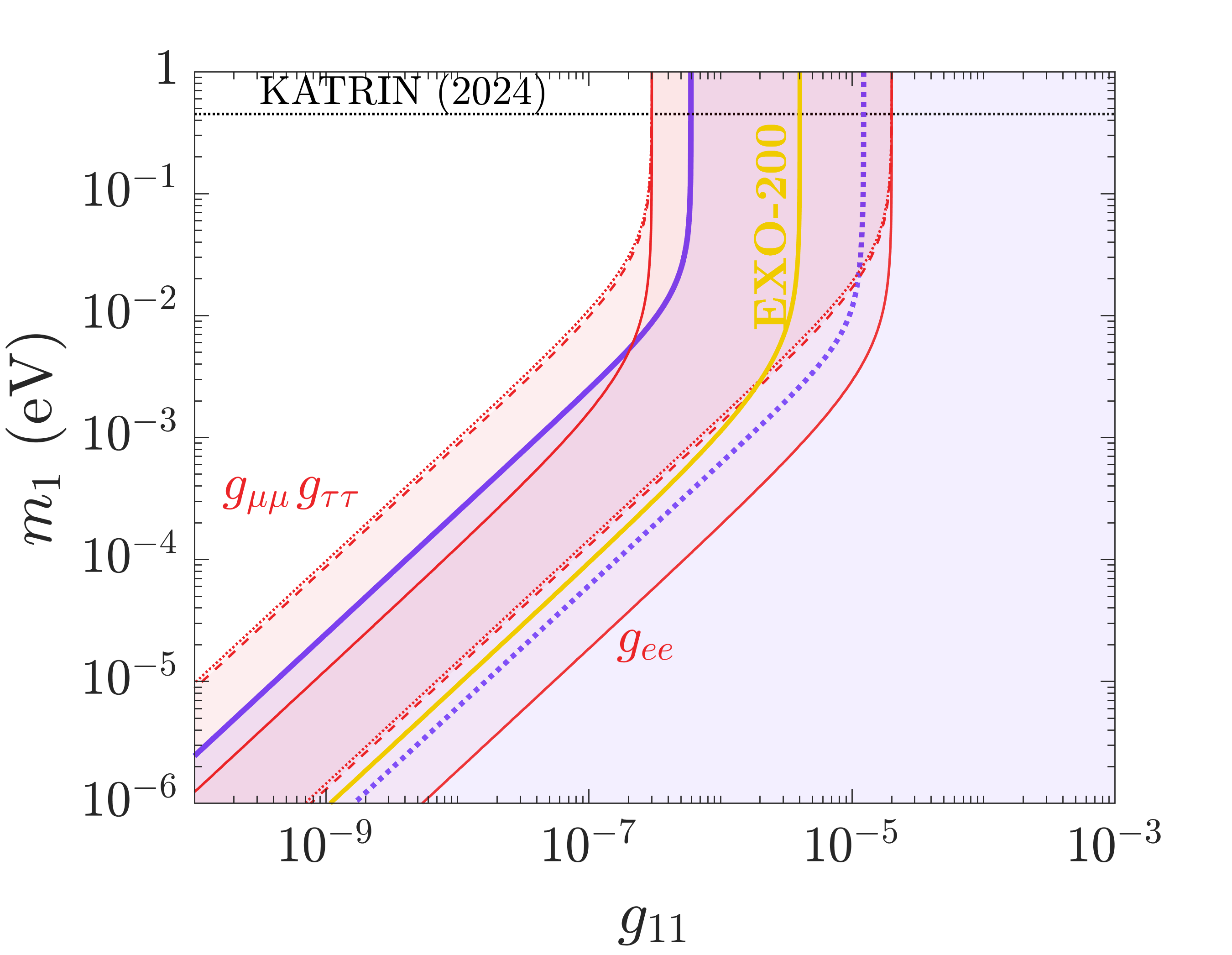}
        \includegraphics[width=0.49\textwidth]{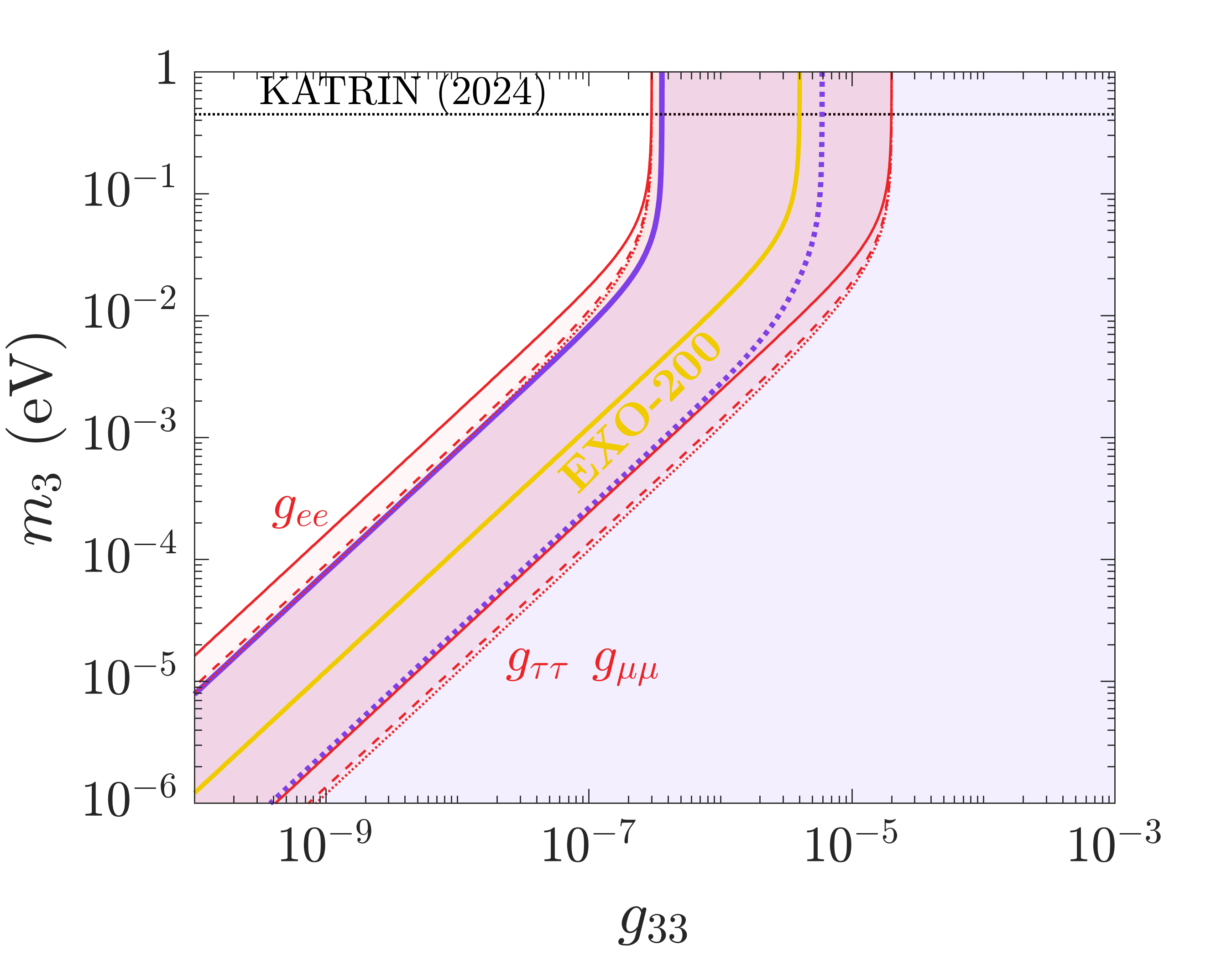}
        \caption{True model: 30T (BH)}
    \end{subfigure}
    \caption{Projected bounds on the neutrino-Majoron coupling in the $g_{ii}$--$m_i$ plane, with $i = 1$ for normal ordering (left panels) and $i = 3$ for inverted ordering (right panels), for a NS-forming collapse (top panels) and a BH-forming collapse (bottom panels) at 10~kpc. Solid blue lines show the $90\%$~CL limits from this analysis, while dotted blue lines indicate the corresponding constraints from our SN1987A reference study. For comparison, we include bounds from the Majoron luminosity argument from ref.~\cite{Kachelriess:2000qc} (red shaded regions and lines), obtained assuming one active flavor-coupling at a time, and the most stringent current limits from neutrinoless double-beta decay experiments (solid yellow line), obtained by EXO-200 with the most optimistic nuclear matrix element~\cite{Kharusi:2021jez}. The upper limit on the absolute neutrino mass from KATRIN is also shown~\cite{Katrin:2024tvg}. Note that the bounds are obtained here by profiling on the other models for the same case (see text).}
    \label{fig:gii-marg}
\end{figure}

For each possible outcome, NS or BH, we found the projected bounds to be nearly equal for the two true models considered. Therefore, in the following, unless otherwise stated, we only show the results for the 1.62SFHx model in the NS case and the 30T model in the BH case.
Figure~\ref{fig:gii-marg} presents a comparison of the projected bounds with current limits on the neutrino-Majoron coupling. Both the results for the NS case in normal and inverted mass ordering and the BH case ones are given. The figure also shows a comparison of our bounds from our SN1987A reference analysis and the current most optimistic limits from EXO-200 \cite{Kharusi:2021jez}. Moreover the limits from the luminosity argument from \cite{Kachelriess:2000qc} applied to the couplings $g_{ee}$, $g_{\mu\mu}$, and $g_{\tau\tau}$ are shown. One can see that the projected bounds from a future supernova improve significantly upon those from SN1987A, especially if a BH is formed. Such improvement is essentially coming from the increased statistics.

\begin{figure}[t]
\centering
        \includegraphics[width=0.5\textwidth]{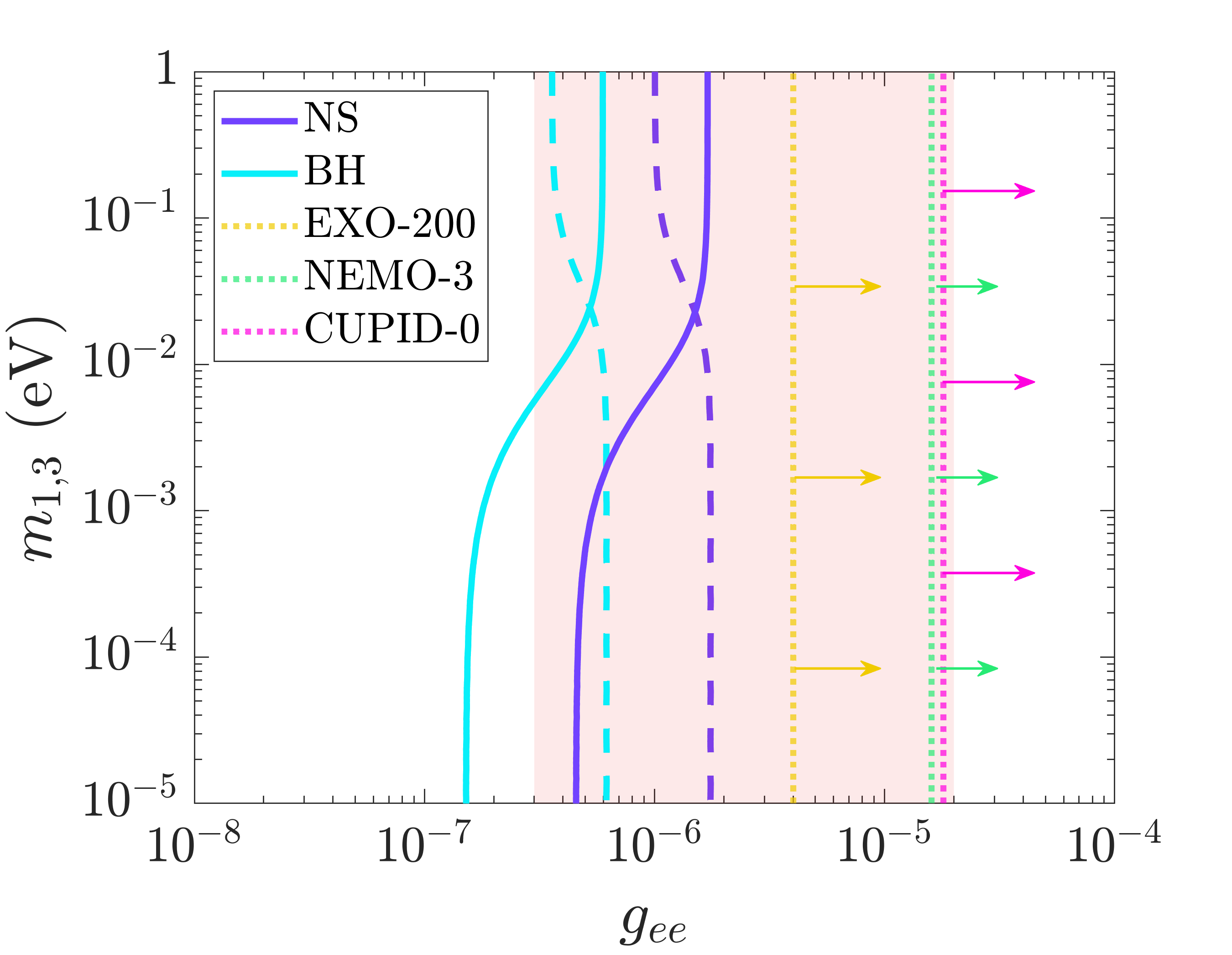}
    \caption{ Projected bounds on the neutrino-Majoron coupling $g_{ee}$ from neutrino decay in matter, as a function of the lightest stable neutrino mass eigenstate $m_1$ for normal (blue solid lines) and $m_3$ for inverted (blue dashed lines) mass orderings. The results shown correspond to the NS case using the 1.62SFHx model and the BH case with the 30T model. For comparison, the most stringent limits for massless Majorons from neutrinoless double-beta decay experiments are also given, for the EXO-200 \cite{Kharusi:2021jez}, NEMO-3 \cite{NEMO-3:2015jgm}, and CUPID-0 \cite{CUPID-0:2022yws} experiments. The uncertainties from the nuclear matrix elements are shown by the arrows. The red band corresponds to the constraint from the supernova cooling argument \cite{Kachelriess:2000qc}.}
    \label{fig:gee-fu}
\end{figure}

To assess how our projected bounds would improve upon current limits from laboratory experiments, we translate them into the $g_{ee}$--$m_{1,3}$ plane. Figure~\ref{fig:gee-fu} shows the results for the NS and the BH cases, considering both mass orderings, in comparison with the ones from the luminosity argument ref.~\cite{Kachelriess:2000qc} and from EXO-200 \cite{Kharusi:2021jez}, NEMO-3 \cite{NEMO-3:2015jgm}, and CUPID-0 \cite{CUPID-0:2022yws} experiments. Also shown is how the bounds vary due to the uncertainty of the neutrinoless double-beta decay nuclear matrix elements.

\begin{figure*}[t]
\begin{center}
\begin{subfigure}{0.49\textwidth}
\includegraphics[width=1\textwidth]{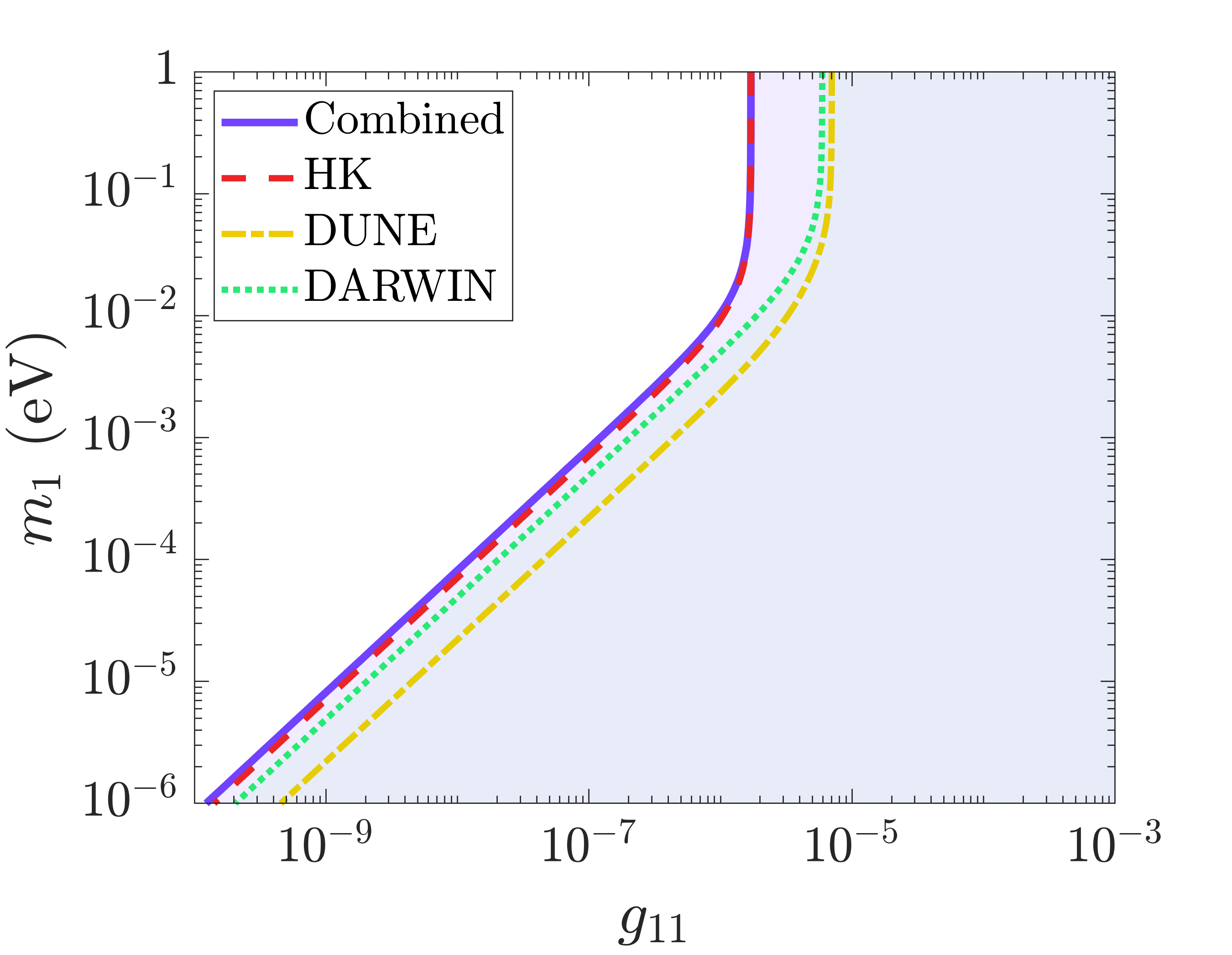}
\end{subfigure}
\begin{subfigure}{0.49\textwidth}
\includegraphics[width=1\textwidth]{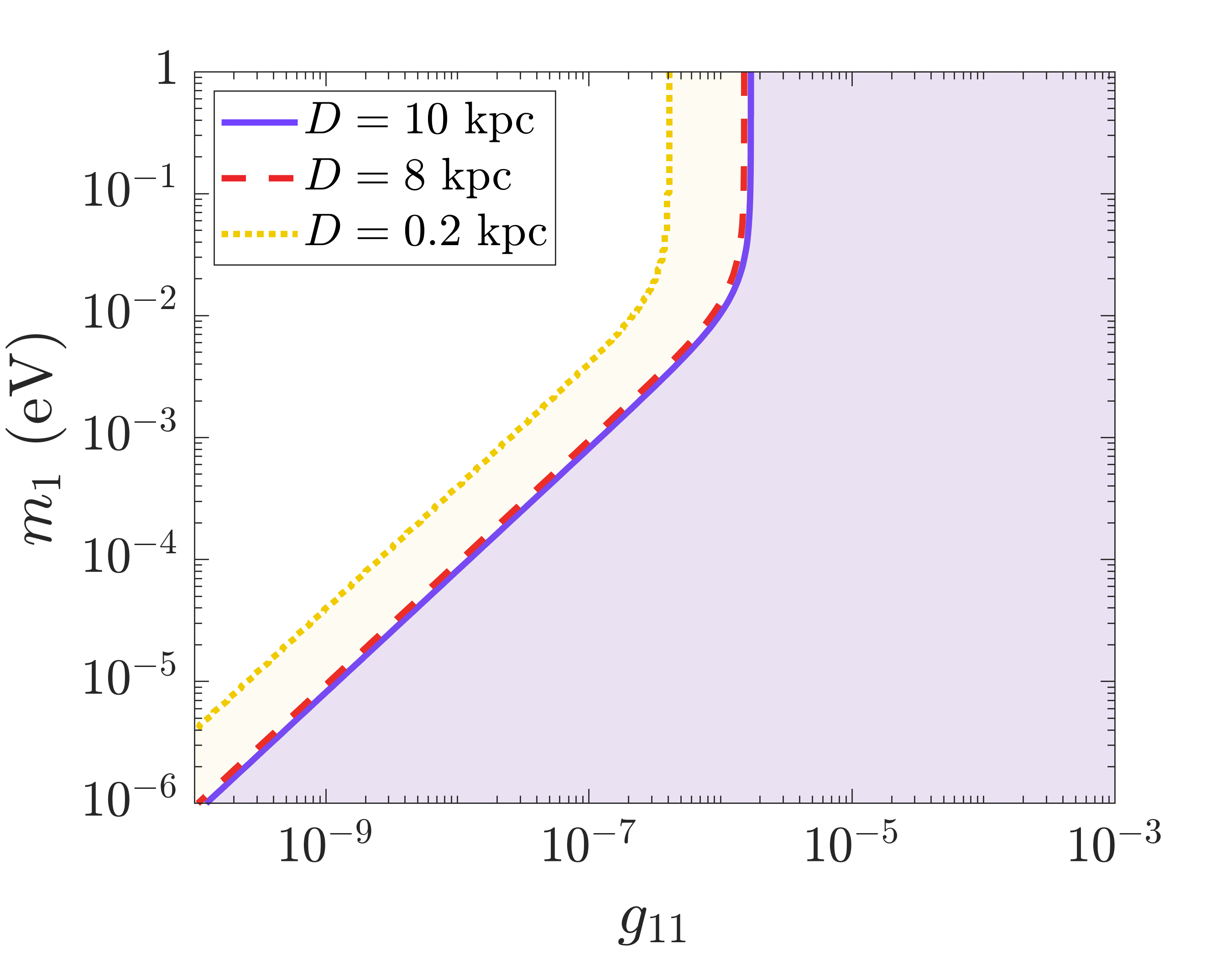}
\end{subfigure}
\caption{Limits on the neutrino-Majoron coupling $g_{11}$ (90$\%$ C.L.) as a function of the lightest mass eigenstate $m_1$ in the case of neutrino normal mass ordering. The results shown are obtained by considering a NS with 1.62SFHx as the true model and profiling over the others. Left figure: Bounds from signals at individual detectors, and from their combination. The results are for a supernova at 10 kpc. Right figure: Bounds vary depending on the supernova core-collapse supernova distance, taken from the nominal value of 10 kpc, to 8 kpc (galactic center), to about 0.2 kpc (distance of Betelgeuse) as examples. }
\label{fig:otherbounds}
\end{center}
\end{figure*}

In order to highlight the role of the different experiments in determining the bounds,  
we show the projected bounds from individual cases in figure~\ref{fig:otherbounds}, along with the result of the combined analysis. The results are in normal ordering for the NS case, assuming a distance of 10~kpc, for HK, DUNE, DARWIN, along with those from the combined analysis. This comparison highlights the dominant role of HK, whose strong sensitivity largely drives the results of the combined analysis. 

Finally, it is interesting to look at the dependence of the limits on neutrino-Majoron couplings from neutrino decay in matter on the possible location of the supernova. This is shown in figure~\ref{fig:otherbounds}, where results for the NS case and a distance of 10~kpc, 8~kpc, and 0.2~kpc are given. Interestingly, a close supernova located at 0.2~kpc would result in bounds nearly an order of magnitude larger than those from an event at 10~kpc.

\section{Neutrino decay in matter and impact on the Diffuse Supernova Neutrino Background} \label{sec:dsnb}
After investigating the prospects of setting improved limits on neutrinos coupling to Majorons from the next supernova, we turn to the case of the DSNB, whose discovery might lie in the foreseeable future. Here we present the results of the first investigation of how neutrino non-radiative two-body decay in matter can influence the DSNB.
The DSNB flux, produced by past core-collapse supernova explosions, needs important information from cosmology, astrophysics, and neutrino physics. 
As such, its observation will bring unique information with respect to individual core-collapse supernovae. 
We will show how this process impacts the DSNB fluxes and event rates in the upcoming JUNO, HK, and near-future DUNE and DARWIN experiments.
But let us first see the theoretical framework of our study and the scenarios we considered. Note that here we employ the parametrization commonly used. 
Recently, a new parametrization based on the compactness has been suggested \cite{Lunardini:2025nvk}.

\subsection{The DSNB flux}
The DSNB flux reads
\begin{equation} \label{eq:flux}
    \begin{aligned}
        F_{\nu_i}(E_\nu) & = c \int_0^{\infty}  \mathrm{d}z \left[ \int_\Omega \mathrm{d}M R_{\text{SN}}(z, M) \phi_{\nu_i, {\rm NS}}(E_\nu', M) \right. \\
        &\left.\quad+ \int_\Sigma \mathrm{d}M R_{\text{SN}}(z, M) \phi_{\nu_i, {\rm BH}}(E_\nu', M) \right] (1+z) \left| \frac{\mathrm{d}t_\mathrm{c}}{\mathrm{d}z} \right|  \ .
    \end{aligned}
\end{equation}
Here $c$ is the speed of light, $z$ the cosmological redshift, $M$ is the progenitor mass,  $E'_{\nu}=(1+z) E_{\nu}$ is the redshifted neutrino energy, and $\left| {\rm d}t_{\rm c}/{{\rm d } z} \right|$ the cosmic time. The two terms in the integral  \eqref{eq:flux} depend on the evolving core-collapse supernova rate $R_{\text{SN}}(z, M)$ and the neutrino yields $ \phi_{\nu_i, {\rm NS}}(E_\nu', M)$ and $ \phi_{\nu_i, {\rm BH}}(E_\nu', M)$ correspond to the contributions from core-collapse supernovae leaving either a neutron-star (NS) or a black-hole (BH). The terms $\Omega $ and $\Sigma $ indicate the relative progenitor mass domains. In our computations, we considered $z \in [0,5]$ and progenitor masses in the range $M \in [8, 125]~M_\odot$. 

For the evolving core-collapse supernova rate, we used 
\begin{equation}\label{eq:rsnz}
    R_{\rm SN}(z,M) = \frac{\dot{\rho}_*(z) \phi(M)}{\int_{0.5M_\odot}^{125M_\odot} \phi(M) M \, dM}\,.
\end{equation}
with the  the initial mass function (IMF) $\phi(M)$ such that $\phi(M){\rm d}M$ is the number of stars in the mass interval $[M, M+{\rm d}M]$. We assume that $ \phi(M) \sim M^\chi$ with $\chi=-2.35$ as introduced by Salpeter in ref.~\cite{Salpeter:1955it}.  

For the star formation rate history, we take the piecewise broken power-law parametrization by Refs.\cite{Yuksel:2008cu,Horiuchi:2008jz}
\begin{equation}
\dot{\rho}_*(z) = \dot{\rho}_0 \left[ (1 + z)^{\alpha \eta} + \left( \frac{1 + z}{B} \right)^{\beta \eta} + \left( \frac{1 + z}{C} \right)^{\gamma \eta} \right]^{1/\eta}
\end{equation}
where $\alpha = 3.4$, $\beta = -0.3$, $\gamma=-3.5$ are the logarithmic slopes, $\eta=-10$ is a smoothing parameter and the parameters $B=5000$ and $C=9$ define the redshift breaks. The quantity $\dot\rho_0 $ is the local star formation rate history whose value is adjusted to obtain the desired local core-collapse supernova rate\footnote{See table~I in ref.~\cite{Ivanez-Ballesteros:2022szu} for explicit values.}.

One can define the fraction of failed supernovae by
\begin{equation}
    f_{\rm BH} = \frac{\int_{\Sigma} {\rm d}M\phi(M)}{\int_{8M_\odot}^{125M_\odot} {\rm d}M}
\end{equation}
which was shown to be important in ref.~\cite{Lunardini:2009ya} since neutrinos from supernovae leaving a black-hole have higher luminosities and hotter spectra because of the compression of baryonic matter to high densities, thus influencing the tail of the DSNB flux. In our analysis, we considered $f_{\rm BH} = 0.21$ and $0.41$ in accordance with the most conservative and optimistic cases found in the extensive detailed supernova simulations of ref.~\cite{Kresse:2020nto}.  The current uncertainty on the local core-collapse supernova rate  $R_{\rm SN}(0)$ is given by
\begin{equation}
R_{\rm SN}(0) = \int_{8M_\odot}^{125M_\odot} R_\text{SN}(0, M) \, {\rm d}M = (1.25\pm 0.5) \times 10^{-4} {\rm  ~y^{-1} Mpc^{-3}} \ .
\end{equation}

Concerning the cosmological model, we assume the $\Lambda$CDM model. For the cosmic time we have
\beq\label{eq:Hz}
\left| \frac{\mathrm{d}t_\mathrm{c}}{\mathrm{d}z} \right|^{-1} = H_0 ( 1 + z) \sqrt{ \Omega_{\Lambda} + (1+z)^3 \Omega_m}
\eeq
with $H_0$ the Hubble constant that we take $H_0 = 70~{\rm km ~s}^{-1} {\rm Mpc}^{-1}$, $\Omega_{\Lambda} = 0.7$ and $\Omega_m = 0.3$ the dark energy and the matter cosmic energy density, respectively.

\subsection{The DSNB model}
In order to model the contribution from different progenitors $M$ to the DSNB flux, we employed a parametric approach as done in e.g.\cite{Priya:2017bmm,Moller:2018kpn,Ivanez-Ballesteros:2022szu,Roux:2024zsv}. More explicitly, we considered supernova templates for different progenitors and equations of state, as in our previous analysis for a future galactic supernova. Note that refs.\cite{Lunardini:2005jf,Horiuchi:2008jz,Vissani:2011kx} also employed information from SN1987A to model the DSNB flux, with different approaches and limitations. 

\begin{figure*}[t]
\begin{center}
\includegraphics[scale=0.7]{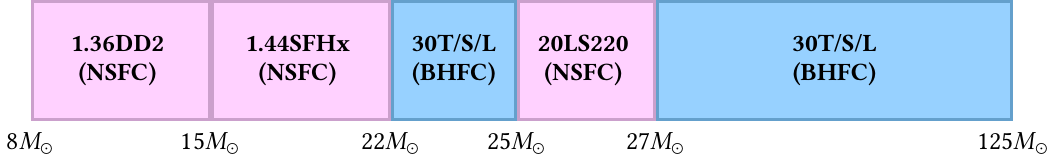}
\caption{DSNB model: The figure shows the reference models taken in each progenitor mass interval, where the core-collapse supernova leaves either a neutron star or a black hole. For the former, the detailed one-dimensional simulations for SN1987A from the Garching group were used \cite{Fiorillo:2023frv} as well as 
the $20~M_{\odot}$ model from \cite{Hudepohl:2013zsj}. For the latter, the BH simulation from Nakazato was employed \cite{Nakazato}. For the BH cases, the three equations of state were considered, namely the Togashi (T), the Lattimer-Swesty (L), and Shen (S). Note that the same BH template was used for the two mass intervals, i.e., $M \in [22, 25]~M_{\odot} $ and $M > 27~M_{\odot}$.}
\label{fig:templates}
\end{center}
\end{figure*}

Figure \ref{fig:templates} presents the progenitor mass intervals and the supernova models used in each of them. For NS-forming collapses, we used models from the Garching group \cite{Garching}: the 1.36DD2 and 1.44SFHx \cite{Fiorillo:2023frv}---with progenitors of $9~M_{\odot}$ and $18.8~M_{\odot}$, respectively---and the s20.0 model with the LS220 EOS, denoted 20LS220 \cite{Bollig:2017lki}. Note that for the NS cases, we also considered the 1.62 $M_{\odot}$, which has an $18.6~M_{\odot}$ progenitor, but the results are insensitive to this change. For BH-forming collapses, we used the 30T, 30S, and 30L models from simulations of a $30~M_\odot$ progenitor with different equations of state~\cite{Nakazato,Nakazato:2012qf, Nakazato:2021gfi}. We considered three separate scenarios, each using one of these models to represent a BH forming case. The model we adopted corresponds to a fraction of BH $f_{\rm BH} = 0.21$ (as the reference model in our previous work \cite{Ivanez-Ballesteros:2022szu}) while we also considered the most optimistic scenario in which $f_{\rm BH} = 0.41$ where we have one NS model with 1.44SFHx for the NS in the range $M \in [8, 15]~M_{\odot} $ and one BH model with $30~M_{\odot}$ for the range in $M > 15~M_{\odot}$. 
We will comment on how our results change when considering the most optimistic scenario for the BH fraction.

To implement the impact of neutrino decay in matter at each individual supernova, we follow the theoretical treatment described in section~\ref{sec:theory}. In particular, we consider that neutrinos start free streaming from the energysphere to the transportspheres in the case of $\nu_x$ and $\bar{\nu}_x$ and from the neutrinosphere for $\nu_e, \bar{\nu}_e$ and undergo flux suppression Eqs.\eqref{eq:survival_nue}-\eqref{eq:survival_nux} and spectral modification due to decay in these dense stellar regions. Then neutrinos propagate and undergo the MSW effect in the outer stellar layers \eqref{eq:MSWNO}-\eqref{eq:MSWIO}. We emphasize that in the inclusion of neutrino decay in matter, we implemented the same time-dependence as in section~\ref{sec:flux}. The impact of decay on the neutrino fluxes at individual supernovae is the one described previously, for each neutrino species. From the time-dependent fluxes with and without decay, we build the DSNB fluences. 

\subsection{Neutrino decay in matter and the DSNB: numerical results}
\begin{figure*}[t]
\begin{center}
\includegraphics[scale=0.4]{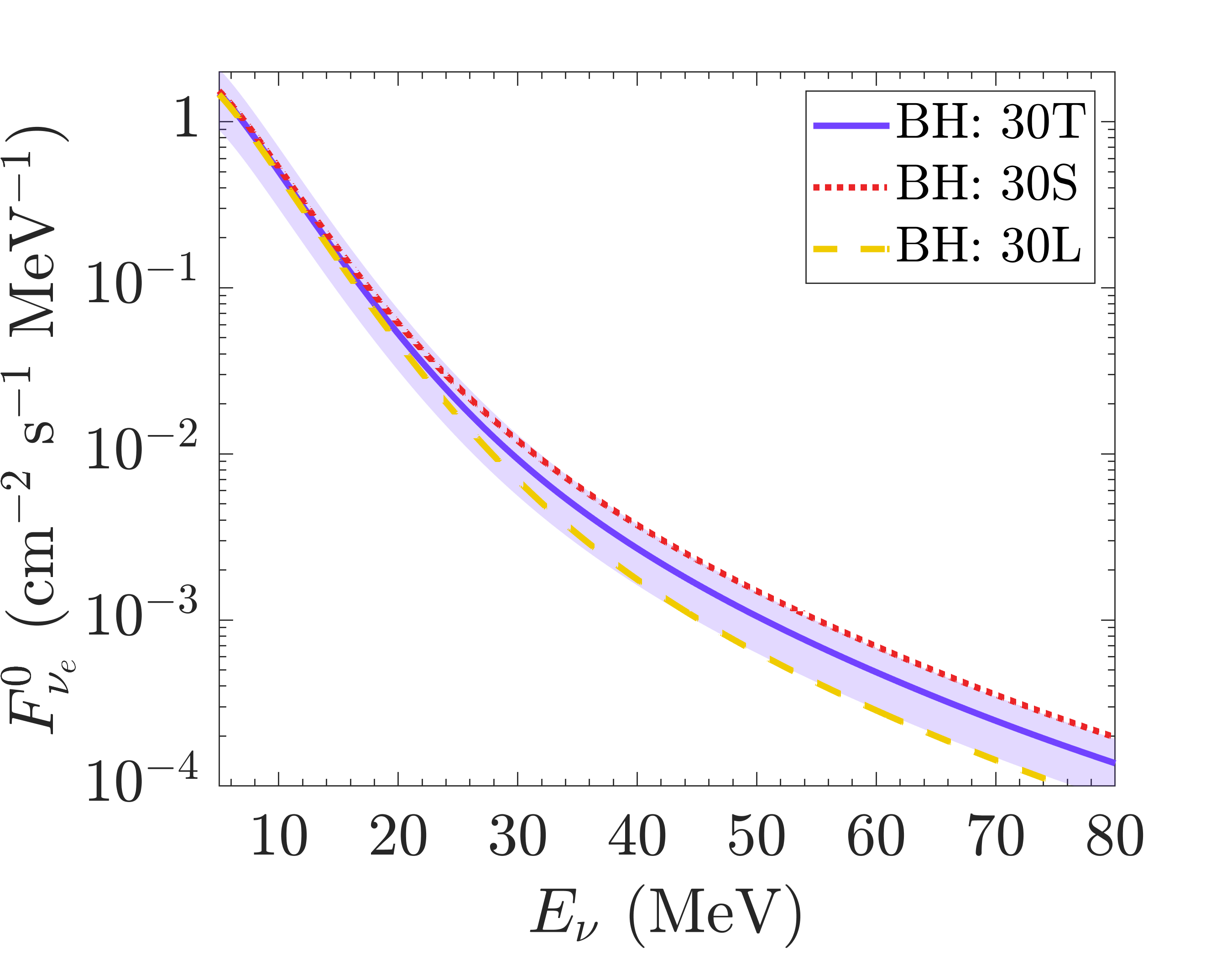}
\includegraphics[scale=0.4]{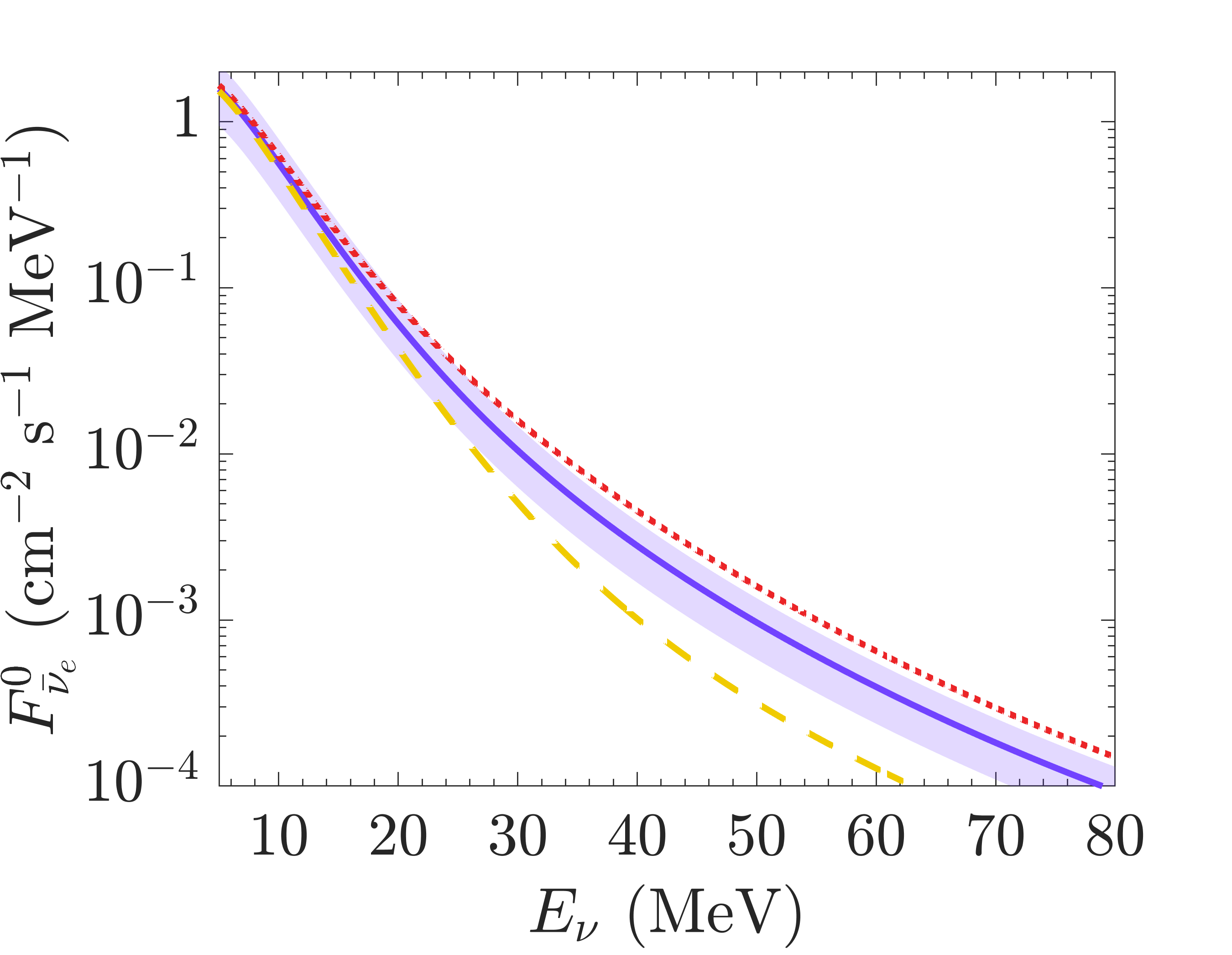}
\includegraphics[scale=0.4]{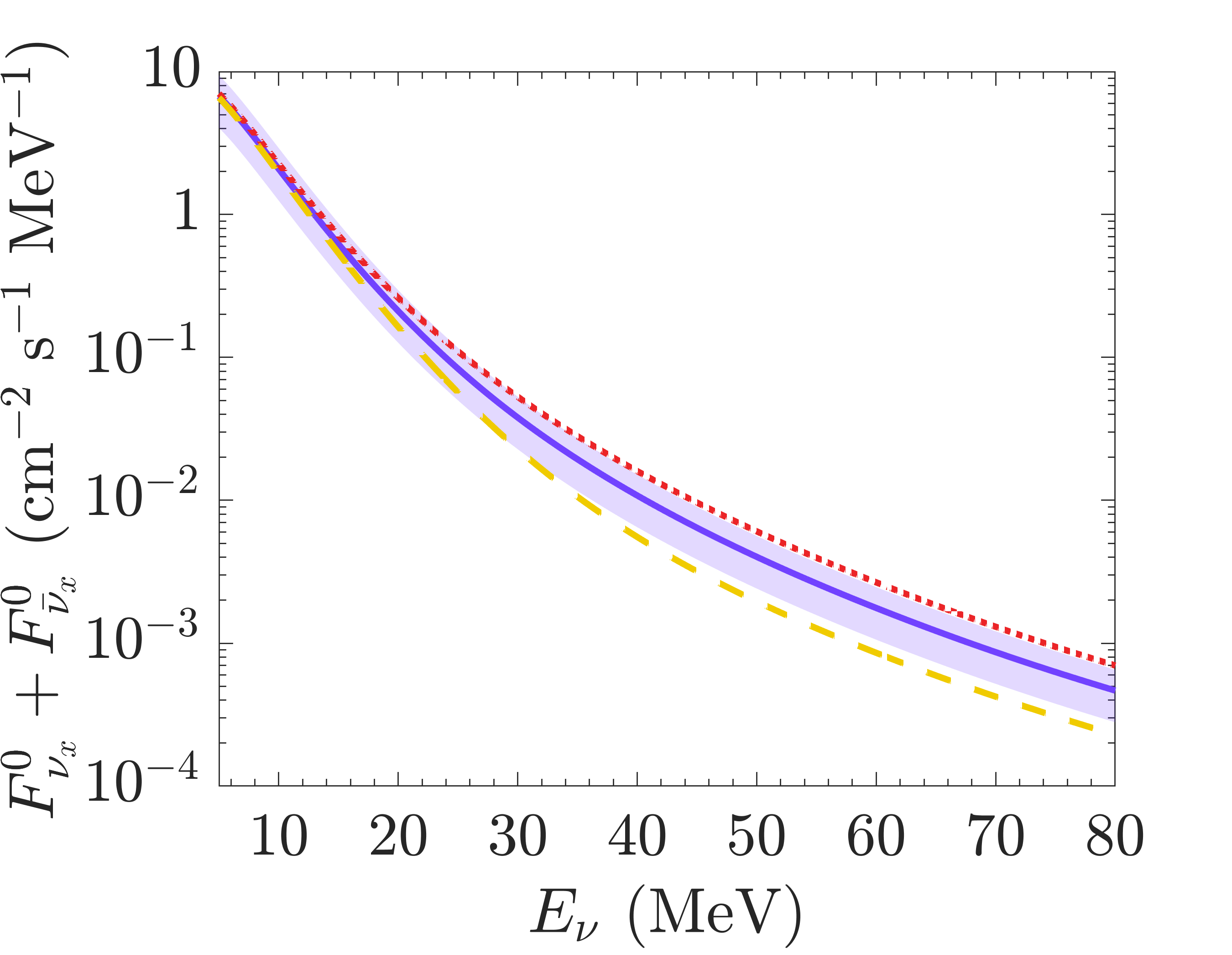}
\caption{DSNB fluxes in the absence of neutrino decay for the different neutrino species for the case of normal neutrino mass ordering. The different curves show the variations due to changes in the BH template when considering different equations of state, namely the Togashi (T), the Lattimer-Swesty (L), and Shen (S). The bands correspond to the current uncertainty on the evolving core-collapse supernova rate associated with the results of the Togashi EOS (see text).}
\label{fig:DSNBfluxes}
\end{center}
\end{figure*}

Before giving the DSNB results in the presence of decay, we show the sensitivity of the DSNB fluxes to the BH contribution when the three EOS are considered.
This is shown in figure \ref{fig:DSNBfluxes} for $\nu_e$, $\bar{\nu}_e$ and the non-electronic neutrino species, for the case of normal ordering. The largest variation reaches up to a factor of a few in the energy range $E_{\nu} \in [10, 35]$ MeV, from the Lattimer-Swesty EOS and for $\bar{\nu}_e$. It is to be noted, though, that this EOS
does not comply with some of the bounds for neutron star matter \cite{Tews:2016jhi}. Table \ref{tab:eventsDSNB} displays the modifications in the DSNB total number of events, in the different experiments, when varying the EOS. The results are for the two mass orderings and our reference calculations with $f_{\rm BH} = 0.21$. Note that in DARWIN, we find only a couple of DSNB events. 

\begin{table}
\centering
\setlength{\tabcolsep}{12pt} 
\renewcommand{\arraystretch}{1.5} 
\begin{tabular}{lcccc}
\toprule
\multirow{2}{*}{Flux} &  \multicolumn{3}{c}{BH template} & \multirow{2}{*}{Current limits}                     \\ \cmidrule{2-4}
    & 30T & 30S & 30L & \\ \midrule 
$\nu_e~$         & \ctightcell{$0.16 \pm 0.07$ \\ ($0.16 \pm 0.07$)} & \ctightcell{$0.21 \pm 0.08$ \\ ($0.22 \pm 0.09$)} & \ctightcell{$0.13 \pm 0.05$ \\ ($0.11 \pm 0.04$)}  & 2.7 (SK)  \\
$\bar{\nu}_e~$      & \ctightcell{$0.58 \pm 0.23$ \\ ($0.52 \pm 0.21$)} & \ctightcell{$0.80 \pm 0.32$ \\ ($0.63 \pm 0.25$)} & \ctightcell{$0.37 \pm 0.15$ \\ ($0.43 \pm 0.17$)} & 19  (SNO)  \\
$\nu_x~$            & \ctightcell{$1.46 \pm 0.58$ \\ ($1.50 \pm 0.60$)} & \ctightcell{$1.93 \pm 0.77$ \\ ($2.03 \pm 0.81$)} & \ctightcell{$0.99 \pm 0.39$ \\ ($0.98 \pm 0.39$)}  &  $(1.3\text{--}1.8)\times 10^3 $ \\
\bottomrule
\end{tabular}
\caption{Integrated DSNB fluxes in normal and inverted ordering  (in parenthesis) with the uncertainties from the core-collapse supernova rate. 
The flux values ($\nu /$cm$^{2}/$s) correspond to  the energy ranges $E_{\nu}> 19.3$ MeV for $\bar{\nu}_e$  $E_{\nu} \in [22.9, 36.9]$ MeV for $\nu_e$,
for different EOS for the BH model. The last column gives the experimental upper limits on the DSNB fluxes  (90$\%$ C.L.) from SK \cite{Super-Kamiokande:2021jaq} and SNO \cite{SNO:2006dke}. 
The loosest bounds are for $\nu_x$ ($x = \mu, \tau$ flavors) and correspond to $E_{\nu} > 19$ MeV \cite{Lunardini:2008xd}.}\label{tab:fluxexp}
\end{table}

\begin{table}
\centering
    \setlength{\tabcolsep}{17pt} 
    \renewcommand{\arraystretch}{1.5} 
\begin{tabular}{lccc}
\toprule
 & \multicolumn{3}{c}{\textbf{BH template}} \\ \cmidrule{2-4}
 & 30T & 30S & 30L \\ \cmidrule{2-4}
HK-Gd   & \ctightcell{55 \\ (48)}& \ctightcell{75 \\ (58)}& \ctightcell{35 \\ (40)} \\
JUNO    & \ctightcell{19 \\ (17)}& \ctightcell{25 \\ (19)}& \ctightcell{14 \\ (15)} \\
DUNE     & \ctightcell{9 \\ (9)}& \ctightcell{10 \\ (11)}& \ctightcell{7 \\ (6)}  \\
\bottomrule
\end{tabular}
\caption{Expected total number of DSNB events, in the absence of neutrino-Majoron interactions, at several detectors (first column) after 20 years of exposure. The results were obtained using fixed templates for the NS cases and varying the templates for the BH cases between the models 30T (second), 30S (third), and 30L (fourth column). The events are given in normal and inverted ordering (in parentheses). The results for HK-Gd are obtained in the energy window $E_{e^+} \in [16, 30] $~MeV and an efficiency of 0.4. The detection energy window for JUNO is $E_{e^+} \in [11.5, 29.5]$~MeV, and we assumed an efficiency of 0.86. For DUNE, we considered $E_{\nu} \in [19, 31]$~MeV and an efficiency of 0.85.}
\label{tab:eventsDSNB}
\end{table}

\begin{figure*}[t]
\begin{center}
\includegraphics[scale=0.42]{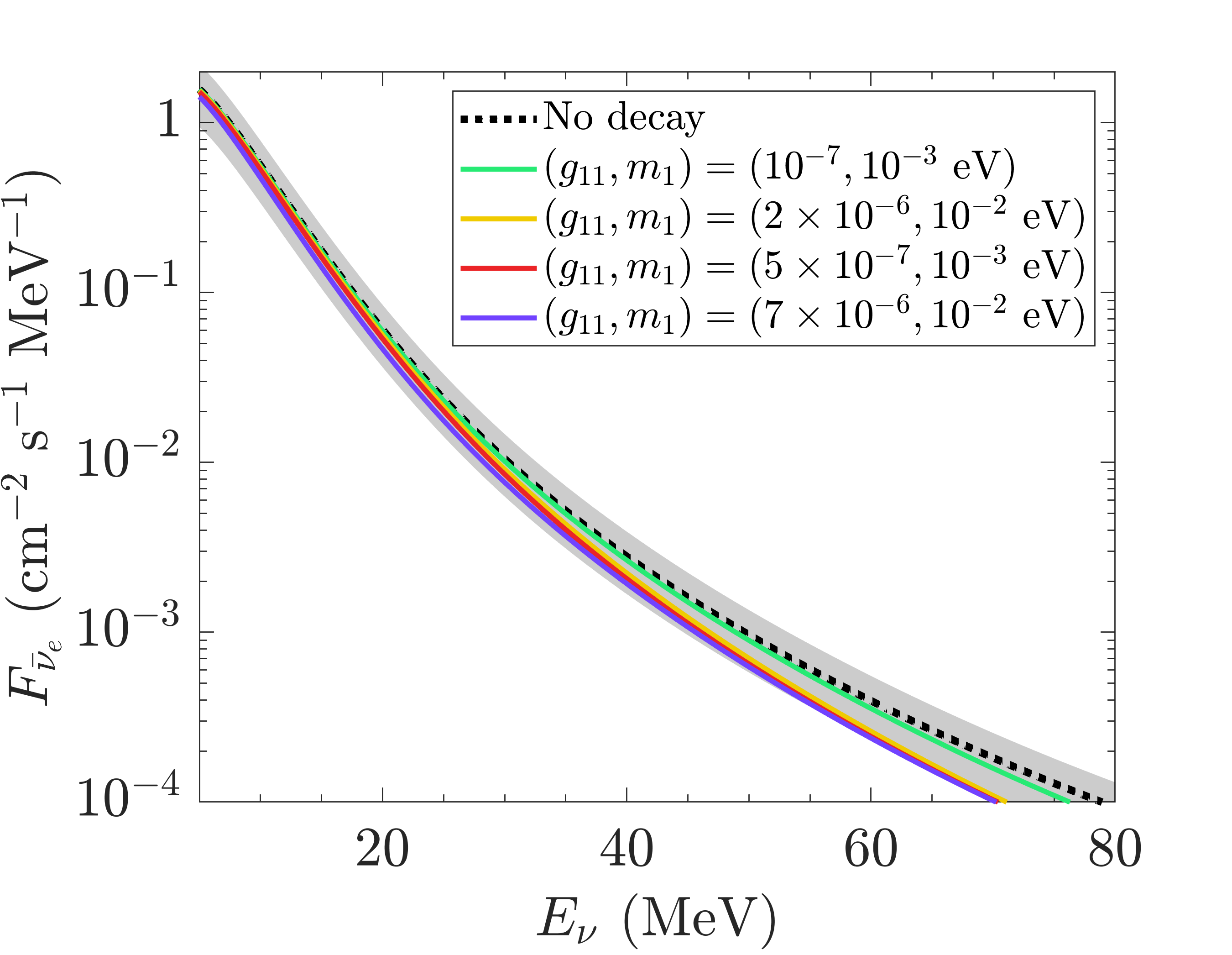}
\includegraphics[scale=0.42]{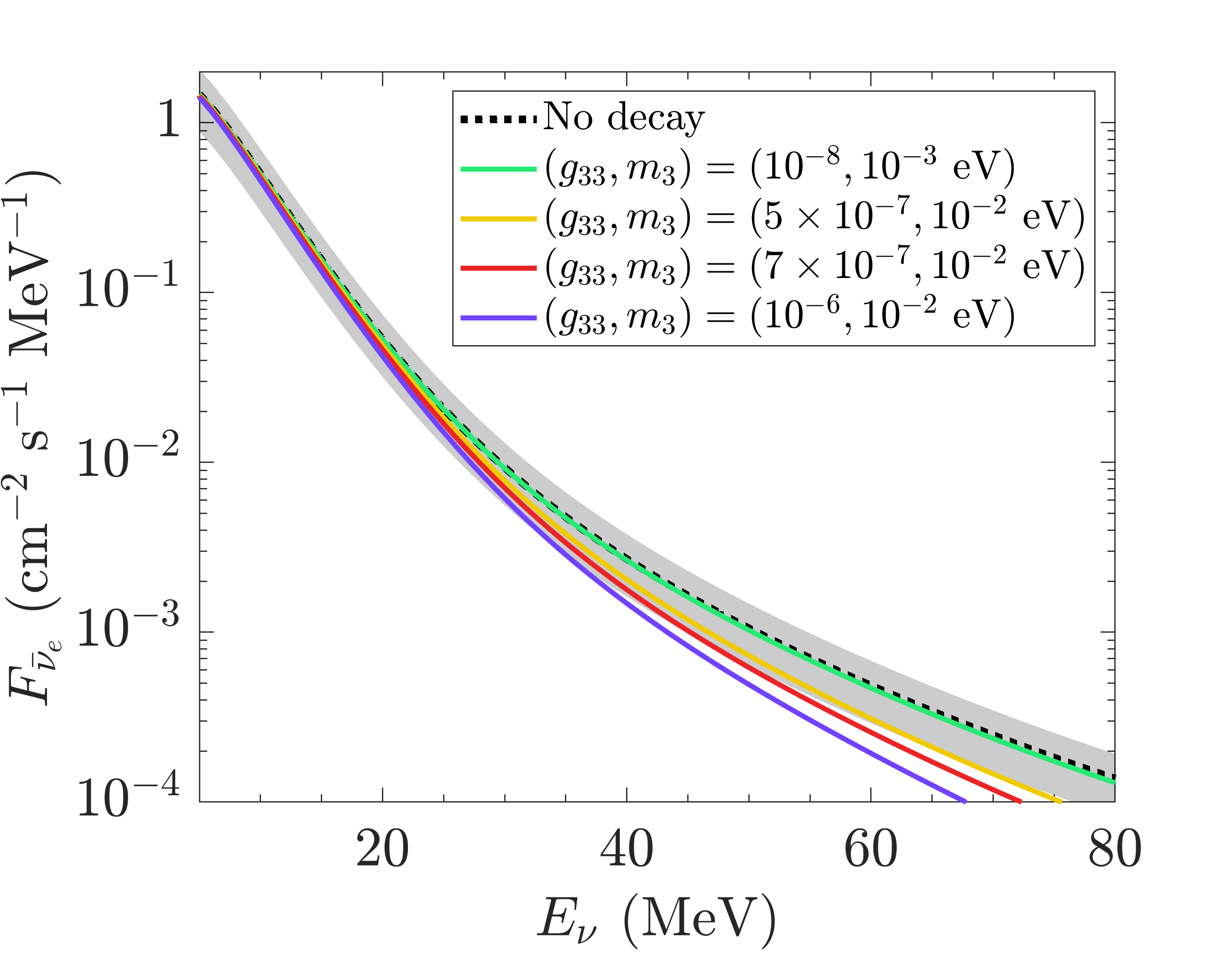}
\caption{DSNB fluxes for $\bar{\nu}_e$ for normal (left) and inverted (right) mass ordering.  The figure compares the fluxes in the absence (dotted) and in the presence (full lines) of decay for different values of neutrino-Majoron coupling $g_{11}$ and lightest neutrino mass $m_1$ (left) or $g_{33}$ and $m_3$ (right). Results were obtained using the 30T model as the BH template. The bands display the uncertainty coming from the evolving core-collapse supernova rate associated with the no decay case (dotted line).}
\label{fig:DSNBfluxesdecay}
\end{center}
\end{figure*}

\begin{figure*}[t]
\begin{center}
\includegraphics[scale=0.42]{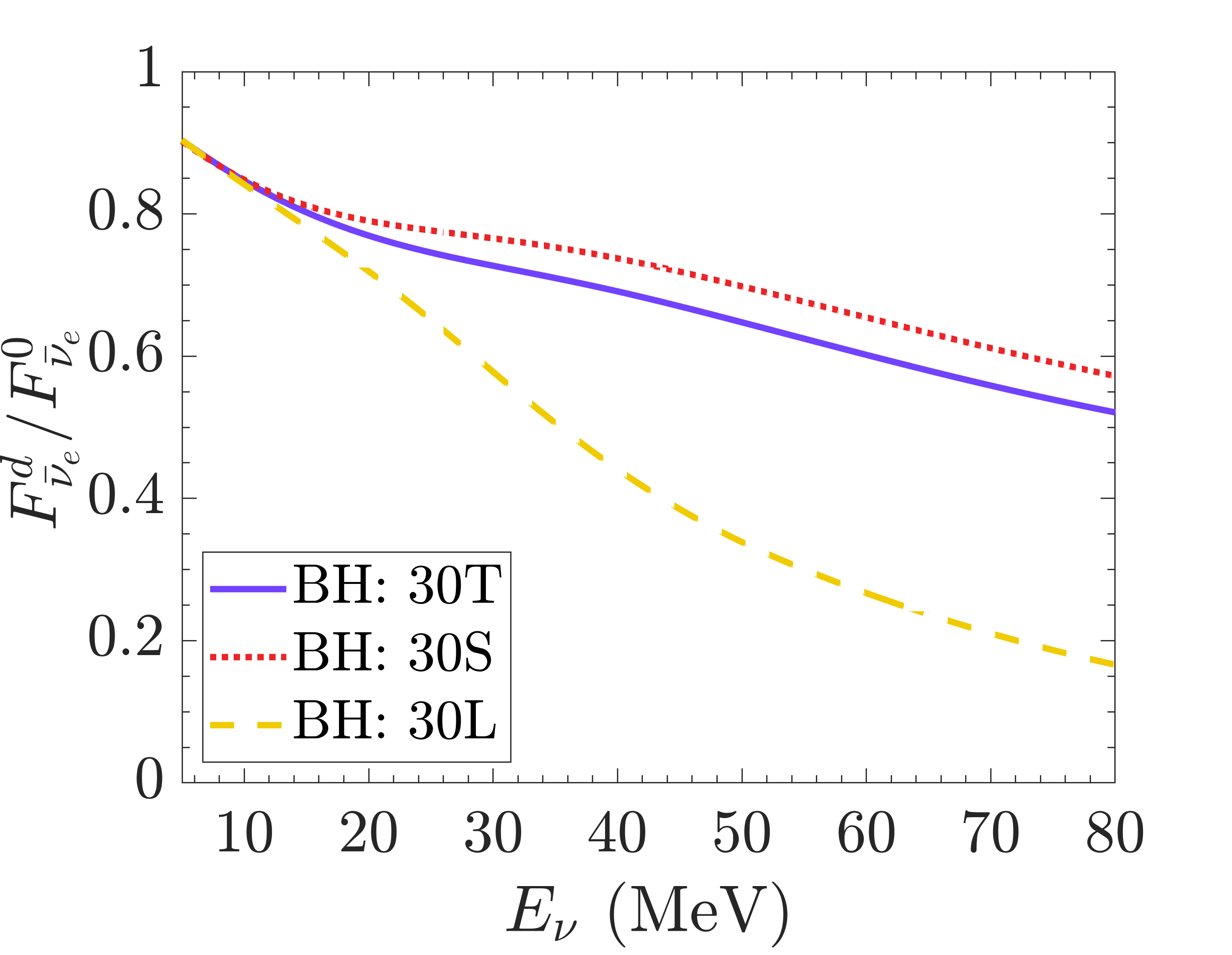}
\includegraphics[scale=0.42]{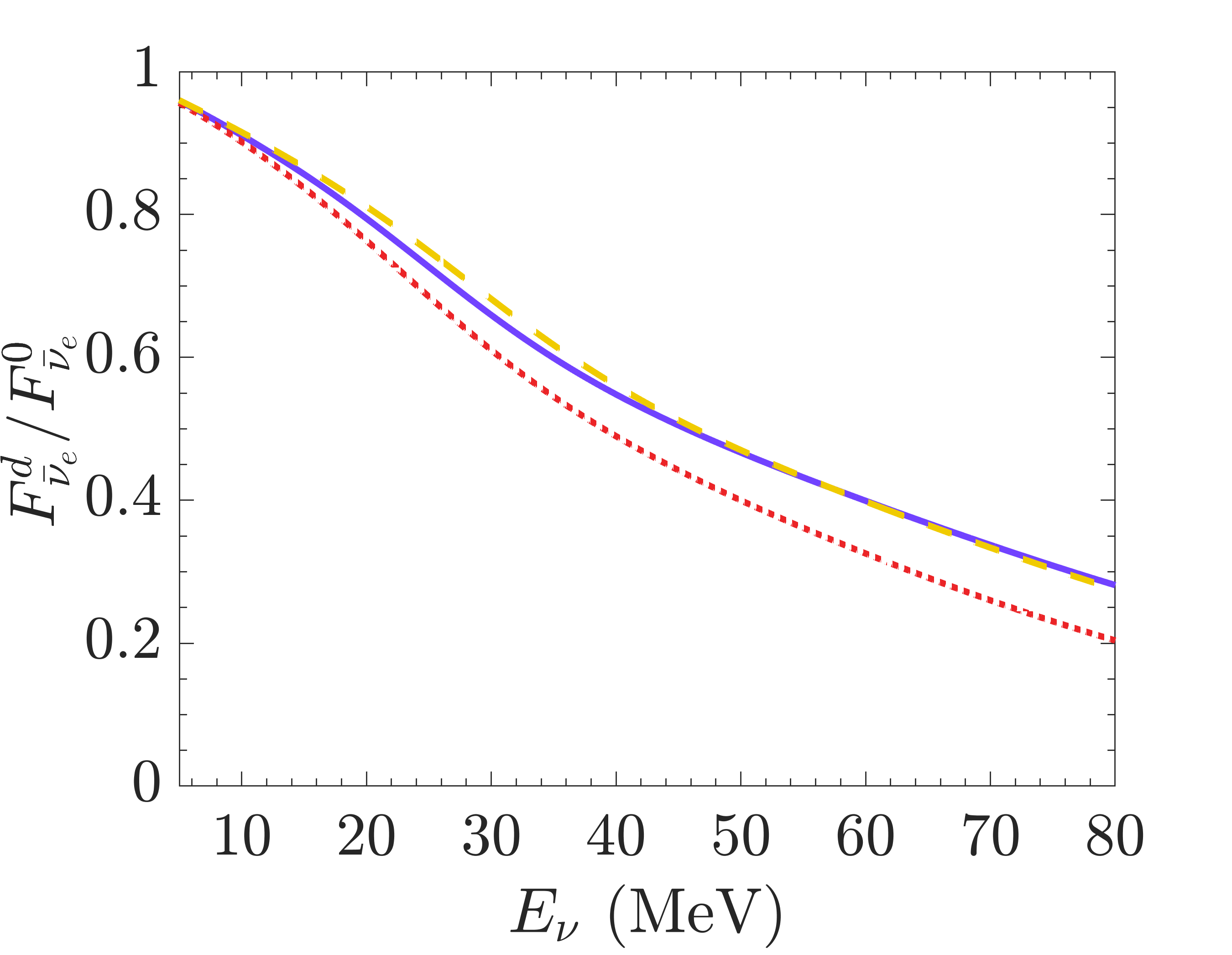}
\includegraphics[scale=0.42]{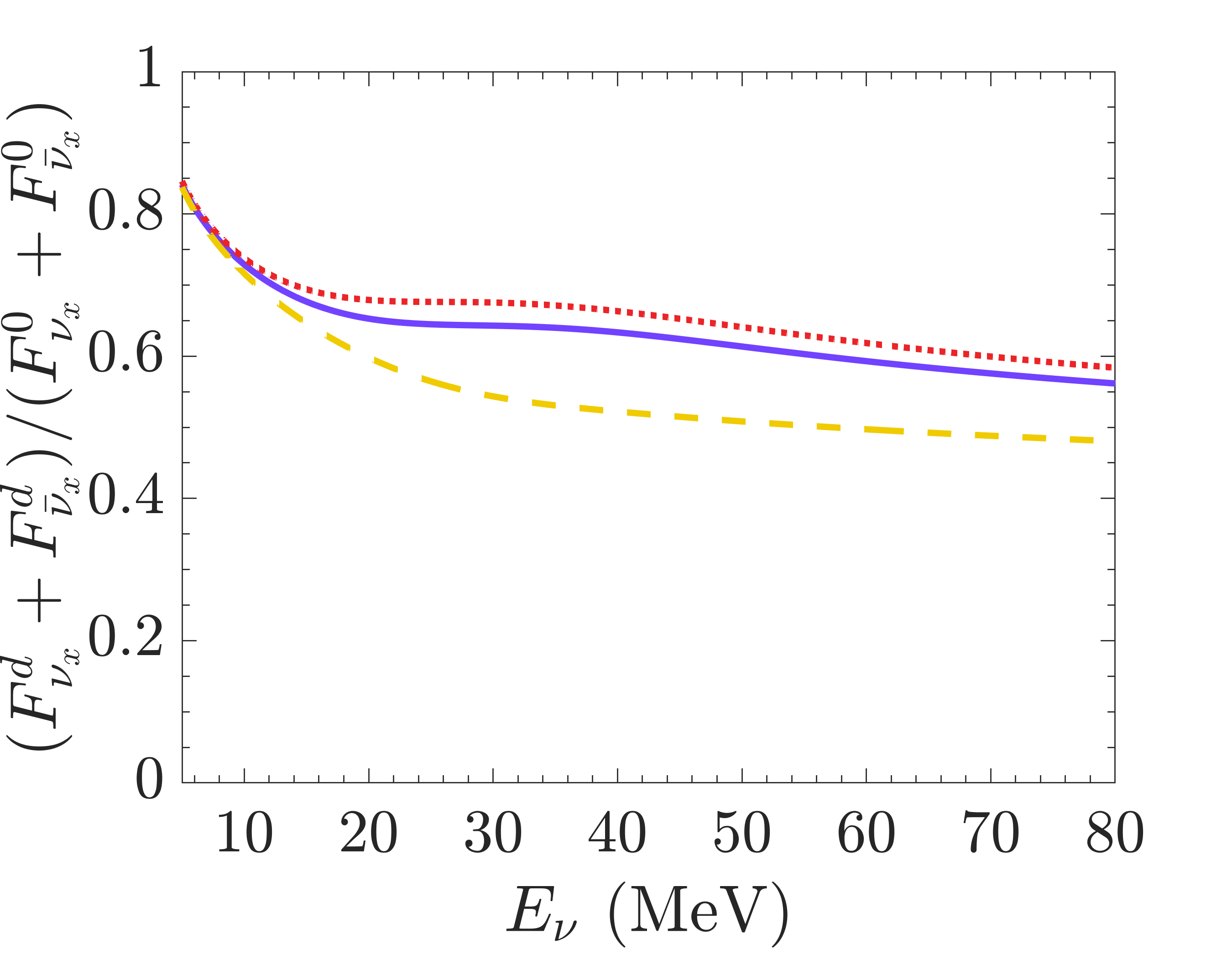}
\caption{Ratios of the DSNB fluxes with decay over the ones without decay for $\bar{\nu}_e$ for normal (top left) and inverted (top right) mass orderings, as well as for $\nu_x$ in normal ordering(bottom figure). The different curves show the variations due to changes in the BH template when considering different equations of state, namely the Togashi (T), the Lattimer-Swesty (L), and Shen (S).}
\label{fig:DSNBfluxesratios}
\end{center}
\end{figure*}

Concerning the impact of decay, figures \ref{fig:DSNBfluxesdecay} and  \ref{fig:DSNBfluxesratios} show,  for normal neutrino mass ordering, the DSNB fluxes for the $\nu_e$ and $\bar{\nu}_e$ and the flux ratios of the $\bar{\nu}_e$ and $\nu_x$ fluxes with decay over those without decay, respectively. (The results for inverted neutrino mass ordering are shown in appendix~\ref{appendix:results}.) As one can see, neutrino nonradiative decay suppresses the fluxes when the neutrino-Majoron couplings vary from $(g_{11}, m_1) = (10^{-7}, 10^{-3})$ eV to $(7 \times 10^{-6}, 10^{-2})$ eV.  
This reduction is of less than 10$\%$ for $\nu_e$, of about $30\%$ for $\nu_x$ for normal ordering and negligible for the inverted, and between 20--30$\%$ for $\bar{\nu}_e$ in the $[10, 30]$ MeV neutrino energy range. 
Thus one can see that the $\bar{\nu}_e$ and $\nu_x$ suppression at high energies found for a single supernova leaving a NS (figure  \ref{fig:fluxesNS}) and a BH (figure \ref{fig:fluxesBH}), are reflected in the DSNB predictions, after integrating over redshift and including different progenitors, in our DSNB model. In inverted ordering, the flux suppression in the DSNB region of interest is larger than for normal ordering. 

Table \ref{tab:fluxexp} presents our integrated DSNB flux predictions, in comparison with the current experimental DSNB bounds on $\nu_e$ and $\bar{\nu}_e$ from the SK and SNO experiments. One can see that our results are more than an order of magnitude smaller
than the current SK upper limit \cite{Super-Kamiokande:2021jaq}, a factor of more than 20--30 below the ones from SNO and thus are
on the conservative side, as the ones in, e.g., refs.\cite{Ivanez-Ballesteros:2022szu,Roux:2024zsv}. The loosest bounds are those for $\nu_x$.

\begin{figure*}[t]
\begin{center}
\includegraphics[scale=0.42]{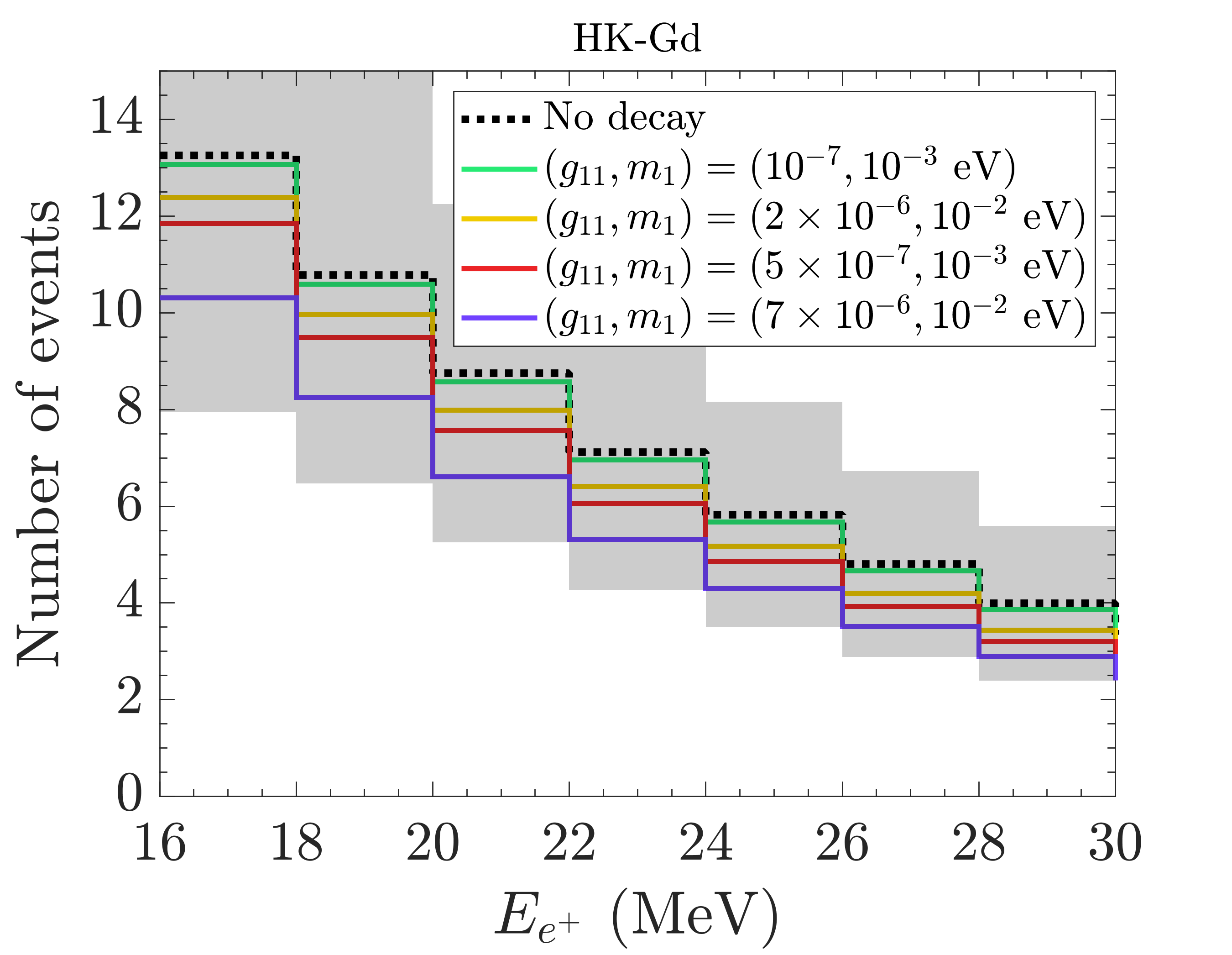}
\includegraphics[scale=0.42]{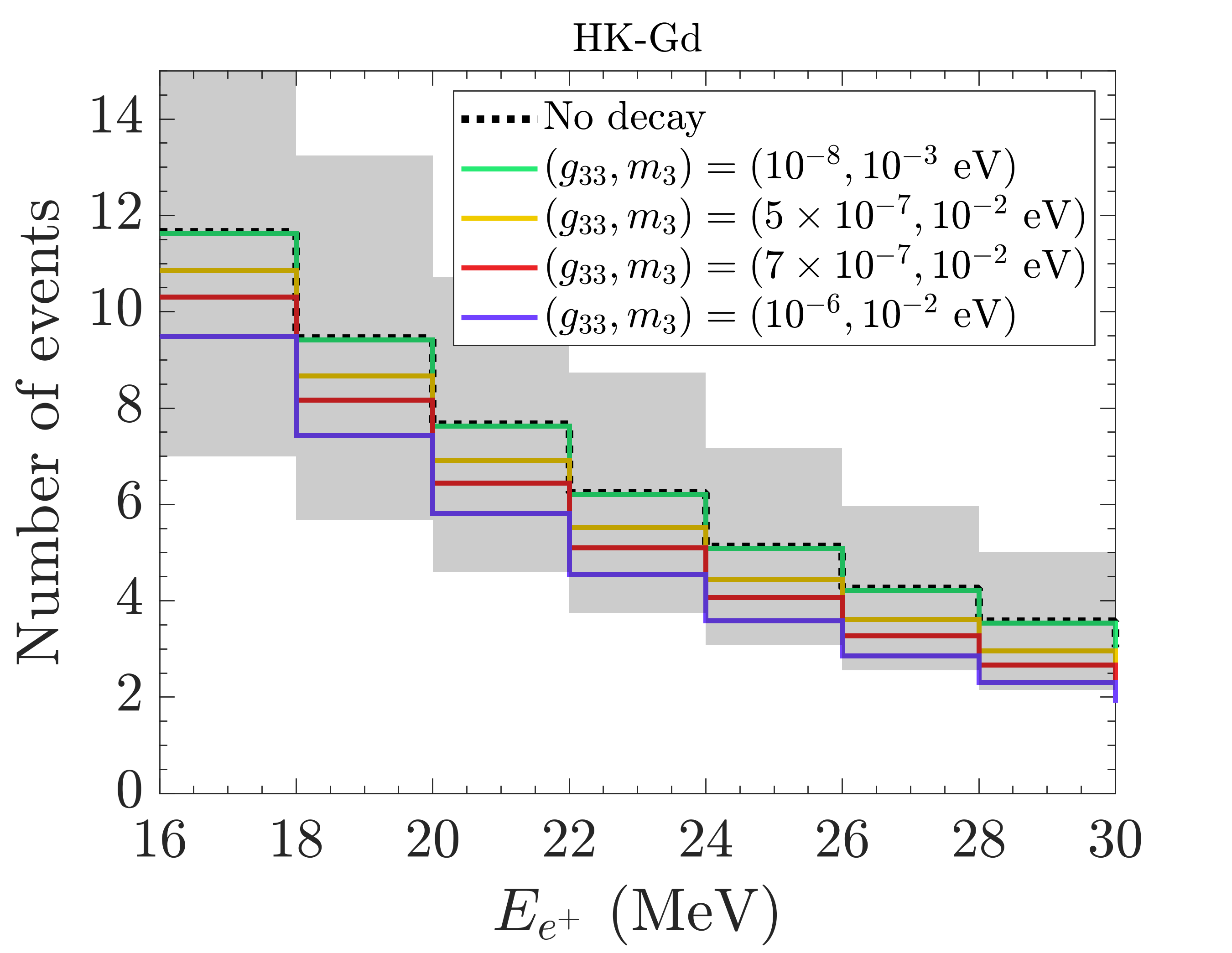}
\caption{DSNB predictions on the number of events as a function of positron energy in HK-Gd, after 20 years running, considering the 30T model as the BH template. The results correspond to the case with decay for different values of the neutrino-Majoron couplings $g_{11}$ ($g_{33}$) as a function of the lightest neutrino mass $m_1$ ($m_3)$ for normal on the left figure (inverted, right figure) mass ordering. The bands display the uncertainty coming from the evolving core-collapse supernova rate associated with the no decay case (dotted line).}
\label{fig:DSNBsignal}
\end{center}
\end{figure*}

To predict the number of events in the four experiments, we use the same cross sections for the main detection channels as in section~\ref{sec:detectors}.
The impact of neutrino nonradiative two-body decay in matter on the DSNB number of events is shown in figure \ref{fig:DSNBsignal} for HK with Gadolinium (HK-Gd). Table \ref{tab:eventsDSNB} presents the DSNB expected number of events in HK after 20 years running, for the detection channels. Our results show a decrease in the number of events as a function of positron energy and in the total number of events for HK, while in the other detectors, the DSNB predictions appear to have little sensitivity to this non-standard process. Table \ref{tab:eventsDSNBHKGd} shows that the impact of decay on HK-Gd ranges at the level of about $25$--$30\%$. While the neutrino decay impact is within the largest uncertainties associated with other factors that determine the DSNB flux, in particular the evolving core-collapse supernova rate, the fluxes at individual supernovae and the fraction of BH (or of binaries, see \cite{Kresse:2020nto,Horiuchi:2020jnc}), its impact is at the same level as aspects pointed out in the literature such 
as, e.g., shock wave or flavor conversion effects  (see \cite{Volpe:2023met,Mathews:2019klh}).  

\begin{table}
\centering
    \setlength{\tabcolsep}{10pt} 
    \renewcommand{\arraystretch}{1.5} 
\begin{tabular}{cccc}
\toprule
BH template & No decay & Largest effect & Ratio \\ 
\midrule
30T   & \ctightcell{55 \\ (48)} & \ctightcell{41 \\ (36)} & \ctightcell{0.76 \\ (0.75)} \\
30S    & \ctightcell{75 \\ (58)}   & \ctightcell{59 \\ (41)}   & \ctightcell{0.78 \\ (0.71)} \\ 
30L & \ctightcell{35 \\ (40)}   & \ctightcell{24 \\ (31)}   & \ctightcell{0.68 \\ (0.77)} \\
\bottomrule
\end{tabular}
\caption{Expected total number of DSNB events at HK-Gd loaded after 20 years of exposure. The results are obtained considering different templates for the BHs. The second column shows the number of events in the absence of decay, while the third and fourth columns correspond to the cases with the largest decay effects and the ratio between the two cases, respectively. Results are given for both normal mass ordering and inverted ordering (in parentheses). The largest decay effects are obtained for $g_{11} = 7 \times 10^{-6}$ and $m_1 = 10^{-2}~$ eV for normal ordering and $g_{33} = 10^{-6}$ and $m_3 = 10^{-2}~ $eV for inverted ordering. The energy window considered for HK-Gd is $E_{e^+} \in [16, 30]~$ MeV, and the efficiency is taken as $\eta = 0.4$.}
\label{tab:eventsDSNBHKGd}
\end{table}

\section{Conclusions} \label{sec:conclusions}
\noindent
In this work, we have investigated the impact of neutrino non-radiative two-body decay into a massless Majoron, in matter, during a core-collapse supernova explosion, while our results are also of interest for massless or close to massless Majoron-like particles (fuzzy dark matter).
The theoretical framework used is the one developed in our previous study ref.~\cite{Ivanez-Ballesteros:2024nws}. In particular, we consider a full 3$\nu$ framework and employ inputs from detailed one-dimensional core-collapse supernova simulations from the Garching group and from Nakazato's simulations. In particular, we have implemented the time-dependence of the neutrino fluxes, the matter densities, and the electron fraction in our microscopic computation of the neutrinospheres, energyspheres, and transportspheres necessary to determine the neutrino flux suppression due to decay.  

Based on such a theoretical framework, our investigation has followed three directions. We have first given new limits on neutrino-Majoron couplings using SN1987A observations. Then we have presented our prospects for a future galactic core-collapse supernova obtained from a likelihood analysis exploiting the neutrino spectral distortion due to neutrino nonradiative two-body decay. Besides, we have explored for the first time its impact
on the DSNB and shown that it is potentially significant for the HK experiment.

After showing the impact of neutrino decay on the neutrino fluxes, we have first presented new bounds on neutrino-Majoron couplings obtained thanks to a 2-dimensional likelihood analysis of the 24 $\bar{\nu}_e$ events from SN1987A in Kamiokande, IMB, and Baksan. The results, obtained both for normal and inverted neutrino mass ordering, include bounds on the $g_{33}$ couplings and on all the flavor-basis couplings, which had not been previously shown in the literature. Moreover, our limits on $g_{ee}$ are competitive with the limits from neutrinoless double-beta decay experiments. Overall, our bounds are robust, varying only slightly when changing the model or the EOS for the cases considered. 
  
Then, we have presented the expected number of neutrino events, from a core-collapse supernova in four upcoming experiments, namely the near future JUNO and HK experiments, and the more distant future DUNE and DARWIN, showing the impact of neutrino decay for values of the neutrino-Majoron couplings still allowed, also considering our limits from SN1987A. By exploiting the spectral distortions due to neutrino decay in matter, we performed a 2-dimensional likelihood analysis, profiling on the models, to give prospects on the neutrino-Majoron coupling if a core-collapse supernova explodes at different locations in our Milky Way.
Such an analysis has shown that a significant improvement on the limits for neutrino-Majoron couplings could be obtained from the next galactic supernova, both if a neutron star or if a black-hole is formed, the latter being the most interesting case. 

In our analysis of the impact of neutrino decay in matter on the DSNB, we have employed a DSNB model using the NS and BH detailed simulations used for an individual supernova. Our results show that neutrino decay can suppress the $\bar{\nu}_e$ DSNB flux up to several tens of percent in the energy region of interest and can impact the DSNB number of events in HK-Gd up to 30$\%$ while JUNO, DUNE, and DARWIN appear to have negligible sensitivity to this nonstandard process. The effect in HK or HK-Gd does not vary in our case if we increase the BH fraction to the most optimistic value. This is likely due to the limited number of templates included.
In our DSNB predictions, the BH contribution is particularly important and dominates the effect from decay.
While our findings on the potential sensitivity of HK or HK-Gd to neutrino decay in matter are encouraging, further investigations will be necessary in particular using a more detailed description of the BH contributions and a larger set of EOS.

Clearly, several improvements of the presented analysis can be envisaged in the future, e.g., with the implementation of multidimensional supernova simulations, or of a more detailed description, in particular of the BH contribution, for the case of the DSNB. Moreover, a more complete description of neutrino evolution in the presence of Majorons in the supernova core, up to the solution of transport equations, could be envisaged, including neutrino decay in matter. Furthermore, extensions of the presented framework could include more sophisticated flavor conversion effects in the dense regions coming from, e.g., neutrino-neutrino interactions.   

While awaiting the next core-collapse supernova in our Galaxy or nearby, the search for the DSNB will continue.
The upcoming results from SK, including Gadolinium, and the start of JUNO, HK, and DUNE, which have the potential to discover the DSNB,
open very exciting times for neutrino astrophysics and the search for new physics. Our investigation unveils new elements of the interesting potential that a future core-collapse supernova and the DSNB have for the process of neutrino nonradiative decay.

\section*{Acknowledgments}
\noindent
The authors wish to thank Ken'ichiro Nakazato for the use of the detailed output from his simulations of the 30 $M_{\odot}$ supernova leaving a black-hole \cite{Nakazato:2012qf,Nakazato:2021gfi,Nakazato} as well as
Thomas Janka, Daniel Kresse, and their collaborators for providing us with information
from the one-dimensional simulations for SN1987A ref.~\cite{Fiorillo:2023frv}, giving us access to the Garching archives \cite{Garching}. The authors
acknowledge financial support from the Masterproject NUFRONT of "CNRS Nucl\'eaire et Particules".

\appendix
\section{Statistical analysis} \label{appendix:statistics}
In this work, we have considered two types of analysis: the analysis of observed data (SN1987A) and the analysis of future simulated data. For the analysis of SN1987A, we followed closely the procedure described in refs.~\cite{Vissani:2014doa, Ivanez-Ballesteros:2023lqa}. Due to the small size of the dataset, we used an unbinned likelihood given by
\begin{equation} \label{eq:lh_unbinned}
\mathcal{L}(x) = e^{-(S_\mathrm{tot}(x) + B_\mathrm{tot})} \times \prod_{i=1}^{N_\mathrm{obs}} dE \left[ \frac{dS}{dE_i}(x) + \frac{dB}{dE_i} \right],
\end{equation}
where $x$ represents the set of parameters describing our model, $S_\mathrm{tot}$ indicates the expected total number of signal events, and $B_\mathrm{tot}$ the total background for each detector. The number of observed events, $N_\mathrm{obs}$, is equal to $11$ for Kamiokande-II, $8$ for IMB, and $5$ for Baksan. The term $dS/dE_i$ represents the expected number of signal events around the observed energies $E_i$, while the background at those energies is denoted as $dB/dE_i$. We followed \cite{Ivanez-Ballesteros:2023lqa} to calculate the expected signal and used the information on the events given there.

In contrast, the next galactic core-collapse supernova is expected to result in a large number of events. Therefore, it is convenient to use a binned Poisson likelihood for the analysis. Dropping the constant terms, the logarithm of the binned likelihood is given by \cite{Cowan:1998ji}
\begin{equation} \label{eq:lh_binned}
    \log \mathcal{L} (x) = - S_\mathrm{tot}(x) + \sum_{i = 1}^{N_\mathrm{bins}} N_i \log \frac{dS}{dE_i}(x),
\end{equation}
where $x$ denotes the model parameters, $S_\mathrm{tot}$ the total number of expected events, and $N_\mathrm{bins}$ is the number of energy bins. The observed number of events in bin $i$ is represented by $N_i$, while $dS/dE_i$ indicates the expected number of events in that bin. Since we are interested in obtaining prospects for a future supernova detection, we simulate the observed data by computing $N_i$ using a chosen "true" model.

In both of our analyses, for normal ordering, we chose $g_{11}$ and $m_1$ as the free parameters, while for inverted ordering, we took $g_{33}$ and $m_3$. The other two couplings were obtained using the relations in Eq.~\eqref{eq:g_NO} and Eq.~\eqref{eq:g_IO} for normal and inverted ordering, respectively. For the analysis of a future supernova, we additionally considered the core-collapse models as discrete nuisance parameters, and we denote them by $\theta_\mathrm{SN}$. As a result, the parameters in the analysis of a future supernova are $g_{ii}$, $m_i$, and $\theta_\mathrm{SN}$, with $i = 1~(3)$ for normal (inverted) ordering. The profile likelihood is obtained by profiling over the core-collapse models, i.e., searching for the model that maximizes the likelihood for each set of $(g_{ii}, m_i)$:
\begin{equation} \label{eq:proflh}
    L_p(g_{ii}, m_i) = \max_{\theta_\mathrm{SN}} \mathcal{L}(g_{ii}, m_i,\theta_\mathrm{SN}).
\end{equation}
The maximum likelihood, on the other hand, is defined as 
\begin{equation} \label{eq:lhmax}
    L_{\max} = \mathcal{L} (\hat g_{ii}, \hat m_i, \hat \theta_\mathrm{SN}) = 
    \max_{g_{ii}, m_i, \theta_\mathrm{SN}} \mathcal{L}(g_{ii}, m_i,\theta_\mathrm{SN}),
\end{equation}
where $\hat g_{ii}$, $\hat m_i$, and $\hat\theta_\mathrm{SN}$ are the maximum likelihood estimators. 

\section{Additional results} \label{appendix:results}
\begin{figure}
    \centering
    \begin{subfigure}{\textwidth}
        \includegraphics[width=0.49\textwidth]{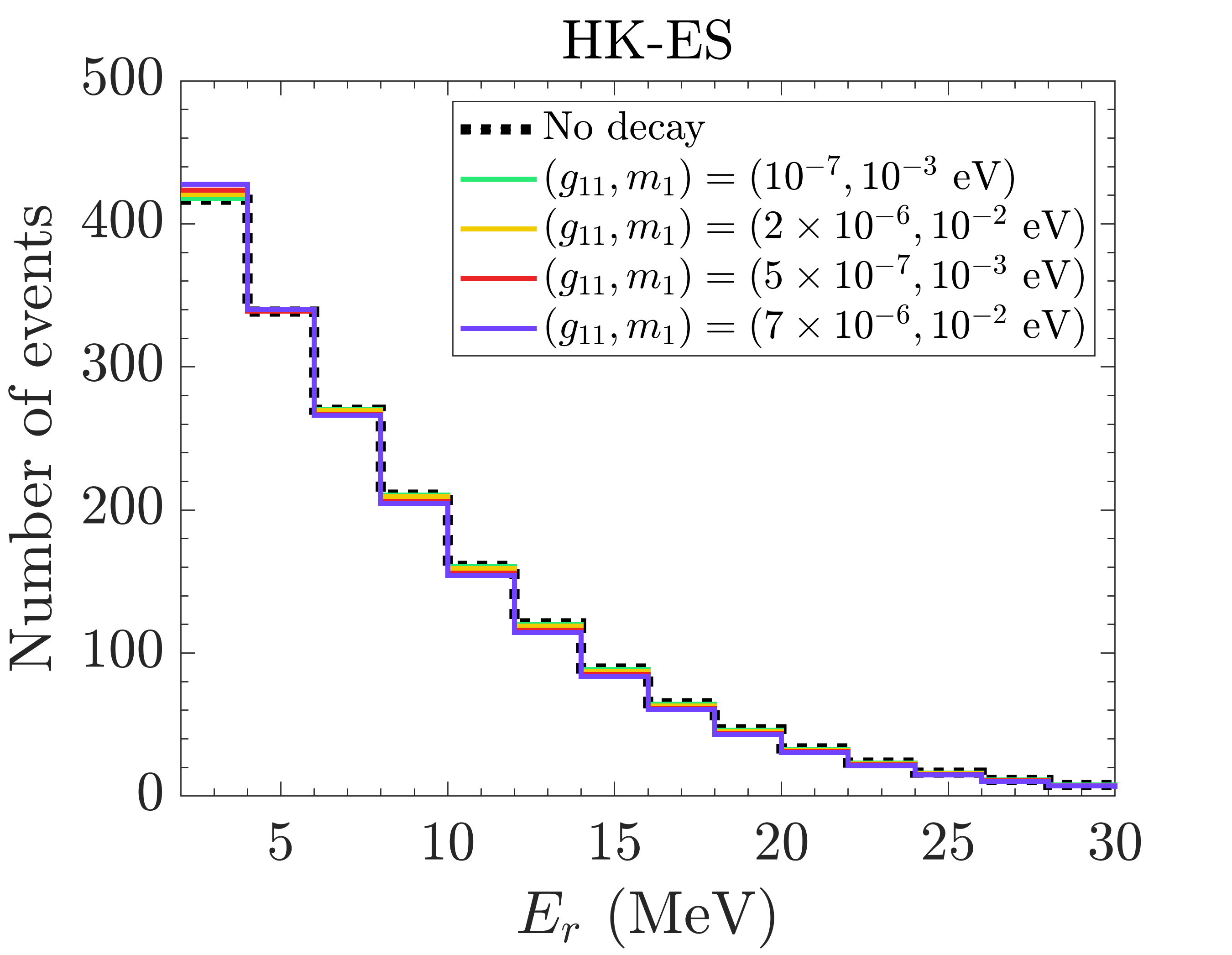}
        \includegraphics[width=0.49\textwidth]{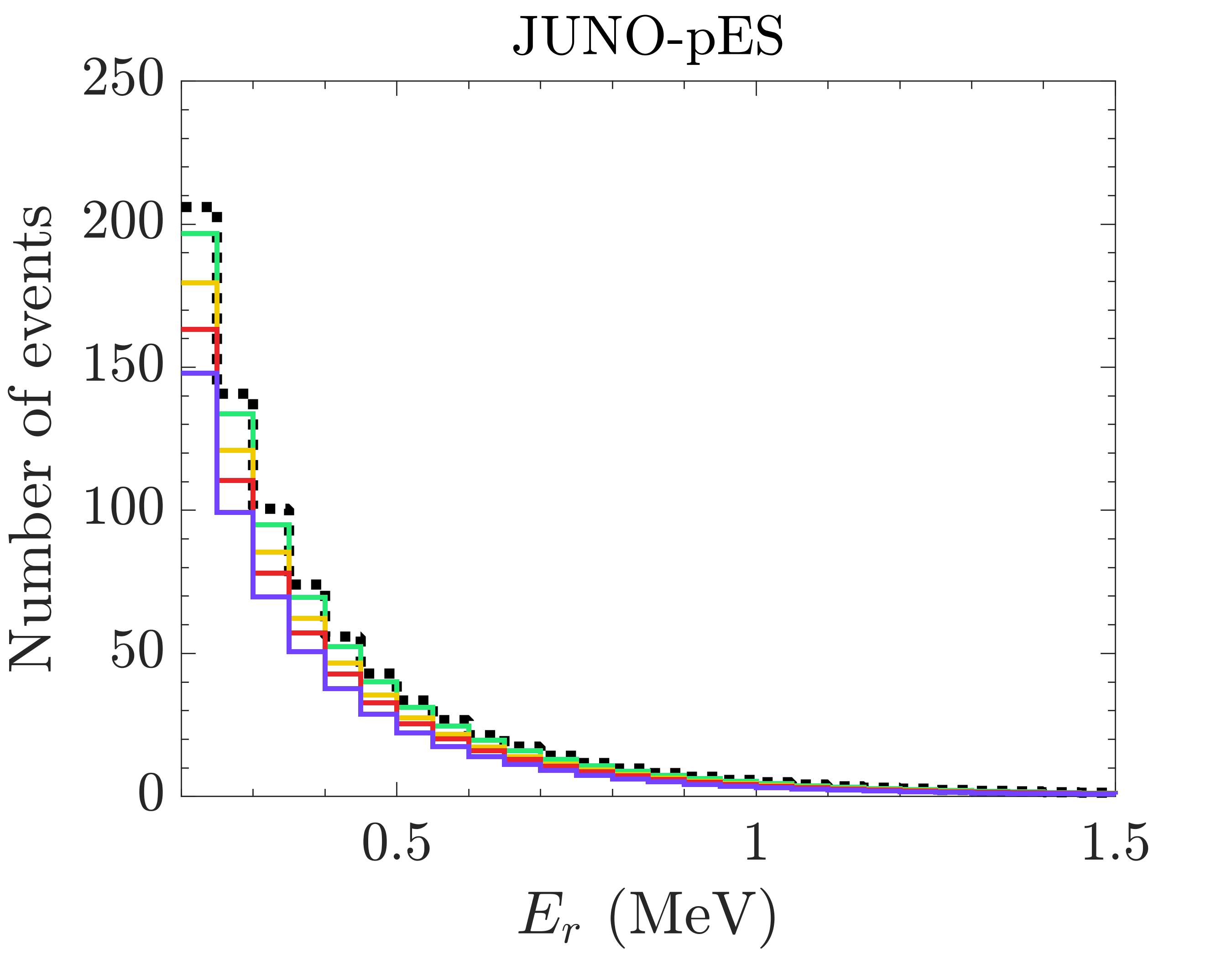}
        \caption{1.44SFHx (NS)}
    \end{subfigure}
    \begin{subfigure}{\textwidth}
        \includegraphics[width=0.49\textwidth]{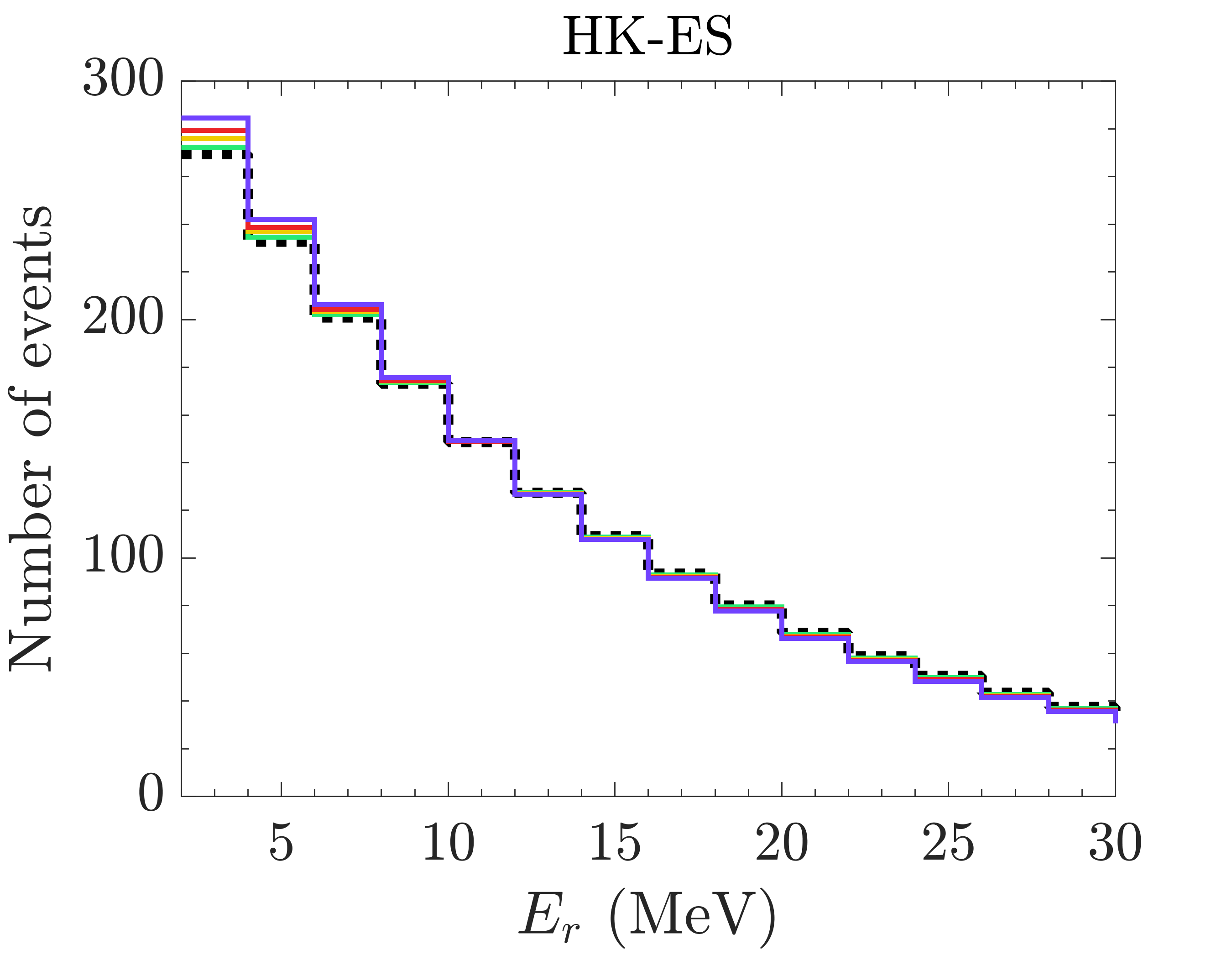}
        \includegraphics[width=0.49\textwidth]{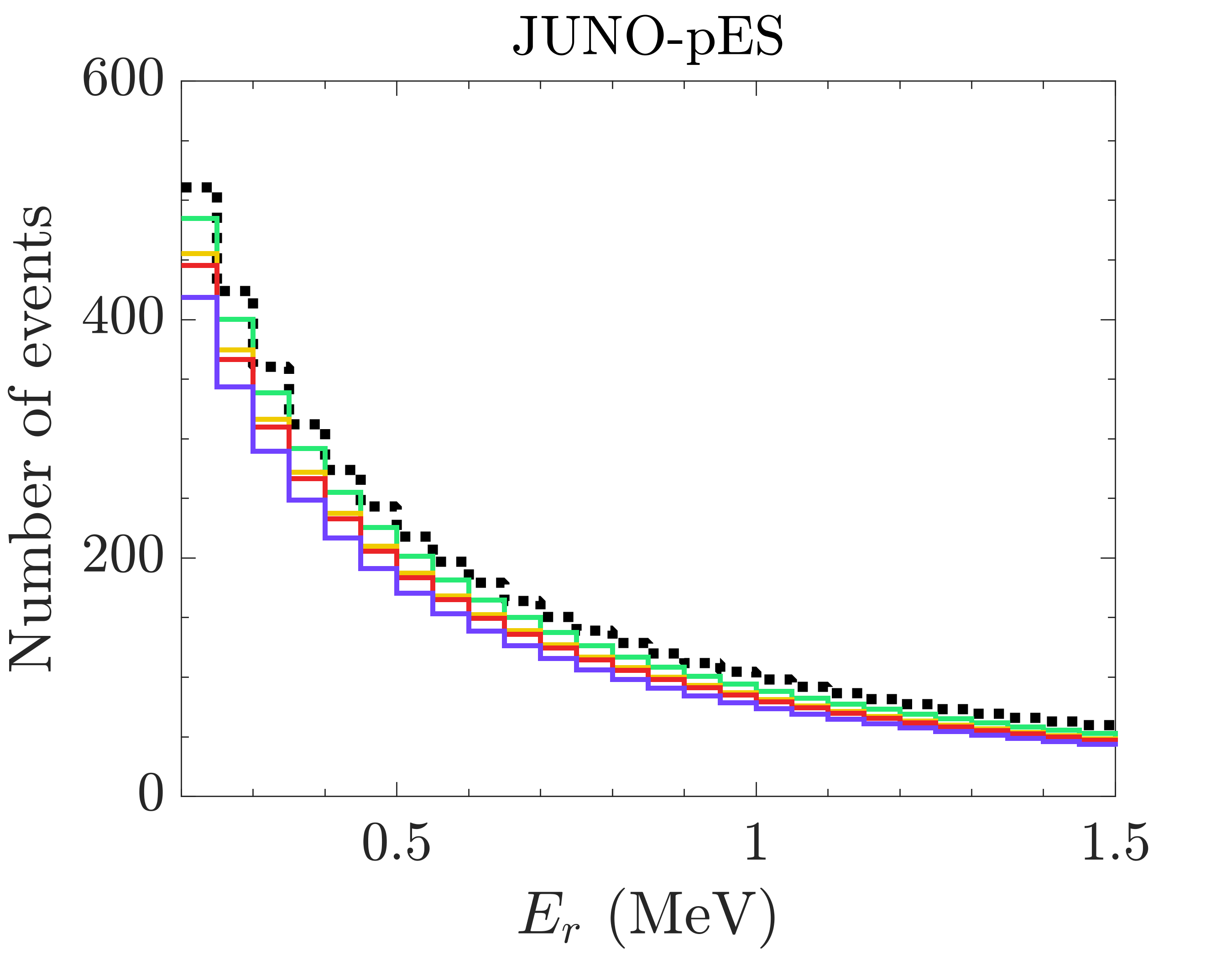}
        \caption{30T (BH)}
    \end{subfigure}
    \caption{Expected events from a future core-collapse supernova located at 10~kpc, assuming normal ordering and including neutrino-Majoron interactions for different values of $g_{11}$ and $m_{1}$ compatible with current constraints. Predictions are given for the elastic scattering signal at HK and the proton elastic scattering signal at JUNO for the NS case using the 1.44SFHx model (top panels) and for the BH case using the 30T model (bottom panels). For comparison, the black dotted line shows the prediction in the absence of neutrino-Majoron interactions.}
    \label{fig:eventsES-pES}
\end{figure}

\begin{figure}[t]
    \centering
    \includegraphics[width=0.49\textwidth]{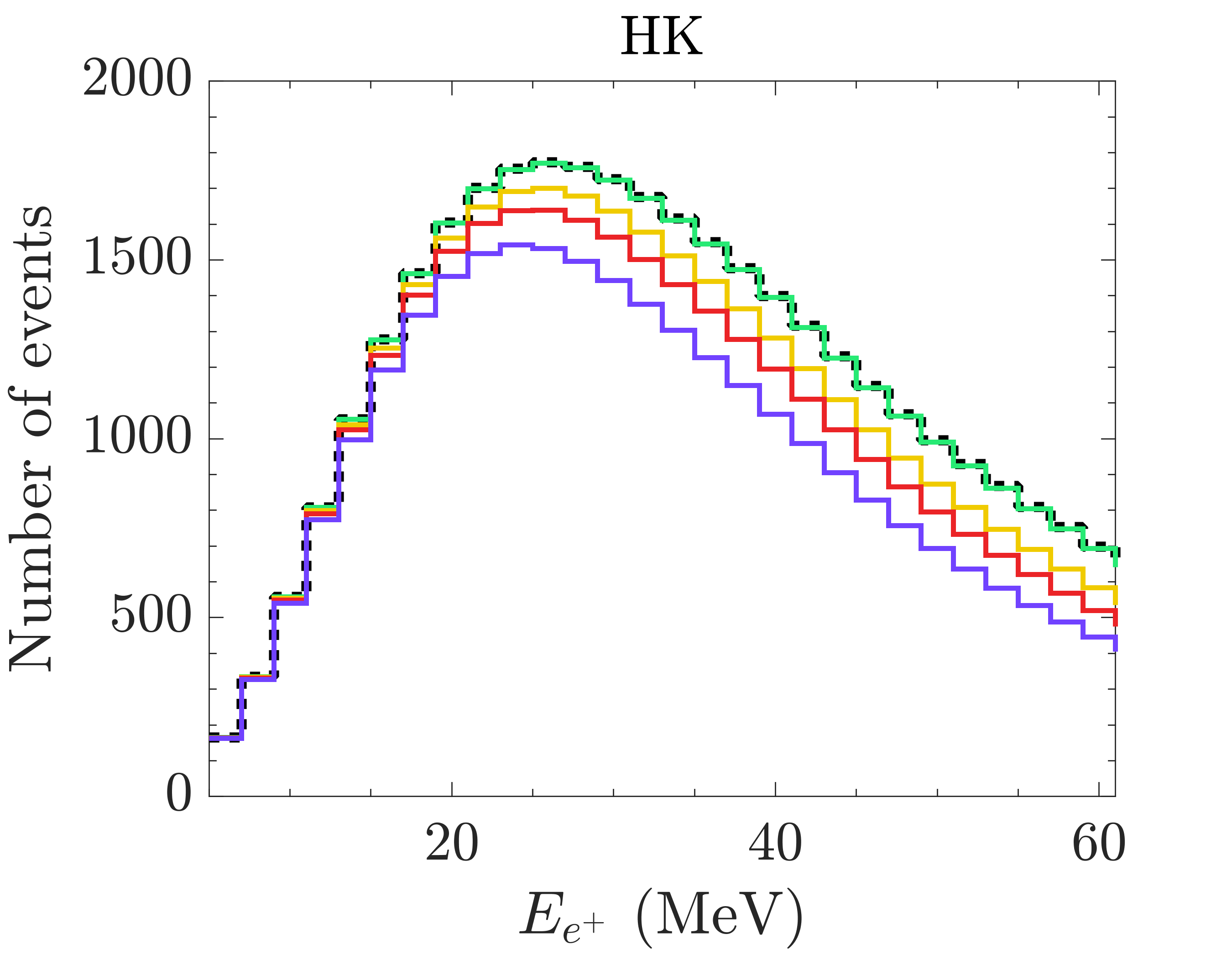}
    \includegraphics[width=0.49\textwidth]{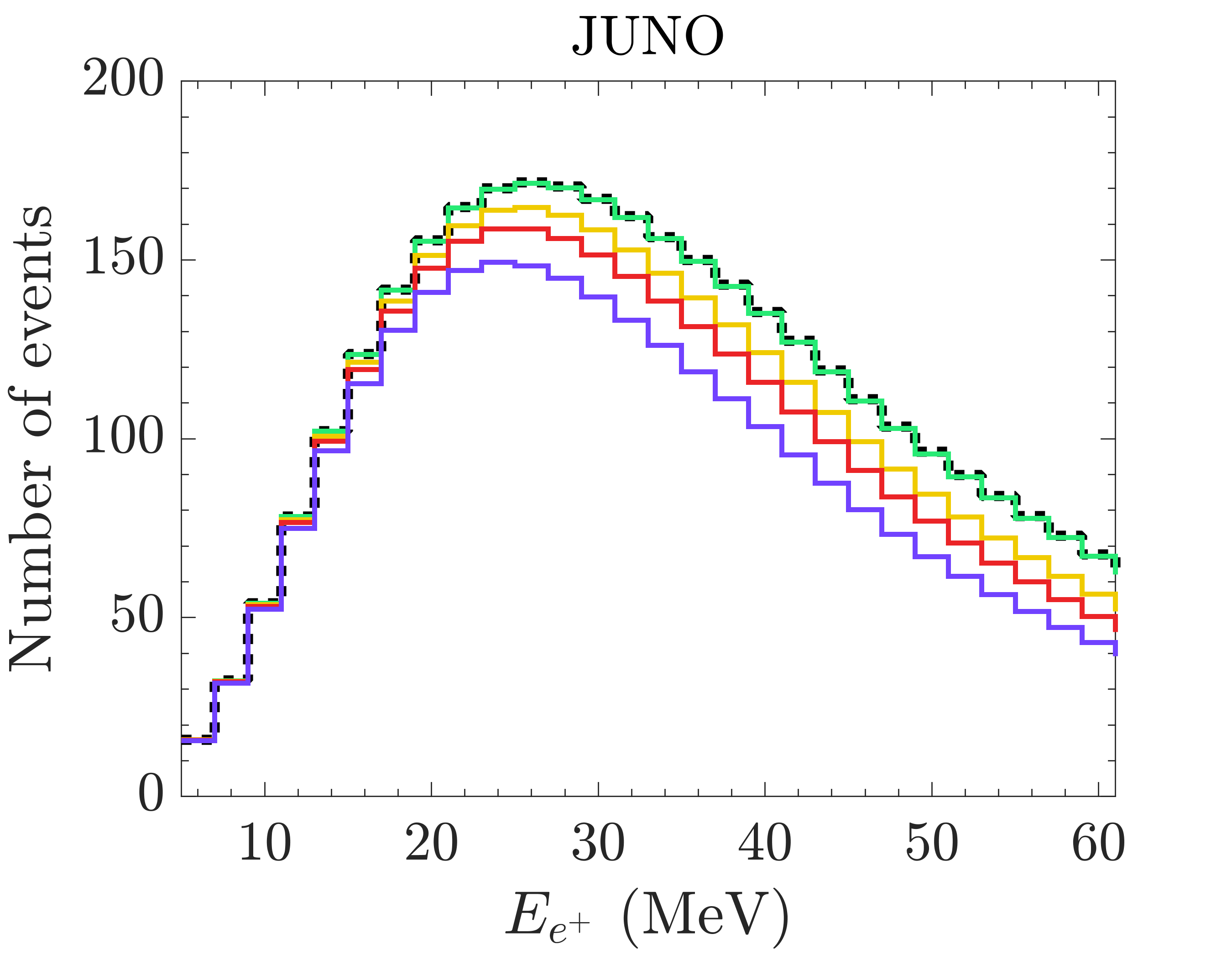}
    \includegraphics[width=0.49\textwidth]{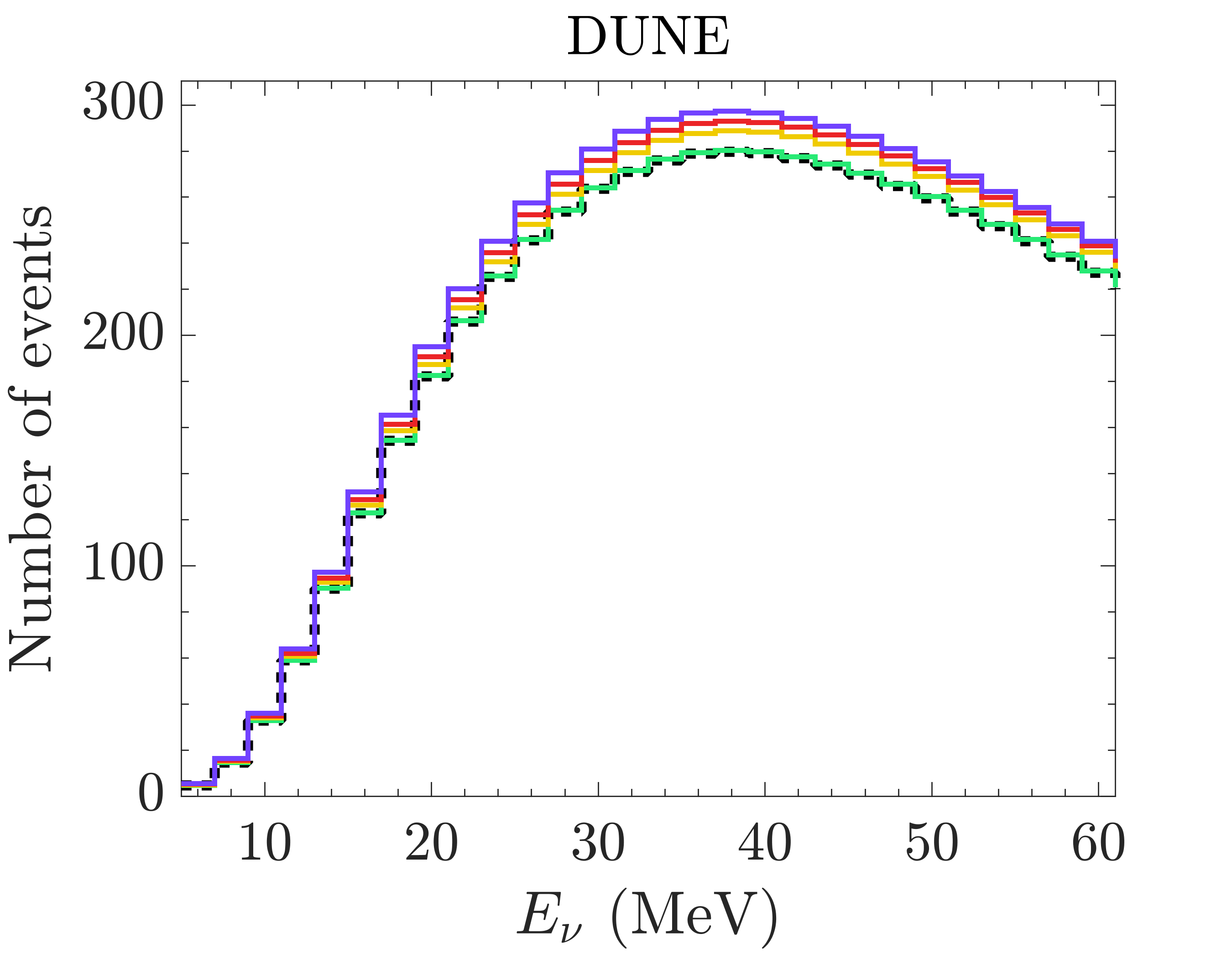}
    \includegraphics[width=0.49\textwidth]{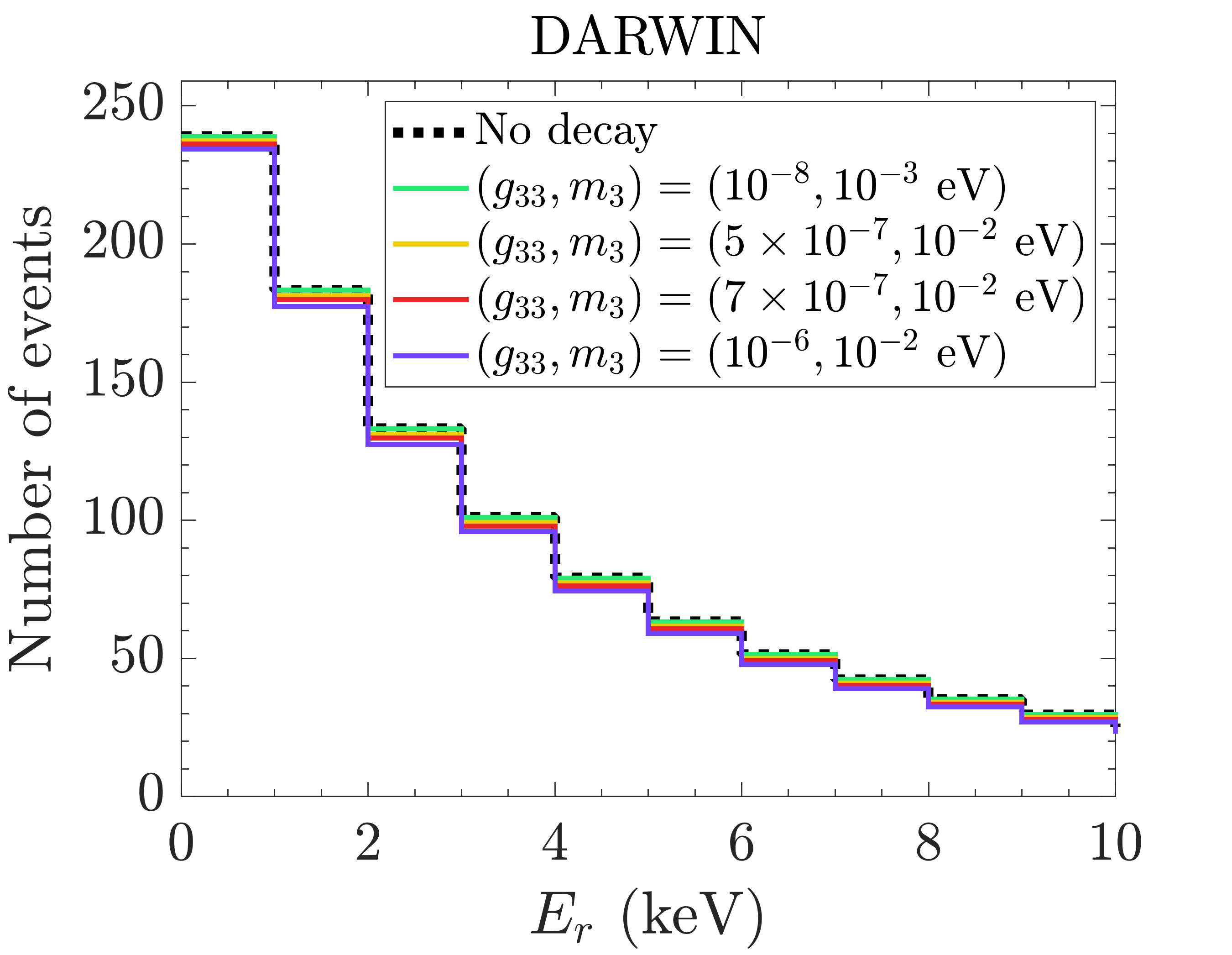}
    \caption{Expected events from a future core-collapse supernova located at 10~kpc, assuming inverted ordering and including neutrino-Majoron interactions for different values of $g_{33}$ and $m_{3}$ compatible with current constraints. For comparison, the black dotted line shows the prediction in the absence of neutrino-Majoron interactions. The results were obtained for the BH case using the 30T model. The predictions correspond to the following cases. Upper figures: inverse beta decay events in HK (left) and JUNO (right).
    Bottom figures: $\nu$-Ar scattering in DUNE (left), coherent neutrino-nucleus scattering in DARWIN (right).}
    \label{fig:eventsIOBH}
\end{figure}

This appendix provides additional results for a future supernova. Figure~\ref{fig:eventsES-pES} shows the expected number of events in normal ordering for the ES channel at HK and the pES channel at JUNO. For inverted ordering, figure~\ref{fig:eventsIOBH} presents the event distributions for the dominant detection channels at HK, JUNO, DUNE, and DARWIN in the BH case. The ES signal at HK and the pES signal at JUNO are omitted in this case, as the impact of neutrino decay on these channels is negligible. Finally, table~\ref{tab:events-IO} summarizes the total expected number of events for all experiments and detection channels assuming inverted ordering.

\begin{table}
\centering
    \setlength{\tabcolsep}{5.3pt} 
        \renewcommand{\arraystretch}{1.2} 
        \begin{tabular}{lrrrrr}
\toprule
        & No decay & \begin{tabular}[c]{@{}c@{}}$g_{33} = 10^{-8}$ \\ $m_3 = 10^{-3}$~eV\end{tabular} & \begin{tabular}[c]{@{}c@{}}$g_{33} = 5\times 10^{-7}$ \\ $m_3 = 10^{-2}$~eV\end{tabular} & \begin{tabular}[c]{@{}c@{}}$g_{33} = 7\times 10^{-7}$ \\ $m_3 = 10^{-2}$~eV\end{tabular} & \begin{tabular}[c]{@{}c@{}}$g_{33} = 10^{-6}$ \\ $m_3 = 10^{-2}$~eV\end{tabular} \\ \midrule
HK-IBD   & \tightcell{27884 \\ (42560)} & \tightcell{27868 \\ (42291)} & \tightcell{27438 \\ (37850)} & \tightcell{27046 \\ (35339)} & \tightcell{26304 \\ (31888)} \\
HK-ES    & \tightcell{1169 \\ (1631)}   & \tightcell{1169 \\ (1631)}   & \tightcell{1168 \\ (1628)}   & \tightcell{1167 \\ (1624)}   & \tightcell{1164 \\ (1617)}   \\
JUNO-IBD & \tightcell{2700 \\ (4120)}   & \tightcell{2698 \\ (4094)}   & \tightcell{2656 \\ (3664)}   & \tightcell{2618 \\ (3421)}   & \tightcell{2547 \\ (3087)}   \\
JUNO-pES & \tightcell{459 \\ (3997)}    & \tightcell{459 \\ (3989)}    & \tightcell{448 \\ (3829)}    & \tightcell{439 \\ (3728)}    & \tightcell{423 \\ (3597)}    \\
DUNE     & \tightcell{2136 \\ (10982)}  & \tightcell{2137 \\ (11018)}  & \tightcell{2146 \\ (11392)}  & \tightcell{2155 \\ (11514)}  & \tightcell{2170 \\ (11631)}  \\
DARWIN   & \tightcell{406 \\ (870)}     & \tightcell{406 \\ (869)}     & \tightcell{402 \\ (845)}     & \tightcell{399 \\ (830)}     & \tightcell{393 \\ (810)}     \\ \bottomrule
\end{tabular}
\caption{Expected total number of events from a future supernova located at 10~kpc, including neutrino-Majoron interactions producing neutrino nonradiative two-body decay in matter. The results are for the case of inverted mass ordering. The first column shows the experiment and the detection channel considered. The second column shows the results in the absence of decay, while the other columns give the expected values for different values of the neutrino-Majoron couplings $g_{33}$ and the lightest neutrino mass $m_{3}$. The values correspond to the NS case with the 1.44SFHx model and to the BH case with the 30T model (in parentheses).}
\label{tab:events-IO}
\end{table}

\bibliographystyle{JHEP}
\bibliography{references}
\end{document}